\documentclass[12pt]{report}
\usepackage{thesis}
\usepackage{graphicx} 
\usepackage[a4paper, total={6in, 8in}]{geometry}
\usepackage{listings}
\usepackage{xcolor}
\usepackage{tabularx}
\usepackage[none]{hyphenat}
\usepackage{microtype}
\usepackage{amsfonts}
\usepackage{subcaption}
\usepackage{multirow}
\newcommand{\apeq}{\mathrel{\vcenter{\offinterlineskip
  \hbox{\tiny$\sim$}
  \hbox{\tiny$=$}}}}
\usepackage[withpage]{acronym}
\usepackage{multirow}
\usepackage{float}
\usepackage{lipsum} 
\usepackage{tikz}
\usetikzlibrary{positioning}
\usepackage{tikz-feynman}
\tikzfeynmanset{compat=1.1.0}
\usepackage{pgfplots}
\pgfplotsset{compat=1.18}
\usepackage{caption} 
\usepackage{amsmath}
\usepackage{slashed} 
\graphicspath{{Images/}}
\usepackage{cite}

\definecolor{codegreen}{rgb}{0,0.6,0}
\definecolor{codegray}{rgb}{0.5,0.5,0.5}
\definecolor{codepurple}{rgb}{0.58,0,0.82}
\definecolor{backcolour}{rgb}{0.95,0.95,0.92}

\lstdefinestyle{mystyle}{
    backgroundcolor=\color{backcolour},
    commentstyle=\color{codegreen},
    keywordstyle=\color{magenta},
    numberstyle=\tiny\color{codegray},
    stringstyle=\color{codepurple},
    basicstyle=\ttfamily\footnotesize,
    breaklines=true,
    numbers=left,
    captionpos=b,
    keepspaces=true,
    showspaces=false,
    showstringspaces=false,
    tabsize=2
}
\begin{document}

\pagenumbering{roman}

\thesistitle
  {One Loop Calculations of Rare B-decays Beyond the Standard Model}
  {Maryam Bibi}
  {Doctor of Philosophy}
  {Department of Physics}
  {October 2025}



\thispagestyle{empty} 
\vspace*{\fill} 

\begin{center}
\Large\textbf{Dedicated }\\[2em]
    \emph{To my beloved parents,\\
    whose unconditional love, support, prayers and sacrifices made this journey possible.}
\end{center}

\vspace*{\fill} 

\vspace*{\fill} 

\addcontentsline{toc}{chapter}{Acknowledgement}
\begin{center}
\textbf{\large Acknowledgements}
\end{center}

First, I would like to thank my supervisors, Prof. Mohammad Ahamdy, Prof. Svetlana Barkanova and Prof. Ruben Sandapen  for valuable guidance, freedom for work, and consistent encouragement throughout my research work. In particular,  Dr. M. Ahmady and Dr. R. Sandapen for deep insights on the subject phenomenology and continuous support and guidance to keep me motivated. I am grateful for their encouragement to explore new topics and to present research at various conferences during my PhD life. 

  Secondly, I would like to extend my gratitude to the entire Physics department of Mount Allison University, they offered me all kind of support and made me feel home, when I was the only graduate student in the department away from my home institution. Most notably; Cathy Pattipas, a fantastic person to be around, helped me and guided me throughout my stay, was one of the reasons I was able to settle down in Sackville.I also want to thank Prof. Ralf Bruening for giving me opportunity to  work as Teaching Assistant  and for the chance of discuss and learn different aspects of physics and its applications. I am grateful to Bob Sorba and Fraser Turner for offering help when I was stuck with the computation of my code. I would also like to thank Turkler for giving me advice from her experience when I needed it the most and Kellie for all her help with the paper work. I am also grateful to all my teachers and mentors who guided me in different stages of my life.

 An Important part of this journey has been the wonderful people I have met along the way. I am truly grateful to my friends, Rasmiya, Kusum, Mahumm, Soumya and Hadiya, who made my life much more enjoyable and were always there when I needed a break from my demanding routine. From traveling together, sharing endless conversations and laughter, to exploring  restaurants and discovering the best cafes, the time we spent together has been one of the most memorable and cherished moments of my life. I also want to express my gratitude to my friend Mirza, whose support and encouragement helped me through some of the most challenging and uncertain times of my academic journey. Their presence made this path easier and reminded me to continue moving forward.
 
 Finally,  I would like to express my deepest gratitude to my family for believing in me and giving me the opportunity to pursue higher studies abroad. My Parents worked tirelessly, overcoming challenges in a broken system, to ensure that my siblings and I could afford an education, and I owe everything to their sacrifices. I am forever grateful to my greatest role model, my mother, whose unconditional love, prayers, and encouragement have been my source of strength throughout this journey. I am also grateful to my sister Samia, who shared my burdens and comforted me in times of homesickness. 
 
 Thank you for always being there for me.

\addcontentsline{toc}{chapter}{Statement of Contribution}
\begin{center}
\textbf{\large Statement of Contribution}
\end{center}

This thesis is the result of independent research conducted by the author as part of a Ph.D. degree at Memorial University of Newfoundland. The author developed the extended theoretical framework incorporating vector-like quark (VLQ) effects into the effective Hamiltonian for rare $B$ decays, performed the analytical derivations of mass diagonalization and extended CKM matrix in Sections 4.1 and 4.2. In Section 4.4, the author calculated the modified Wilson coefficients ($C_7$, $C_9$, $C_{10}$, and $C_L$) and further evaluated the branching ratios for $B \rightarrow K \nu \bar{\nu}$ and $B \rightarrow X_s \nu \bar{\nu}$, and performed a combined analysis using experimental data from $B_s \rightarrow \mu^+ \mu^-$, $B \rightarrow X_s \mu^+ \mu^-$, and $B^+ \rightarrow K^+ \nu \bar{\nu}$.

All numerical routines, including the computation of observables and the  numerical integration of differential branching ratios present in Chapter 5 were performed by the author, and the generation of $\chi^2$ contour plots was written by the author in Python. Figures, diagrams, and plots were created by the author using \texttt{TikZ-Feynman} and \texttt{Matplotlib}.

In summary, the conceptual development, theoretical derivations, coding, numerical analysis, and interpretation of the results presented in Chapters 4 and 5 represents the contribution of author.
\addcontentsline{toc}{chapter}{Abstract}
\begin{center}
\textbf{\large Abstract}
\end{center}

Rare $B$-meson decays provide a sensitive window into potential physics beyond the Standard Model (SM), as they occur only through loop-level processes and are heavily suppressed by the Glashow-Iliopoulos-Maiani (GIM) mechanism. These suppressed transitions, such as the $b \to s$ processes, are particularly useful in the investigation of new particles that may contribute through virtual effects. One compelling possibility is the existence of vector-like quarks (VLQs), which can mix with Standard Model quarks and modify flavor-changing neutral current (FCNC) processes.

VLQs, being singlets under the electroweak symmetry and not requiring symmetry breaking for mass generation, can be much heavier than SM quarks. Although difficult to detect directly at colliders, their effects may be observed indirectly through precise measurements in rare decays. In this work, we explore the influence of down-type iso-singlet VLQs on rare $B$ decays, focusing on decay $B^+ \rightarrow K^+ \nu \bar{\nu}$, which is both theoretically clean and sensitive to new physics.Recent results from the Belle-II collaboration report a branching ratio of $[2.3 \pm 0.5\ (\text{stat})] \times 10^{-5}$~\cite{Belle-II:2023esi}, which provides evidence at the $2.7\sigma$ level and lies significantly above the Standard Model prediction of $[0.45 \pm 0.7] \times 10^{-5}$\cite{Altmannshofer:2009ma}. This discrepancy motivates the study of VLQ contributions to this process.

To constrain the new physics parameter $U_{sb}$, we also examine related decays such as $B_s \rightarrow \mu^+ \mu^-$ and $B \rightarrow X_s \mu^+ \mu^-$. Using these constraints, we compute modified Wilson coefficients ($C_7$, $C_9$, $C_{10}$ and $C_L$) within an effective field theory framework, incorporating VLQ contributions. These coefficients are then used to calculate the branching ratios for $B \rightarrow X_s \nu \bar{\nu}$ and $B \rightarrow K \nu \bar{\nu}$.

Our analysis shows that the inclusion of VLQs can significantly enhance these branching ratios. The resulting parameter space, illustrated by $\chi^2$ contour plots, highlights VLQs as a viable and testable candidate to explain anomalies in rare $B$-meson decays.
\tableofcontents

\addcontentsline{toc}{chapter}{List of Tables}
\listoftables

\addcontentsline{toc}{chapter}{List of Figures}
\listoffigures

\chapter*{List of Acronyms}
\addcontentsline{toc}{chapter}{List of Acronyms}
\begin{acronym}
  \acro{SM}{Standard Model}
  \acro{CKM}{Cabibbo–Kobayashi–Maskawa}
  \acro{VLQ}{Vector-Like Quark}
  \acro{FCNC}{Flavor-Changing Neutral Current}
  \acro{BR}{Branching Ratio}
  \acro{NP}{New Physics}
  \acro{BSM}{Beyond Standard Model}
  \acro{VQM}{Vector Quark Model}
  \acro{GIM}{Glashow–Iliopoulos–Maiani}
  \acro{EFT}{Effective Field Theory}
  \acro{LHC}{Large Hadron Collider}
  \acro{CP}{Charge Parity Invariance}
  \acro{EW}{Electroweak}
  \acro{SSB}{Spontaneous Symmetry Breaking}
  \acro{DR}{Dimensional Regularization}
  \acro{LO}{Leading Order}
  \acro{NLO}{Next to Leading Order}
  \acro{QED}{Quantum Electrodynamics}
  \acro{QCD}{Quantum Chromodynamics}
  \acro{MS}{Minimal Subtraction}
  \acro{MSbar}{Modified Minimal Subtraction} 
  \acro{RGE}{Renormalization Group Equations}
  \acro{OPE}{Operator Product Expansion}
  \acro{CC}{Charged Current}
  \acro{NC}{Neutral Current}
\end{acronym}



\pagenumbering{arabic}

\chapter{Introduction}
\label{chap:intro}

The \ac{SM} is a renormalizable field theory that is mathematically consistent and compatible with most experimental observations. The SM describes the properties of all elementary particles and three of the four fundamental forces in nature: the weak, electromagnetic, and strong forces. Despite its outstanding achievements, some shortcomings preclude recognizing SM as a fundamental theory. For instance, the following unresolved challenges highlight the limitations of the Standard Model. 
\begin{itemize}
    \item \textbf{Matter-Antimatter Asymmetry:} The SM cannot account for the observed dominance of matter over antimatter in the universe, as its sources of CP violation are insufficient to explain the baryon asymmetry.
    
    \item \textbf{Dark Matter and Dark Energy:} These mysterious components, which constitute approximately 95\% of the universe's energy content, are not represented within the SM framework \cite{Farnes:2017gbf}.
    
    \item \textbf{Particle Generations:} The SM does not provide an explanation for the existence of exactly three generations of quarks and leptons.
    
    \item \textbf{Neutrino Masses:} Neutrinos are massless in the SM, yet neutrino oscillation experiments \cite{T2K:2023smv}, have confirmed that they possess small but non-zero masses, necessitating physics beyond the SM.
    
    \item \textbf{Quantum Gravity:} The SM does not include a quantum theory of gravity, leaving a fundamental gap in unifying all known forces.
\end{itemize}

These issues suggest that \ac{NP} exists beyond the Standard Model (SM). Among the proposed extensions, the introduction of the fourth generation of \ac{VLQ} offers a compelling avenue to address the shortcomings of SM \cite{Handoko:1994xw}. The possible fourth generation plays a critical role in understanding the flavor structure of the standard model theory and in resolving anomalies in quark- and lepton-mixing patterns.

VLQs are spin 1/2 particles with the left- and right-handed components defined by the same color and electroweak quantum numbers. Under a given gauge group, their left- and right-handed projections belong to the same representation. Critically, their masses are not tied to electroweak symmetry breaking, allowing them to exist at energy scales far exceeding those of known particles.

VLQs can be much heavier than the SM quarks, since their masses do not require weak symmetry breaking, and couple to SM fermions through Yukawa coupling. Unfortunately, no new particles have yet been found by direct search at the \ac{LHC} which indicates that the energy of these new particles might be too high or their couplings to SM particles might be too weak to be detected by current experiments. Furthermore, VLQs could inferred from indirect searches. VLQs leave detectable imprints on low-energy phenomena, particularly through \ac{FCNC}. In the SM, FCNC processes, e.g., transitions like $b\rightarrow s$ are highly suppressed, occurring only via loop diagrams governed by the \ac{GIM} mechanism. This suppression makes them exceptionally sensitive probes of new physics, as even small contributions from beyond SM particles could yield observable deviations. VLQs, for instance, could induce tree-level FCNCs mediated by the weak Z or Higgs bosons, introduce novel sources of \ac{CP}, or modify loop-level processes with additional Higgs loops in ways that challenge SM predictions. Thus, any experimental evidence for sizable CP-violating effects in the B system would hint at the NP scenario. The FCNC transitions in VLQs contain much fewer parameters and possibly have simultaneously sizeable effects in the K and B meson systems compared to other NP models such as supersymmetry.

Rare B decays are essential probes for physics beyond the Standard Model. These decays are intrinsically rare in the SM because of loop suppression, which makes them acutely sensitive to NP contributions. By analyzing observables like branching ratios and forward-backward asymmetries, we can disentangle potential VLQ effects from SM backgrounds. Such precision studies not only constrain NP parameter spaces but naturally induce non-unitarity in the \ac{CKM} matrix through mixing between SM quarks and VLQs, fundamentally modifying the structure of FCNCs.

In this work, we investigated the impact of down-type VLQs on rare B-decay processes by calculating the branching ratio of $B\rightarrow K \nu^+ \nu^-$, which has shown sizeable deviation from the SM prediction. The branching fraction of the decay $B^+\rightarrow K^+ \nu^+ \nu^-$ from the recent Belle-II results is $[2.3 \pm 0.5(stat)] \times 10^{-5}$ \cite{Belle-II:2023esi}, providing the first evidence at $3.5\sigma$ and combined result with $2.7\sigma$ above the SM expectation $[0.45\pm0.7]\times10^{-5}$ \cite{Altmannshofer:2009ma}. To better understand the contribution of VLQs in FCNC processes, we explored various B meson decays for constraining the NP parameters using $B_s \rightarrow \mu^+ \mu^-$,and $B\rightarrow X_s \mu^+ \mu^-$. Once we constrains the NP parameters we calculated the new Wilson coefficients $(C_7, C_9 , C_{10}, C_L)$ including the contribution from vector like quarks. These Wilson Coefficients are then used to calculate the key physical observables, here branching ratios of $B\rightarrow X_s \nu^+ \nu^-$ and $B\rightarrow K \nu^+ \nu^-$, we aim to identify signatures of the effects of adding the extra generation of iso-singlet down-type VLQs.

Several works in the literature have investigated the impact of vector-like quarks (VLQs) on  FCNCs in rare $B$-meson decays. A brief overview is given below:

\begin{itemize}
    \item An effective theory analysis of VLQ models has been performed with calulation of Wilson coefficients for $b \to s \ell^+ \ell^-$ including the loop-level penguin diagrams \cite{Handoko:1994xw,Morozumi:2018cnc}. They showed how the additional quark modifies the Wilson coefficients relevant for radiative decays. Their results indicated potentially small deviations but were limited to radiative channels only. 

    \item A framework where vector-like quarks generate light quark masseswas discussed in \cite{Botella:2016ibj}. They emphasized tree-level $Z$-mediated FCNCs and explored implications for rare $B$ decays, though loop-level VLQ effects were not systematically included. 

    \item Investigation of new vector-like fermions in flavor physics has been done in \cite{Ishiwata:2015cga}, while focusing on their contributions to $b \to s \ell^+ \ell^-$ and related electroweak penguin processes. Their work highlighted model-independent constraints, but loop-induced VLQ corrections were discussed only qualitatively. 

    \item A model-independent study of $Z$-penguin contributions, with emphasis on $b \to s \ell^+ \ell^-$ transitions has been studied in \cite{Bobeth2017}. They showed strong correlations with electroweak precision tests, but their analysis concentrated on tree-level couplings. 

    \item Inclusive dileptonic rare $B$ decays with an extra generation of VLQs has been explored in \cite{Ahmady:2001qh}. Both penguin and box diagrams were mentioned, the treatment of loop contributions was not considered for the systematic exploration of parameter space. 

    \item Analyzed flavor signatures of an isosinglet down-type VLQ, with a focus on $B_s \to \mu^+ \mu^-$ and $b \to s \ell^+ \ell^-$ in \cite{Alok:2012xm, Alok:2014yua}.They included tree-level $Z$-mediated couplings and found only mild enhancements in branching ratios.
    
    \item The contribution of VLQs to $B$-meson radiative decay ($b \to s\gamma$) has been studied in \cite{Chang:1998sk}. Their work concentrated on radiative penguins and did not extend the analysis to semileptonic FCNC processes such as $b \to s \ell^+\ell^-$.

    \item Updated constraints on quark mixing matrices in the presence of VLQs has been provided in \cite{Vatsyayan:2020jan}. Their analysis placed strong bounds from neutral meson mixing and electroweak observables, but did not address loop-induced rare $B$ processes in detail.
\end{itemize}

From this overview, one observes that the majority of studies have emphasized tree-level $Z$-mediated FCNCs arising from VLQ--SM mixing. Although some work briefly touched on loop contributions (e.g., radiative $b \to s \gamma$ or benchmark penguin diagrams), a comprehensive analysis of VLQ effects in loop-induced processes such as $b \to s \ell^+ \ell^-$ and $b \to s \nu \bar{\nu}$ remains missing. Since in the SM these FCNC transitions originate entirely from loop diagrams, incorporating VLQ contributions at the loop level could significantly alter the branching ratios and enhance the sensitivity of these processes to new physics.

In contrast, our study systematically includes both tree- and loop-level VLQ contributions. By incorporating VLQ effects into loop diagrams, we explore the full potential for branching ratio enhancements in processes such as $b\to s\ell^+\ell^-$ and $B_s\to \mu^+\mu^-$. This comprehensive treatment enables us to identify parameter regions where VLQs could produce significant deviations from SM predictions, offering novel windows into new physics.

In this study, we go beyond previous treatments by including full loop-level contributions from VLQs, demonstrating that they can lead to more pronounced deviations from SM predictions. By quantifying the additional diagrams generated by the VLQs, and their impact on key Wilson coefficients ($C_9$, $C_{10}$, $C_L$), we establish a direct connection between the VLQ parameters and the branching ratios. Next, we find the deviation in the observables like branching ratios for $B\rightarrow X_s \nu^+ \nu^-$ and $B\rightarrow K\nu^+ \nu^-$, final-state mesons for the semileptonic B decays under different BSM scenarios, and compare with LHC and Belle-II results. Our combined analytical and numerical framework reveals how the precision measurements of processes $B_s \rightarrow \mu^+ \mu^-$ and $B\rightarrow K \nu^+ \nu^-$ constrain the parameters of VLQ model, i.e, $U_{sb} = (V_{CKM}V^{\dagger}_
{CKM})_{sb}$ and $\theta_{sb}$ .

This thesis is structured to progressively build the theoretical and phenomenological framework for studying rare $B$-meson decays, with a focus on FCNCs within and beyond the Standard Model. The opening chapter is the Introduction, and the second chapter provides a detailed review of the SM, with emphasis on the electroweak sector and the role of the CKM matrix in flavor transitions. The third chapter introduces FCNCs, highlighting their absence at the tree level due to the GIM mechanism, and examining how such processes arise at the loop level in the SM. This includes a discussion of \ac{EFT} techniques and their application to $b \to s$ transitions, particularly those mediated by the $Z$ boson. The fourth chapter extends the analysis to beyond the SM scenarios by introducing VLQs, detailing their theoretical motivations, interactions, and loop-level contributions to rare $B$ decays through the same $Z$-mediated channel. Both inclusive and exclusive decay modes are considered, and new model parameters are constrained by analytical calculations. The fifth chapter presents a comprehensive numerical analysis in which we employ \texttt{Python} to explore the parameter space, visualize branching ratios, and compare theoretical predictions with experimental data. This structure allows for a coherent investigation of VLQs as potential windows into new physics.

The final chapter presents the conclusion of the thesis and summarizes the analytical and numerical investigation of rare B-decays in the presence of down-type VLQs. The key finding highlights how the extended quark mixing and loop contribution of VLQs in FCNC modifies the branching ratios, offering insights into possible signatures of NP. Evidence of NP would hopefully be a directional guide in order to address some of the most fundamental questions that remain unanswered about our universe.

\chapter{ The Standard Model}
\label{chap:SM}

The SM is the most accurate theory that describes fundamental particles and interactions. The elementary constituents of SM are classified as spin-1/2 fermions, which form all observed matter, and spin-1 gauge bosons, which are the force carriers responsible for interactions between particles. The fermions are further divided into quarks and leptons, each a family of six particles known as flavors, and the flavors are divided into up-type $(u,c,t)$ and down-type $(d,s,b)$ quarks. Quarks can be classified into three generations: the two lightest quarks, $u$ and $d$, comprise the first generation, the $c$ and $s$ quarks – the second one, and finally the two heaviest quarks, $t$ and $b$, enter the third generation given in Table \ref{table:1}. Similarly, there are six types of lepton and they can be classified into three generations (e,$\mu$,$\tau$) with their corresponding neutrinos ($\nu_e$,$\nu_\mu$,$\nu_\tau$). In each generation, the quarks are electrically charged, with up-type quarks having a charge of $Q=+2/3$ and the down-type quarks have $Q=-1/3$. Regarding leptons, they have charge $ Q=-1$ and neutral leptons or neutrinos $Q=0$. Unlike leptons, quarks carry color charge
under the $SU(N_c)$ gauge group in the fundamental representation and are equipped with $N_c = 3$ separate color charges. Thus, the sole diﬀerentiating factor between quark generations is, in fact, their masses.

 All surrounding matter is made of first-generation \(u,d\) particles. Only stable fermions \(u,d\), neutrinos, and electrons are observed in nature, and the rest of the unstable fermions decay into lighter particles. The SM incorporates electromagnetic, strong, and weak sectors, whereas all known natural phenomena can be attributed at the microscopic level to one of these interactions, except gravity. For example, the strong interaction mediates the forces that bind protons and neutrons in the atomic nuclei. The binding of electrons to nuclei in atoms or of atoms in molecules (therefore, the entire variety of chemical phenomena) is caused by electromagnetism. Finally, the radioactive beta decay and the energy production in the Sun involve processes induced by weak interactions. Gravity is not incorporated in the SM; however, compared to the other three forces, the gravitational interaction is so weak at the scale of elementary particles that it can be neglected.

\section{Fundamental Interactions}

The SM of Particle Physics is based on the gauge group $SU(3)_C \times SU(2)_L \times U(1)_Y$. After spontaneous symmetry breaking (SSB), $SU(2)\times U(1)$ is broken to a single unbroken $U(1)_{em}$, symmetry of QED. The quantum numbers of $SU(3)_C$, $SU(2)_L$, and $U(1)_Y$ are called color C, weak isospin I, and hypercharge Y, respectively. The fermion part of the SM involves three families of quarks and leptons.
Each family consists of
\begin{equation}
\begin{aligned}
L\text{-doublets}: &\quad 
Q_L = \begin{pmatrix} u \\ d \end{pmatrix}_L, 
\quad 
\ell_L = \begin{pmatrix} \nu \\ e \end{pmatrix}_L \\[6pt]
R\text{-singlets}: &\quad  u_R, d_R, e_R, \nu_R
\end{aligned}
\end{equation}
where left-handed particles are $SU(2)_L$ doublets, and right-handed particles do not participate in weak interaction and appear as $SU(2)_L$ singlets. The $SU(2)$ subscript L refers to the fact that the $SU(2)$ gauge bosons only couple to left-handed fermions.
\begin{table}[h!]
\centering
\begin{tabularx}{\textwidth}{ |X|c|c|c|c|c|c|c| }
 \hline
 Name & Fields & Content & charge & Spin & $SU(3)_c$ & $SU(2)_L$ & Y\\
 \hline
\multirow{3}{6em}{Quarks (3-generations)}& $Q_L$ & $(u,d)_L$ &(2/3, -1/3) & 1/2 & 3 & 2  & 1/6\\
&$u_R$ & $u_R$ & (2/3) & 1/2 & 3 & 1 & 2/3\\
&$d_R$ & $d_R$ & (-1/3) & 1/2 & 3 & 1 & -1/3\\
\hline
\multirow{2}{6em}{Leptons (3-generations)}& $L_L$ & ($\nu_e , e)_L$ & (0, -1)& 1/2 & 1 & 2 & -1/2\\
&$e_R$ & $e_R$ & -1 & 1/2 & 1 & 1 & -1\\
\hline
\multirow{3}{6em}{Gluons W-Bosons B-Bosons} & $G^a_{\mu}$ & g & 0 & 1 & 8 & 1 & 0\\
&$W^b_{\mu}$ & $W^{\pm}$ , $Z$, $\gamma$& ${\pm}1, 0, 0$ & 1& 1&3&0\\
&$B_{\mu}$& $Z$, $\gamma$ & 0,0 & 1 &1& 1&0\\
\hline
Higgs Boson & ($\Phi^+$, $\Phi^0$)& h & 0 & 0 & 1 & 2 & 1/2\\ \hline
\end{tabularx}
\caption{Particle Content of SM: for each particle the corresponding fields, charge, spin and representation under SM gauge group. The
gauge fields $W^b_\mu$ and $B_\mu$ mix to form the electromagnetically neutral mass eigenstates Z boson and $\gamma$ (photon) after Electroweak symmetry breaking \cite{Buras:2020xsm}}.
\label{table:1}
\end{table}
The $SU(3)_C$ symmetry represents Quantum Chromodynamics (QCD) and the strong interaction. The term $SU(2)_L \times U(1)_Y$ describes the electroweak (EW) interaction, i.e. a combination of weak and electromagnetic interactions.
The symmetries under the transformations of this gauge group determine the interactions and the number of gauge bosons that correspond to the group's generators. (All these gauge bosons have spin 1).

The gauge bosons mediating each of the interactions are:
\begin{itemize}
    \item \textbf{Strong Interaction}.  Eight massless neutral gluons g, each carrying a different combination of color and anti-color.
    \item \textbf{Electromagnetic interaction}. A massless neutral colorless photon ($\gamma$)
    \item \textbf{Weak interaction}. Two massive colorless charged bosons $W^\pm$, and one neutral massive colorless boson $Z$.
\end{itemize}
This interaction is characterized by a coupling constant, and is not precisely constant but depends on the energy scale. 
The strong interaction dominates over the electromagnetic interaction at low energies, exceeding it by approximately two orders of magnitude \cite{Griffiths:1987tj}. Although the coupling constant of the weak interaction is roughly four times larger than that of the electromagnetic interaction, the presence of massive gauge bosons significantly suppresses its effective strength at low energies \cite{Peskin:1995ev}. The effective strength of a force is also distance-dependent: for the strong interaction, the coupling increases with distance due to confinement, whereas for the electromagnetic interaction, the coupling decreases with distance. At the same time, the short-range property of the weak interaction is due to the vast mass of the $W^\pm$ and $Z$ bosons (compared to massless gluons or photons).

The gauge bosons $W^\pm$ and $Z$ are capable of self-interactions, and similarly, gluons carry a non-zero color charge, which allows them to interact with each other. In contrast, photons are electrically neutral and therefore do not self-interact \cite{Griffiths:1987tj}. The Standard Model also predicts a spin-zero particle, the Higgs boson, arising from the Higgs mechanism. This mechanism allows particles to acquire mass through the spontaneous symmetry breaking of the electroweak gauge group.

\section{Gauge Symmetry}

The gauge symmetry ensures that quantum field theories remain renormalizable, preventing unmanageable infinities and preserving the predictive power of theories like the Standard Model \cite{Peskin:1995ev}.

Given a Lagrangian, a global symmetry group describes a class of transformations that leave the Lagrangian invariant. Global symmetries are non-local, meaning they are independent of the coordinates of space and time in the Lagrangian formulation of a theory. Local symmetries are the group, where transformation is coordinate dependent; hence, we can enforce invariance under said local gauge transformation. This procedure achieves local gauge invariance under a symmetry group \cite{Cheng:1984vwu}.

Consider the example of a fermion labeled by a Dirac spinor $\psi$, and mass, m. The Dirac Lagrangian of a free particle,
\begin{equation}
    \mathcal{L_{D}}=\bar{\psi}(i\slashed{\partial}-m)\psi
\end{equation}
where $\bar{\psi}=\psi^{\dagger}\gamma^{0}$, $\slashed{\partial}=\gamma^{\mu}\partial_{\mu}$ and $\gamma^{\mu}$ are Dirac matrices.
The infinitesimal local symmetry group $U(1)$ transformation,
 \begin{equation}
     \psi\rightarrow e^{-U}\psi
     \label{eq2}
\end{equation}
where $U=ir\theta(x)$, $x = (t, r)$ is the spacetime coordinate \cite{Buras:2020xsm}, and 
\begin{equation}
    \partial_\mu \psi = \partial_\mu \left( e^{i\theta(x)r} \psi(x) \right) = e^{i\theta(x)r} \partial_\mu \psi(x) - ir \partial_\mu \theta(x) e^{i\theta(x)r} \psi(x)
    \label{partial}
\end{equation}
With infinitesimal coordinate-dependent parameters, $\theta(x)$ is a phase and $r$ is a parameter that measures the strength of the phase transformation and can then be interpreted as a conserved charge. If $\theta(x)$ is constant, then the Dirac Lagrangian is invariant under the symmetry transformation, called global symmetry. So, by promoting $\theta(x)$ as coordinate dependent, we are now considering a local symmetry. In this case, the Dirac Lagrangian is no longer invariant and transforms as
\begin{equation}
   \mathcal{L^{'}_{D}} \rightarrow \mathcal{L_{D}}+\bar{\psi}(r\partial_\mu\theta(x))\gamma^{\mu}\psi
\end{equation}
The reason why this Lagrangian is not invariant is due to the derivative, which does not transform covariantly under the local gauge transformation given in eq.(\ref{partial}). Thus, we employ Noether’s theorem, which states that the new term appearing under the local transformation
is a conserved current with $\partial_\mu J^{\mu}=0$. In order to get invariant Lagrangian under local U(1) gauge transformation, implies the existence of a new vector field $A_\mu(x)$, defined by its ability to cancel $J^\mu$ once the local gauge transformation is applied. 
The Lagrangian under $U(1)$ transformation is,
\begin{equation}
   \mathcal{L}'_{U(1)} \rightarrow \bar{\psi} \left(i\gamma^\mu \partial_{\mu} - m - r g \gamma^\mu A_{\mu} + r \gamma^\mu \partial_\mu \theta(x)\right) \psi
\end{equation}

We replace the derivative $\partial_\mu$ by a covariant derivative $D_\mu$ , which is given by 
\begin{equation}
    D_\mu = \partial_\mu + ir g A_\mu 
\end{equation}
where $g$ is a real parameter, the gauge coupling that characterizes the strength of the interaction, and $r$ being the charge of $\psi$. In order to satisfy the transformation law in eq.(\ref{partial}), $A_\mu$ has to transform under $U(1)$ as \cite{Buras:2020xsm,Cheng:1984vwu}:
\begin{equation}
   A'_\mu \rightarrow A_\mu + \frac{1}{g} \partial_\mu \theta(x)
    \label{A}
\end{equation}
The vector field, $A_\mu(x)$, is known as a gauge field of the group. For $A_\mu$ to be interpreted as a physical particle, a kinetic term describing the particle's motion has to be added to the original Lagrangian and must be invariant under the transformation in eq.(\ref{A})\cite{Cheng:1984vwu}. The additional term is
\begin{equation}
\mathcal{L}_{\text{gauge}} = -\frac{1}{4} F_{\mu\nu} F^{\mu\nu}, \quad F_{\mu\nu} = \partial_\mu A_\nu - \partial_\nu A_\mu.
\end{equation}
 The new Lagrangian obtained by replacing the partial derivative with a covariant derivative exhibits invariance under local symmetry transformations, giving the total Lagrangian for QED.
\begin{equation}
    \mathcal{L}_{QED} = \bar{\psi} \left( i \slashed{D} - m \right) \psi- \frac{1}{4}F_{\mu\nu} F^{\mu\nu}.
      \label{eq: L}
\end{equation}
where, $\slashed{D}$ is a shorthand for $D_{\mu} \gamma^\mu$.
In the case of a $U(1)$ symmetry, a global symmetry can be promoted to a local one by replacing the ordinary derivative with the covariant derivative and introducing the vector field $A_\mu$. This procedure ensures that the system remains invariant under local gauge transformations.

Now, the transformation of $D_\mu$ for $U(1)$ is given by
\begin{equation}
      D'_\mu \psi' = (\partial_\mu + ir g A'_\mu)\psi' 
\end{equation}
inserting the value of $A_\mu$ and $\psi$, from eqs.(\ref{A}) and (\ref{eq2}),
\begin{align}
    D'_\mu \psi' = (\partial_\mu + ir g (A_\mu + \frac{1}{g} \partial_\mu \theta(x))(e^{-ir\theta(x)}\psi)
    = e^{-ir\theta(x)}(\partial_\mu + ir g A_\mu)\psi
\end{align}
Thus,
\begin{equation}
      D'_\mu \psi' = U(D_{\mu}\psi) 
\end{equation}
Hence, the Lagrangian in eq.(\ref{eq: L}) can be transformed in parts under a local \( U(1) \) gauge transformation involving the vector field \( A_\mu \), and by direct inspection it remains invariant. This invariance under local transformations exemplifies the \textit{gauge principle}, which asserts that fundamental interactions arise from the requirement of local symmetry invariance of the Lagrangian. While this example focuses on the Abelian \( U(1) \) symmetry relevant to electromagnetism, the same principle can be extended to non-Abelian gauge groups, which form the foundation of the Electroweak theory in the Standard Model. In the following section, we explore how the unification of electromagnetic and weak interactions is achieved through the gauge group \( SU(2)_L \times U(1)_Y \), leading to the rich structure of electroweak interactions mediated by the \( W^\pm \), \( Z \), and photon fields.

\section{ Standard Model Electroweak Theory}

The Electroweak (EW) sector of the SM involves all fermions in the SM and not simply quarks, that transform non trivially under $SU(2)_L \times U(1)_Y$. These gauge fields are coupled to the characteristic charges of the electroweak interaction: the weak isospin $I$, with its third component $T_3$, and the weak hypercharge $Y$, related to the electric charge Q via the relation $Y = Q-T_3$.
The left and right representations then transform under $SU(2)_L \times U(1)_Y$ as
\begin{equation}
    \psi_L \rightarrow U_L U_Y \psi_L = \exp[i(\theta_i \tau_i + \rho Y_L \mathbf{1})] \psi_L
\end{equation}
 \begin{equation}
     \psi_R \rightarrow U_Y \psi_R = \exp[i\rho Y_R] \psi_R
 \end{equation}
 with $Y$ corresponding to the weak hypercharge, $\tau_i = \sigma_i/2$ are $SU(2)_L$ group generators, and $\sigma_i$ are Pauli matrices with $i = 1,2,3$. As this is a finite transformation, we take the parameters $\theta_i$ and $\rho$ to be finite. To obtain an invariant EW Lagrangian under such transformations, new massless gauge fields $W_i$ and $B$ are introduced, which are the generators of the electroweak group and transform as
 \begin{equation}
     W_\mu^i \tau_i \rightarrow U_L W_\mu^i \tau_i U_L^\dagger + \frac{1}{g_2} (\partial_\mu U_L) U_L^\dagger 
 \end{equation}
 \begin{equation}
     B_\mu \rightarrow B_\mu - \frac{1}{g_1} \partial_\mu \rho
 \end{equation}
 in which we have introduced new parameters, $g_1$ and $g_2$, which correspond to the couplings of the $U(1)_Y$ and $SU(2)_L$. Moreover, constructing the kinetic terms of the Lagrangian requires the field strength tensors, which are defined as
 \begin{equation}
    W_{\mu \nu}^{i} \rightarrow \partial_{\mu} W_{\nu}^{i} - \partial_{\nu} W_{\mu}^{i} - g_2 \epsilon^{ijk} W_{\mu}^{j} W_{\nu}^{k},
\end{equation}
\begin{equation}
    B_{\mu \nu} \rightarrow \partial_{\mu} B_{\nu} - \partial_{\nu} B_{\mu}.
\end{equation}
Here, $\epsilon_{ijk}$ is the antisymmetric Levi-Chevita tensor and the structure constant of $SU(2)_L$. Now that the gauge fields are defined and their properties understood, we may construct the covariant derivatives for the left- and right-handed spinors,
\begin{equation}
    D_{\mu} \psi_L = \partial_{\mu} + i g_2 W_{\mu}^{i} \tau^{i} + i g_1 Y B_{\mu} \psi_L,
\end{equation}
\begin{equation}
    D_{\mu} \psi_R = \partial_{\mu} + i g_1 Y B_{\mu} \psi_R.
\end{equation}
 The EW Lagrangian can be divided into four parts. 
 \begin{equation*}
     \mathcal{L}_{EW}= \mathcal{L}_f +\mathcal{L}_{g}+\mathcal{L}_{\phi}+\mathcal{L}_{Y}
 \end{equation*}
 which refer to the theory's fermion, gauge, Higgs, and Yukawa sectors.
 The kinetic terms of the SM fermions;
\begin{equation}
    \mathcal{L}^{SM}_{f} =\sum_{i=1,2,3} i\overline{L}_j \slashed{D} L_j+i\overline{e}_{Rj}\slashed{D}e_{Rj}+i\overline{Q}_{j}\slashed{D}Q_j + i\overline{u}_{Rj}\slashed{D}u_{Rj} +i\overline{d}_{Rj}\slashed{D}d_{Rj}
    \label{eq:Lag1}
\end{equation}
where we used the shorthand notation $\slashed {D}= D_{\mu}{\gamma}^{\mu}$. The covariant derivatives are given as \cite{Buras:2020xsm},
\begin{align}
    D_{\mu} &= \partial_{\mu} - ig_{1} Y B_{\mu} - ig_{2} \frac{\sigma^a}{2} W^{a}_{\mu},
    && \text{for LH lepton doublets } L_{j}, \\
    D_{\mu} &= \partial_{\mu} - ig_{1} Y B_{\mu},
    && \text{for RH lepton singlets } e_{Rj}, \\
    D_{\mu} &= \partial_{\mu} - ig_{1} Y B_{\mu} - ig_{2} \frac{\sigma^a}{2} W^{a}_{\mu}
              - ig_{s} \frac{\lambda^a}{2} G^{a}_{\mu},
    && \text{for LH quark doublets } Q_{j}, \\
    D_{\mu} &= \partial_{\mu} - ig_{1} Y B_{\mu} - ig_{s} \frac{\lambda^a}{2} G^{a}_{\mu},
    && \text{for RH quark singlets } u_{Rj}, d_{Rj}.
\end{align}

where $j=1,2,3$ are three generations. $Y$ is the quantum number of hypercharge, $\sigma^a$ are Pauli matrices, $\lambda^a$ are Gell-Mann matrices, $G^{a}_{\mu}$ are the gluon fields, and $g_1$, $g_2$ and $g_s$ are the couplings of $U(1)_Y$, $SU(2)_L$ and $SU(3)_c$ respectively.
The kinetic term of the gauge bosons is
\begin{equation}
    \mathcal{L}^{SM}_{g}= - \frac{1}{4}G_{\mu\nu}^aG^{\mu\nu,a} - \frac{1}{4}W_{\mu\nu}^aW^{\mu\nu,a} - \frac{1}{4}B_{\mu\nu}B^{\mu\nu}
    \label{eq:LagB}
\end{equation}
where,
\begin{equation}
    G^a_{\mu\nu}=\partial_{\mu}G^a_{\nu}-\partial_{\nu}G^a_{\mu}+ g_{s}f^{abc}G^b_{\mu}G^c_{\nu},\quad a,b,c=1,.... 8
    \end{equation}
    \begin{equation}
    W^a_{\mu\nu}=\partial_{\mu}W^a_{\nu}-\partial_{\nu}W^a_{\mu}+ g_{2}\epsilon^{abc}W^b_{\mu}W^c_{\nu},\quad a,b,c= 1....3
    \end{equation}
    \begin{equation}
    B_{\mu\nu}=\partial_{\mu}B_{\nu}-\partial_{\nu}B_{\mu}
\end{equation}
Explicit mass terms for both fermions and gauge bosons are not allowed in the SM because $-m\bar{\psi}\psi= -m(\bar{\psi}_L\psi_R+\bar{\psi}_R\psi_L)$ couples left and right-handed fields, that transform differently under $SU(2)_L$, and $M^2A_\mu^aA^{\mu, a}$ for gauge bosons breaks the gauge symmetry. Thus, from eqs.(\ref{eq:Lag1}) and (\ref{eq:LagB}) alone, we do not get any mass terms because they would break the gauge invariance and spoil the renormalizability. The solution comes from spontaneous symmetry breaking in the SM, first introduced by Weinberg and Salam\cite{Salam:1968rm,Weinberg:1967tq}. 

\subsection{Spontaneous Symmetry Breaking}

 Spontaneous symmetry breaking (SSB) occurs when the Lagrangian of interest is invariant under a symmetry group transformation but contains a vacuum state that is not invariant. In the Weinberg-Salam model \cite{Salam:1968rm,Weinberg:1967tq}, a complex scalar doublet, $\phi(x)$, is included, which transforms non-trivially under the fundamental representation of $SU(2)_L \times U(1)_Y$ with weak hypercharge, $Y = \frac{1}{2}$.
 
\begin{figure}[h!]
    \centering
    \includegraphics[width=0.9\textwidth]{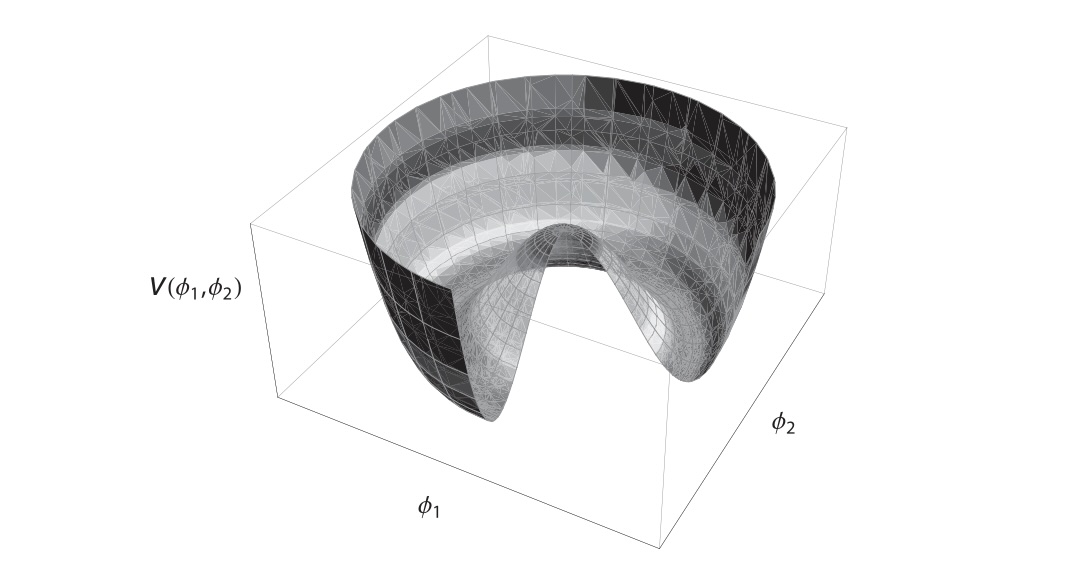}
    \centering
    \caption{Potential for a complex scalar field with $\mu^2 < 0$\cite{Buras:2020xsm}}.
    \label{fig: Mexican Hat}
\end{figure}

Consider a U(1) gauge theory with a single complex scalar field $\phi$, with
\begin{equation}
\mathcal{L} = \frac{1}{2}\partial_\mu \phi^* \, \partial^\mu \phi - V(\phi^*, \phi)
\label{LagV}
\end{equation}
where the potential is given by
\begin{equation}
V(\phi^*, \phi) = \mu^2 \phi^* \phi + \frac{\lambda}{4} (\phi^* \phi)^2
\end{equation}
where \( \lambda > 0 \).
This Lagrangian is invariant under the \( U(1) \) global symmetry transformation:
\begin{align*}
\phi &\to e^{i\theta} \phi, \\
\phi^* &\to \phi^* e^{-i\theta}
\end{align*}
For $\mu^2 > 0$, $V (\phi)$ has a unique minimum at $\phi = 0$, which is the ground state of the theory and is symmetric under gauge theory.
For $\mu^2 < 0$, this symmetry is spontaneously broken and acquired by vacuum expectation value ($vev$),
\begin{equation}
\langle 0 | \phi | 0 \rangle = \frac{\nu}{\sqrt{2}}, \quad \nu = \sqrt{\frac{-2\mu^2}{\lambda}}
\end{equation}
The resulting potential will look like a Mexican hat in Fig.(\ref{fig: Mexican Hat}). To have a ground state, we will choose one of the vacua
\begin{equation}
|\phi_{\text{vac}}|^2 =  \frac{- 2 \mu^2}{\lambda}\equiv \frac{\nu^2}{2}
\end{equation}
and this spontaneously breaks the U(1) symmetry. Let us find the mass spectrum after SSB. The complex scalar field,
\begin{equation}
\phi(x) = \frac{1}{\sqrt{2}}(\phi_1(x) + i\phi_2(x))
\end{equation}
with $\phi_1$ and $\phi_2$ being real. We  choose the vacuum to be
\begin{equation}
(\phi_1 , \phi_2)_{\text{vac}} = (\nu , 0)
\end{equation}
and we introduce $\eta$ and $\xi$ real fields, which describe the fluctuations around the vacuum,
\begin{equation}
    \phi(x) = \frac{1}{\sqrt{2}}(v + \eta(x) + i \xi(x))
\end{equation}
Inserting into the Lagrangian in eq.(\ref{LagV}) we get,
\begin{equation}
    \mathcal{L} = \frac{1}{2}(\partial_\mu \eta)(\partial^\mu \eta) - \frac{1}{2} m^2 \eta^2 + \frac{1}{2}(\partial_\mu \xi)(\partial^\mu \xi) + \text{interactions}
\end{equation}
\begin{equation}
m^2 _ \eta = \frac{\partial^2 V}{\partial \phi_1^2} \bigg|_{(v, 0)} = -2\mu^2
\end{equation}
\begin{equation}
m^2 _ \xi = \frac{\partial^2 V}{\partial \phi_2^2} \bigg|_{(v, 0)} = 0
\end{equation}
 The appearance of a massless particle can easily be understood by noting that the potential V is flat in the $\phi_2$ direction; thus, we have a massless particle. This massless particle is called the Goldstone boson \cite{Goldstone:1962es}. In the $\phi_1$ direction, the potential is not flat, and it costs some energy to move along it, so the particle $\eta$ has a mass.

 This important result of the appearance of a massless particle as the consequence of a spontaneous breakdown of a continuous global symmetry is a special case of the Goldstone theorem, which states that for each broken symmetry there is one Goldstone boson. As the $U(1)$ symmetry has only one generator, we have only one massless boson. For a global symmetry, the spontaneous symmetry breaking would give rise to massless states (Goldstone theorem), but in the case of a local (gauge) symmetry this gives masses to gauge bosons through the Higgs mechanism.

\subsection{Higgs Mechanism}

The local gauge symmetries on which the SM are built require that all fundamental particles be massless. However, as this is not the case in nature, the electroweak gauge symmetry must be spontaneously broken. This is achieved by the Higgs Mechanism \cite{Higgs:1964pj}, which proposes the existence of a scalar Higgs field which has a non-zero vacuum state. This additional field causes the electroweak neutral gauge bosons to mix with each other through a rotation of the weak-mixing angle ($\theta_W$) resulting in four mass eigenstates: one neutral and massless (the photon), one neutral and massive ($Z$), and two charged and massive ($W^\pm$) gauge bosons.

The gauged Lagrangian for a complex scalar field with $U(1)$ symmetry is
\begin{equation}
    \mathcal{L}_{\text{gauged}} = (D_\mu \phi)^\dagger D^\mu \phi - V(\phi) - \frac{1}{4} F_{\mu \nu} F^{\mu \nu}
\end{equation}
where the covariant derivative is $D_\mu = \partial_\mu + i g A_\mu$, and the potential is
\begin{equation}
    V(\phi) = -\mu^2 \phi^\dagger \phi + \lambda (\phi^\dagger \phi)^2, \quad \mu^2, \lambda > 0.
\end{equation}
This Lagrangian is invariant under simultaneous $U(1)$ transformations of $\phi$ and $A_\mu$ \cite{Buras:2020xsm}:
\begin{equation}
    \phi \rightarrow \phi e^{i \theta(x)}, \quad A_\mu \rightarrow A_\mu + \frac{1}{e} \partial_\mu \theta(x)
\end{equation}
After SSB, we can parameterize the field as \cite{Peskin:1997ez}
\begin{equation}
    \phi(x) = \frac{1}{\sqrt{2}} [v + H(x) + i \chi(x)] = \frac{1}{\sqrt{2}} (v + H(x))\, e^{i \chi(x)/v}
\end{equation}
where $v$ is the vacuum expectation value, $H(x)$ is the physical Higgs field, and $\chi(x)$ is the Goldstone boson.

The covariant derivative becomes
\begin{equation}
    D_\mu \phi = \frac{1}{\sqrt{2}} \left[ \partial_\mu H - i g (v + H) A_\mu - i (v + H) \frac{\partial_\mu \chi}{v} \right] e^{i \chi(x)/v}
\end{equation}

By performing a gauge transformation with $\theta(x) = -\chi(x)/v$, the Goldstone boson $\chi(x)$ is ``gauged away'', and the field simplifies to
\begin{equation}
    \phi'(x) = \frac{1}{\sqrt{2}} (v + H(x))
\end{equation}
with the gauge field shifted as $A_\mu' = A_\mu - \frac{1}{g v} \partial_\mu \chi(x)$.

The Lagrangian in terms of the physical fields is,
\begin{equation}
    \mathcal{L}_{\text{gauged}} = \frac{1}{2} \partial_\mu H\, \partial^\mu H - \frac{1}{4} F_{\mu \nu} F^{\mu \nu} - \frac{1}{2} M_H^2 H^2 + \frac{1}{2} m_A^2 A_\mu A^\mu + \cdots
\end{equation}
where
\begin{equation}
    M_A^2 = g^2 v^2, \quad M_H^2 = 2 \lambda v^2
\end{equation}
The gauge boson is now massive. Its mass depends on the gauge coupling $g$ and $v$.

\subsection*{Higgs Mechanism in the Standard Model}

In the Standard Model, the Higgs field is responsible for the breakdown of the electroweak to the electromagnetic gauge symmetry $SU(2)_L \times U(1)_Y \rightarrow U(1)_{em}$. The masses of the $Z$ and $W^\pm$ gauge bosons are acquired through symmetry breaking, by the interaction with the Higgs field. The Higgs field is an $SU(2)_L$ doublet with hypercharge $Y=1/2$:
\begin{equation}
    \Phi = \begin{pmatrix} \phi^+ \\ \phi^0 \end{pmatrix}
\end{equation}
After SSB, the neutral component is expanded as
\begin{equation}
    \phi^0(x) = \frac{1}{\sqrt{2}} [v + H(x) + i \chi^0(x)]
\end{equation}
with four real fields Higgs, each corresponding to degrees of freedom of the Higgs doublet. $H(x)$ is a gauge-invariant fluctuation of the vacuum state and corresponds to the physical Higgs field. The three remaining degrees of freedom are not gauge invariant, $\phi^+(x) = \chi^+(x)$, $\chi^{-}(x)=(\chi^+(x))^\dagger$ where $\chi^\pm$ and $\chi^0$ are Goldstone bosons \cite{Goldstone:1962es}. Due to the local $SU(2)_L$ gauge invariance of the Lagrangian,  choosing the unitarity gauge \cite{Weinberg:1973ew}, one can “rotate away” these fields.

The Higgs potential is
\begin{equation}
    V(\Phi) = -\mu^2 \Phi^\dagger \Phi + \lambda (\Phi^\dagger \Phi)^2
\end{equation}
Here $\mu^2 > 0$, the Lagrangian of the theory is invariant under the symmetry transformation, but the ground state is not. Hence, the vacuum expectation value is chosen as
\begin{equation}
    \langle 0 | \Phi | 0 \rangle = \frac{1}{\sqrt{2}} \begin{pmatrix} 0 \\ v \end{pmatrix}, \qquad v = \sqrt{\frac{\mu^2}{\lambda}}
\end{equation}
Bosonic and Higgs part of the SM Lagrangian:
\begin{equation}
    \mathcal{L}_\phi = -\frac{1}{4} W_{\mu \nu}^i W^{\mu \nu, i} 
- \frac{1}{4} B_{\mu \nu} B^{\mu \nu} 
+ (D_\mu \Phi)^\dagger(D^\mu \Phi) - V(\Phi)
\end{equation}
\begin{equation}
\mathcal{L}_\phi = \frac{g_2^2 v^2}{4} \left( (W_\mu^1)^2 + (W_\mu^2)^2 + (\frac{g_1}{g_2} B_\mu - W_\mu^3)^2 \right)
\end{equation}
The covariant derivative is
\begin{equation}
    D_\mu = \partial_\mu - i \frac{g_2}{2} \sigma^a W^a_\mu - i \frac{g_1}{2} B_\mu
\end{equation}
where $g_1$ and $g_2$ are the $U(1)_Y$ and $SU(2)_L$ couplings, and $\sigma^a$ are the Pauli matrices.
The kinetic term yields mass terms for the gauge bosons $ W$ and $Z$,
\begin{equation}
\mathcal{L}_{\text{mass}} = \frac{1}{2} m_W^2 W^+_\mu W^{-\mu} + \frac{1}{2} m_Z^2 Z_\mu Z^\mu
\end{equation}

The second derivative of the potential provides the mass squared parameter for the Higgs mass $m_H= \sqrt{2\lambda}\, v$ and shows that one field becomes massive while others remain massless. The physical fields are 
\begin{align}
    W^\pm_\mu &= \frac{1}{\sqrt{2}} (W^1_\mu \mp i W^2_\mu)
\end{align}
The diagonalization of the mass matrix of the fields $W^3$ and B provides two new physical fields. Since $U(1)_{em}$ is unbroken these mass eigenstates identified as massless Photon $A_{\mu}$ and massive one $Z_{\mu}$,
\begin{align}
    \begin{pmatrix} Z_\mu \\ A_\mu \end{pmatrix} &= \begin{pmatrix} \cos\theta_W & -\sin\theta_W \\ \sin\theta_W & \cos\theta_W \end{pmatrix} \begin{pmatrix} W^3_\mu \\ B_\mu \end{pmatrix}
\end{align}
where the Weinberg angle is defined by
\begin{equation}
    \tan\theta_W = \frac{g_1}{g_2}, \qquad e = g_2 \sin\theta_W = g_1 \cos\theta_W
\end{equation}
 such that $\sin^2 {\theta_W} = 0.231$. From the four generators of the $SU(2)_L \times U(1)_Y$ gauge symmetry, three are spontaneously broken, which will grant a mass to three ($Z, W^\pm$) of the four physical EW gauge bosons; three degrees of freedom of the Higgs doublet will now correspond to the longitudinal polarization of the massive gauge fields. The remaining $U(1)_{em}$ symmetry is associated with the fourth unbroken generator and the corresponding gauge field, the photon ($\gamma$), which will thus remain massless. The masses of the Higgs and guage bosons at tree level are given as
\begin{align}
    M_W &= \frac{1}{2} g_2 v \\
    M_Z &= \frac{1}{2} \sqrt{g_1^2 + g_2^2}\, v \\
    M_A &= 0 \\
    M_H &= \sqrt{2\lambda}\, v
\end{align}
The redefinition of $W^\pm$ and $Z$ fields 
\begin{equation}
    W^\pm _\mu \mp \frac{i}{M_W} \partial_\mu \chi^\pm \rightarrow W^\pm
\end{equation}
\begin{equation}
    Z _\mu  - \frac{1}{M_Z} \partial_\mu \chi^0 \rightarrow Z _\mu 
\end{equation}
Expanding $W^i_{\mu\nu}W^{\mu\nu,i}$ in Lagrangian, $W^\pm$ exhibits electromagnetic charge of $\pm1$. The mass of $W^\pm$ is related to $Z$
\begin{equation}
    \frac{M_W}{M_Z} = \cos\theta_W
\end{equation}
The Fermi constant relates to the vev as
\begin{equation}
    G_F = \frac{1}{\sqrt{2} v^2} \implies v = (\sqrt{2} G_F)^{-1/2} 
\end{equation}
and
\begin{equation}
     \frac{G_F}{\sqrt{2}}=\frac{g_2^2}{8M_W^2}
\end{equation}
In contrast to the photon, the mediators of weak interactions are massive: $M_{W^\pm} = (80.379 \pm 0.012)\text{GeV}/c^2$, and $M_Z= 91.18760.0021\text{GeV}/c^2$ \cite{ParticleDataGroup:2024cfk}.  The Higgs was discovered at the LHC in 2012, with mass $M_H \approx125 GeV$, completing the experimental verification of the standard model . The $\frac{v}{2}$ arises from the Higgs field $vev$ is \cite{ParticleDataGroup:2024cfk}. 
\begin{equation}
v = (\sqrt{2} G_F)^{-1/2} \approx 246\, \text{GeV}
\end{equation}

\section{Flavor Structure of Standard Model}

\subsection{Rotation from Flavor to Mass Eigenstates}
The SM Yukawa sector is responsible for generating the fermion masses by the rotation of flavor to mass eigenstates and  after rotation it contains free parameters: six quark masses, three charged lepton masses, three mixing angles, and one phase \cite{Buras:2020xsm}.

\begin{equation}
\mathcal{L}_Y = Y^d \bar{d}_L \frac{v}{\sqrt{2}} d_R 
+ Y^u \bar{u}_L \frac{v}{\sqrt{2}} u_R 
+ Y^l \bar{l}_L \phi \frac{v}{\sqrt{2}} e_R + \text{h.c.} + \cdots
\end{equation}

\begin{equation}
    Q_L \rightarrow A_L^u Q_L,\hspace{0.5cm} u_R \rightarrow A_R^u u_R,\hspace{0.5cm}d_R \rightarrow A_R^d d_R \label{eq:Trans1}
\end{equation}
    \begin{equation}
L \rightarrow A_L^eL,\hspace{0.5cm} e_R \rightarrow A_R^ee_R\label{eq:Trans2}
\end{equation}
where $A_{L,R}^i$ are unitary 3×3 matrices. 
Generally, Yukawa matrices $Y^d, Y^u, Y^l$ can be diagonalized through transformations:
\begin{equation}
    (A_L^d)^\dagger Y_DA_R^d = \hat{Y}^D,\hspace{0.2cm} (A_L^u)^\dagger Y_UA_R^u = \hat{Y}^U, \hspace{0.2cm} (A_L^e)^\dagger Y_EA_R^e = \hat{Y}^E.\label{eq: Yuk T}
\end{equation}
Because left-handed fields are embedded into doublets $Q$ and $L$, we have to rotate the members of the doublets with the same matrix, i.e, $u_L$,$d_L$ $v_L$ and $e_L$ are all rotated the same way. $Q_L$ rotates with $A_L^u$ and $L$ with $A_L^e$ as written in eqs.(\ref{eq:Trans1}) and (\ref{eq:Trans2}). Once this is done, these rotations will be fixed. As $A_L^d$ is missing in eq.(\ref{eq:Trans1}), the matrices $Y^D$ and $Y^U$ cannot be simultaneously diagonalized by rotations, which leaves the interaction Lagrangian invariant. After these rotations, we get
\begin{align}
\mathcal{L}_Y &= -\overline{Q}\phi(A_L^u)^\dagger(A_L^d)\hat{Y}^Dd_R - \overline{Q}\phi^c\hat{Y}^Uu_R - \overline{L}\phi\hat{Y}^Ee_R + \text{h.c.}
\label{eq:LagYR}
\end{align}
with the first term being non-diagonal. We must perform an additional rotation of the down quarks, to diagonalize this term and consequently get the mass eigenstate basis, 
\begin{align}
d_L &\rightarrow d_L' = (A_L^u)^\dagger A_L^d d_L
\end{align}
where $d_L$ are the original flavor eigenstates and $d'_L$ the mass eigenstates. The matrix $(A_L^u)^\dagger A_L^d$ is just the CKM matrix, named after Cabibbo, Kobayashi, and Maskawa \cite{cabibbo1963,kobayashi1973} and is responsible for the flavor transition in the SM
\begin{align}
 V_{CKM} = (A_L^u)^\dagger A_L^d\\
 d_L'=V_{CKM} d_L
 \label{Vckm}
\end{align}

Using the CKM matrix, the relation between flavor and mass eigenstates
\begin{equation}
    \begin{pmatrix}
        d_L'\\
        s_L'\\
        b_L'\\
    \end{pmatrix}=\begin{pmatrix}
        V_{ud} &V_{us}& V_{ub} \\
        V_{cd} &V_{cs}& V_{cb} \\
        V_{td}& V_{ts}& V_{tb}\\ 
    \end{pmatrix}\begin{pmatrix}
         d_L\\
        s_L\\
        b_L\\
    \end{pmatrix}
\end{equation}
Here, we resolved $d_L$ into its three components $(d, s, b)_L$, the same as we can do $u_L = (u, c, t)_L$ and the corresponding right-handed fields. The mass eigenstate is where all Yukawa couplings are simultaneously diagonal. For convenience, one can define mass matrices, which absorb Yukawa couplings, the $vev$, and a constant factor $\frac{1}{\sqrt{2}}$. By expanding the Higgs field around non zero vaccum, from eq(\ref{eq:LagYR}) we get diagonal $3\times3$ mass matrices:
\begin{align}
\hat{M}^U &= \text{diag}(m_u, m_c, m_t) = \frac{v}{\sqrt{2}} \hat{Y}^U = \frac{v}{\sqrt{2}} \cdot \text{diag}(y_u, y_c, y_t)
\end{align}
\begin{align}
\hat{M}^D &= \text{diag}(m_d, m_s, m_b) = \frac{v}{\sqrt{2}} \hat{Y}^D = \frac{v}{\sqrt{2}} \cdot \text{diag}(y_d, y_s, y_b)
\end{align}
\begin{align}
\hat{M}^E &= \text{diag}(m_e, m_\mu, m_\tau) = \frac{v}{\sqrt{2}} \hat{Y}^E = \frac{v}{\sqrt{2}} \cdot \text{diag}(y_e, y_\mu, y_\tau)
\end{align}

\subsection{Electroweak Interactions}

The Lagrangian of electroweak interactions is 
\begin{equation}
    \mathcal{L}^{EW}_{int} =\mathcal{L}_{CC}  +\mathcal{L}_{NC}  
\end{equation}
with $\mathcal{L}_{CC}$ and $\mathcal{L}_{NC}$ are charged and neutral current interactions, respectively. 

\subsubsection{Weak Charged Currents}

The redefinition of the fields also implies changes in the interaction part of the SM Lagrangian. This naturally describes a weak interaction between quarks from different families. The couplings of charged gauge bosons $W_\mu^\pm$ to fermions result from the covariant derivative in eq.(\ref{eq:Lag1}).
\begin{align}
\mathcal{L}_{CC} = \frac{g_2}{2\sqrt{2}}\left(J_W^{\mu\dagger}W^+_{\mu} + J^\mu W^-_{\mu}\right)
\end{align}
Here, $W^\pm$ terms are the charged current interaction in the weak eigenstate basis given as
\begin{equation}
   J_W^{\mu\dagger} = \bar{\nu}_L \gamma^\mu (1-\gamma_5)  e_L + \bar{u}_L \gamma^\mu (1-\gamma_5) d_L 
\end{equation}
\begin{equation}
   J_W^{\mu} = \bar{e}_L \gamma^\mu (1-\gamma_5) \nu_L + \bar{d}_L \gamma^\mu (1-\gamma_5) u_L \label{eq:CC}
\end{equation}
denote the charged current, and $g_2$ is the coupling constant of $SU(2)_L$.

using eq.(\ref{Vckm}) rewriting  the currents in terms of mass eigenstates for quarks $(u,d)$ \cite{Langacker:2017uah},
\begin{equation}
   J_W^{\mu\dagger} =  2\bar{u}_L \gamma^\mu A_L^{u\dagger}A_L^{d} d_L 
\end{equation}
\begin{equation}
   J_W^{\mu} =  2\bar{d}_L \gamma^\mu V_q^\dagger u_L \label{eq:CC}
\end{equation}
The unitary quark mixing matrix is $V_q \equiv A_L^{u\dagger} A^d_L$, and describes the mismatch between the weak and mass eigenstates for the up- and down-type quarks.

\bigskip
\begin{figure}
    \centering
    \begin{tikzpicture}
     \begin{feynman}
      \vertex (a) {\(u_j\)};
      \vertex [below=of a] (b) {\(d_k\)};
      \vertex [right=of $(a)!0.5!(b)$] (c);
      \vertex [right=of c] (d) {\(\frac{ig_2}{\sqrt{2}} V_{jk} \gamma^\mu P_L\)};
      
      \diagram* {
        (b) -- [fermion] (c) -- [fermion] (a),
        (c) -- [boson, edge label=\(W_\mu\)] (d),
      };
\end{feynman}
\end{tikzpicture}
\begin{tikzpicture}
\begin{feynman}
      \vertex (a) {\(\nu_j\)};
      \vertex [below=of a] (b) {\(l_k\)};
      \vertex [right=of $(a)!0.5!(b)$] (c);
      \vertex [right=of c] (d) {\(\frac{ig_2}{\sqrt{2}} V^*_{kj} \gamma^\mu P_L\)};
      
      \diagram* {
        (b) -- [fermion] (c) -- [fermion] (a),
        (c) -- [boson, edge label=\(W_\mu\)] (d),
      };
\end{feynman}
\end{tikzpicture}
    \caption{Feynman rules for charged W-fermion couplings}
    \label{fig:W coupling}
\end{figure}
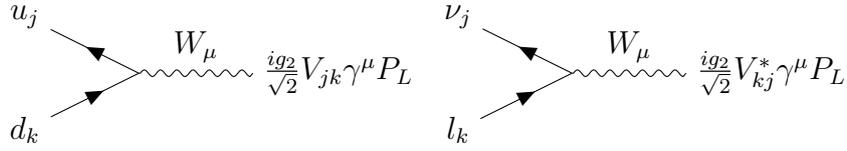

Fig.(\ref{fig:W coupling}) shows flavor violation; the origin of this comes from Yukawa interactions.The Lagrangian is invariant under rotation in flavor space, as in eq.(\ref{eq:Trans1}) and eq.(\ref{eq:Trans2}) . Only after diagonalizing all Yukawa matrices, i.e., on a mass eigenstate basis, is flavor violation transferred to the charged currents of the gauge sector.

\subsubsection{Weak Neutral Currents}
\begin{table}
  \centering
  \begin{tabular}{|c|c|c|c|c|c|c|c|}
    \hline
    & $v_L^e$ & $e_L^-$ & $e_R^-$& $u_L$& $d_L$& $u_R$ & $d_R$ \\
    \hline
    Q &0 & -1 & -1 & $\frac{2}{3}$ & $-\frac{1}{3}$ & $\frac{2}{3}$& -$\frac{1}{3}$\\
    \hline
   $T_3$ & $\frac{1}{2}$& $-\frac{1}{2}$ & 0 & $\frac{1}{2}$ & -$\frac{1}{2}$  &0& 0 \\
    \hline
   Y & -1 & -1 & -2 &  $\frac{1}{3}$ & $\frac{1}{3}$  & $\frac{4}{3}$  & $-\frac{2}{3}$  \\
    \hline
  \end{tabular}
  \caption{Electroweak Quantum Numbers}
  \label{tab:example1}
\end{table}
The weak neutral current interaction (along with W and Z bosons) is an ingredient of SU(2) × U(1) unification.
\begin{align}
\mathcal{L}_{NC} = -eJ_\mu^{em}A_\mu + \frac{g_2}{2\cos\theta_W} J_\mu^0Z_\mu
\end{align}
where e is the QED coupling constant and $\theta_W$ is the Weinberg angle. The neutral electromagnetic and weak currents are given by
\begin{align}
J_\mu^{em} = \sum_f Q_f \bar{f}\gamma_\mu f \label{eq:em}
\end{align}
\begin{align}
J_\mu^{0} = \sum_f Q_f \bar{f}\gamma_\mu (v_f-a_f\gamma_5)f
\end{align}
where
\begin{align}
v_f = T_3^f - 2Q_f \sin^2\theta_W , \hspace{1cm}a_f= T_3^f
\end{align}
Here, $Q_f$ is the electric charge unit of $e$ and $T_3^f$ ($T_3 =0$ for the right-handed (RH) and $T_3=\pm\frac{1}{2}$ for the left-handed (LH) fermions) denote the charge and the third component of the weak isospin of the fermion $f$ respectively. These electroweak charges are given in Table(\ref{tab:example1}). 

From eq.(\ref{eq:em}) we can write, 
\begin{align}
 \bar{f '}\gamma_\mu f'= \bar{f '}_L\gamma_\mu f'_L + \bar{f'}_R\gamma_\mu f'_R
 \end{align}
 \begin{align}
 &=\bar{f }_L\gamma_\mu A_L^{f\dagger} A_L^f f_L + \bar{f}_R\gamma_\mu A_R^{f\dagger} A_R^ff_R\\ 
& = \bar{f}\gamma_\mu f 
\end{align}
This is because only fields of the same charge and chirality can mix with each other. The photon couples in the same manner as left- and right-handed particles. Therefore, $J^{em}_\mu$ is the flavor diagonal. We can write 
\begin{align}
\mathcal{L}_{NC} = \frac{e}{\sin\theta_W \cos\theta_W}(T_3-\sin^2\theta_W )\bar{f}\gamma^\mu Z_\mu f+e Q_f\bar{f}\gamma^\mu A_\mu f\label{eq:LagNC}
\end{align}
\begin{figure}
    \centering
    \includegraphics[width=6cm]{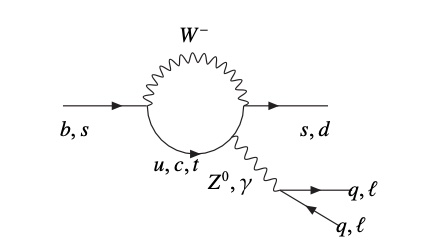}
    \caption{Flavor changing neutral current at one loop level}
    \label{fig:Loop1}
\end{figure}
This expression is valid for all three generations. Multiplying Eq.~(\ref{eq:LagNC}) by $i$ yields the Feynman rules for the interactions of $Z_\mu$ and $A_\mu$ with the SM fermions. The explicit Feynman rules are provided in Appendix~A.

\section{Cabibbo–Kobayashi–Maskawa (CKM) Matrix}

 The quark mixing matrices $V_q$ in eq.(\ref{eq:CC}) arise from the mismatch between the fermion gauge and Yukawa interactions, i.e., between the weak and mass eigenstates.
The unitary matrices, which transform left-handed up- and down-type quarks, are not the same. For this reason, the combination of up and down diagonalization matrices is not equal to the identity. This combination defines the Cabibbo–Kobayashi–Maskawa matrix (CKM) \cite{cabibbo1963,kobayashi1973}, 
\begin{equation}
(V_{\text{CKM}}) = A_L^{u} A_L^{d \dagger} 
\end{equation}
This matrix connects down-type quarks in the flavor and mass basis
\begin{equation}
    d^{'}=V_{CKM}d,\quad  u^{'}=u
\end{equation}
 Up-type quarks in the mass representation are chosen for convenience, like those in flavor one. Further, all quarks need to be understood in the mass basis. It is often convenient to employ the approximate Wolfenstein parametrization\cite{Wolfenstein:1983yz} of CKM matrices.
\begin{equation}
    V_{CKM}=\begin{pmatrix}
        V_{ud}& V_{us}& V_{ub} \\
        V_{cd} &V_{cs}& V_{cb} \\
        V_{td}& V_{ts}& V_{tb}\\ 
    \end{pmatrix}=\begin{pmatrix}
        1-\frac{\lambda^2}{2}& \lambda& A\lambda^3(\rho-i\eta) \\
        -\lambda &1-\frac{\lambda^2}{2}& A\lambda^2 \\
        A\lambda^3(1-\rho-i\eta)& -A\lambda^2& 1\\ 
    \end{pmatrix}
\end{equation}
where $\lambda \sim \sin \theta_c$ and $A \sim 0.811$ (from $V_{cb}$), $\rho$ and $\eta$ are real and of $\mathcal{O}(1)$. Here, $\theta_c$ refers to the Cabibbo angle, which describes the mixing between the first and second generations of quarks (i.e., $u\longleftrightarrow d$, and $c \longleftrightarrow s$. It is crucial to verify that $V_{CKM}$ is unitary, i.e., 
\begin{equation}
    V_{\text{CKM}} V_{\text{CKM}}^{\dagger} = V_{\text{CKM}}^{\dagger} V_{\text{CKM}} = I
\end{equation}

A violation of CKM matrix unitarity would signal the presence of new physics. This could arise, for example, from a \textit{fourth generation of quarks}, in which case the observed $3 \times 3$ CKM submatrix would not be unitary by itself. Unitarity tests include studies of weak universality which involve the diagonal elements, especially the condition
\begin{equation}
    (V_{CKM}V_{CKM}^\dagger)_{11} = |V_{ud}|^2 + |V_{us}|^2 + |V_{ub}|^2 = 1,
\end{equation}
which expresses \textit{first-row unitarity}, often referred to as weak universality, because it reflects that the total coupling strength of the up quark to all down-type quarks sums to unity. A value significantly different from 1 would suggest the presence of additional quark generations or other new physics.

Orthogonality (off-diagonal) conditions provide another 12 relations, six from $V_{CKM} V_{CKM}^\dagger = I$ and six from $V_{CKM}^\dagger V_{CKM} = I$. One such unitarity relation is
\begin{equation}
    (V_{CKM}V_{CKM}^\dagger)_{31} = V_{ub}^* V_{ud} + V_{cb}^* V_{cd} + V_{tb}^* V_{td} = 0,
\end{equation}
which is used to construct the so-called \textit{unitarity triangles} in the complex plane.

The hierarchical structure of the CKM matrix helps explain the relatively long lifetime of $b$ hadrons. Although the matrix element $V_{tb} \approx 1$ allows for a strong coupling between the $b$ and $t$ quarks, decay $b \to t$ is \textit{kinematically forbidden} due to the large top quark mass. Therefore, the $b$ quark decays predominantly to lighter quarks ($c$ or $u$), and the corresponding CKM elements $V_{cb}$ and $V_{ub}$ are small, leading to suppressed decay rates. In contrast, the charm quark decays via the CKM-favored $V_{cs} \approx 1$ transition, so charm hadrons generally have shorter lifetimes.

The CKM matrix appears in charged current interactions of the Yukawa sector but does not appear in neutral current interactions involving the $Z$ boson or photon. This is a consequence of the CKM matrix's unitarity: in the basis where quark masses are diagonal, neutral current couplings remain flavor diagonal, known as the \textit{Glashow--Iliopoulos--Maiani} (GIM) mechanism ~\cite{Glashow:1970gm}, and it implies that there are no flavor-changing neutral currents at tree level in the SM. FCNCs do appear at the loop level, for instance, in penguin diagrams involving transitions such as $b \to s$ or $s \to d$  Fig.(\ref{fig:Loop1}), where all three up-type quarks contribute in the loop.

A key feature of the GIM mechanism is that it not only forbids tree-level FCNCs but also suppresses loop-level FCNCs, especially in observables dominated by light quarks. However, this suppression is less effective when contributions from the top quark dominate, due to the hierarchy $m_t \gg M_W \gg m_{u,c}$, so the GIM suppression breaks down in such cases.

\section{Shortcomings of the SM and searches beyond SM}

Despite its enormous success in passing numerous precision tests, the SM is considered a low-energy effective approximation of a more global theory. The SM does not explain the matter-antimatter asymmetry of the Universe, and does not incorporate Dark Matter, which is believed to dominate over the usual matter in the Universe. It has many free parameters (18 or 19), notably the masses of quarks and charged leptons, but does not explain the mass hierarchy between the different generations. 

These unanswered questions remain among the core topics of the current research activities in particle physics. Searches for signs of NP, i.e., effects beyond the SM, can be performed in two ways. The so-called direct searches profit from an increase in the energy of the accelerator collisions, allowing them to produce heavier particles and probe a higher mass range. This allows to set direct limit on the masses of NP particles (and existence). Alternatively, one could exploit the indirect method by investigating the effects of yet unknown particles in quantum loops and rare B decays. In this case, the masses of probed virtual mediators can be an order of magnitude larger than the scale of the SM. Indirect measurements provide constraints related to the mass of NP particles and their coupling to the SM ones. Under a given assumption of the coupling value, the effects of NP particles on the SM observables decreases with increasing mass.
Flavour physics probes the decays and interactions of b and s hadrons, investigating the decay rates and branching ratios of the processes which could be affected by NP mediators. This technique is most successful when the SM process is suppressed or forbidden so that even tiny NP effects become prominent.

We now turn to a key feature of its tree-level structure: the absence of FCNCs, which are transitions that change quark flavor without altering the electric charge. Processes such as $b \rightarrow s \ell^+ \ell^-$ are forbidden at tree level due to the GIM mechanism, but can still occur via loop diagrams. These rare transitions, suppressed within the SM, are especially sensitive to the presence of new physics and serve as powerful probes of its effects. In the following chapter, we develop the effective theory describing FCNCs, compute their Standard Model predictions, and compare them with precision measurements. Particular emphasis is placed on the theoretical structure and phenomenological significance of FCNCs in rare $B$-meson decays.
\chapter{Flavor Changing Neutral Current (FCNC)}
\label{chap:FCNC}

\section{Loop Calculations}
Particular decays are absent at the tree level, and the leading contributions come from diagrams
involving one loop. Some of the Feynman diagrams at the loop level are divergent. To solve these, we will consider the subdiagrams, which are responsible for the divergences in question. We use Dirac algebra in $D\neq4$ dimensions. This is crucial to address the divergences in loop diagrams.

\subsection{Dimensional Regularization}

To deal with divergences that appear in loop diagrams we have to regularize the theory to have an explicit parametrization of the singularities. Different regularization methods yield the same final result; we used Dimensional Regularization (DR), where  Feynman diagrams are evaluated in \( D = 4 - 2\epsilon \) space-time dimensions, where \(\epsilon\) is a small parameter. Ultraviolet (UV) divergences appear as poles when \(\epsilon \to 0\). 
A typical one-loop calculation then has the general structure:
\[
\text{One-loop result} = \frac{a_1}{\epsilon} + b_1,
\]
where \(a_1\) and \(b_1\) are finite constants. The term \(\frac{a_1}{\epsilon}\) represents the divergent part, while \(b_1\) is the finite remainder.

To maintain consistent mass dimensions in arbitrary \(D\) dimensions, it is necessary to introduce an arbitrary mass scale \(\mu\), often referred to as the \textit{renormalization scale}. This is because coupling constants acquire nontrivial mass dimensions when \(D \neq 4\).  For example, consider the one-loop scalar integral from \cite{Buras:2020xsm}
\begin{equation}
I \equiv \lambda \, \mu^{4 - D} \int \frac{d^D k}{(2\pi)^D} \frac{1}{k^2 - m^2 + i\varepsilon}.
\end{equation}
Here, the prefactor \(\mu^{4 - D}\) ensures that the integral \(I\) has the correct mass dimension, since \(d^D k\) has dimension \(D\). The loop integral evaluates to:
\begin{equation}
\begin{aligned}
\int \frac{d^D k}{(2\pi)^D} \frac{1}{k^2 - m^2 + i\varepsilon} &= \frac{i}{(4\pi)^{2 - \epsilon}} (m^2)^{-\epsilon} \Gamma(\epsilon)= \frac{i}{(4\pi)^2} \left( \frac{1}{\epsilon} - \gamma_E + \ln(4\pi) - \ln m^2 + \mathcal{O}(\epsilon) \right),
\end{aligned}
\end{equation}
where \(\gamma_E\) is the Euler–Mascheroni constant. More generally, it is useful to redefine the coupling constant so that it remains dimensionless in \(D\) dimensions. This is achieved by absorbing the appropriate powers of \(\mu\),
\begin{equation}
g \to g\, \mu^\epsilon, \quad \text{with} \quad \epsilon = \frac{4 - D}{2}.
\end{equation}

In this expression:
\begin{itemize}
    \item \(g\) on the left-hand side is the dimensional coupling in \(D\) dimensions,
    \item \(\mu\) is an arbitrary energy scale introduced by dimensional regularization,
    \item \(g\) on the right-hand side is the dimensionless redefined coupling at the scale \(\mu\).
\end{itemize}

This rescaling ensures that the coupling retains the correct dimensionality and allows for a consistent renormalization procedure. Furthermore, it facilitates the study of scale dependence and running couplings through the renormalization group equations.

\subsection{Renormalization}

The Lagrangian formulation contains interactions and propagating terms that can be translated diagrammatically to Feynman diagrams. Higher-order quantum corrections, also known as radiative corrections, are required for high precision studies and can be distinguished from Leading order (LO) or tree-level processes by containing loops in their diagrammatic representation. A curious fact about going beyond LO is that amplitudes from loop diagrams contain divergences from the UV regions of the momentum integrals being taken. The purpose of renormalization is to eliminate said UV divergences by absorption into the bare parameters of the QFT Lagrangian. Every parameter, including coupling constants, masses and field content, is needed to absorb divergences, and the physical Lagrangian used for predictions exhibits no UV divergence.

One-loop diagrams generally lead to ultraviolet divergences, which also appear in physical quantities such as decay amplitudes. For example, the amplitude for the process \(b\to sZ\), contains such divergences. The procedure of renormalization allows us to systematically remove these infinities and obtain finite, physically meaningful results. Specifically, renormalization provides a prescription for expressing physical observables in terms of a finite set of parameters, allowing comparison with experimental data.

The core idea behind the renormalization program is to replace the bare fields and parameters (unrenormalized) with corresponding renormalized quantities so that the resulting Green functions and decay amplitudes are finite. For instance, the fields and parameters in the Lagrangian are renormalized as follows:
\begin{align}
    A_{0, \mu} &= Z_A^{1/2} A_\mu, \\
    q_0 &= Z_q^{1/2} q, \\
    g_{0,s} &= Z_g g_s \mu^{\epsilon}\\
    m_0 &= Z_m\, m,
\end{align}
where quantities with a subscript '0' denote bare fields and parameters (unrenormalized). The quantities \(A_\mu\), \(q\), and \(m\) are the corresponding renormalized photon field, fermion field, and mass, respectively, and $g_s$ is the renormalized QCD coupling. The factors \(Z_A\) is gauge field, \(Z_q\) quark field, \(Z_g\) coupling, and \(Z_m\) mass renormalizations. These are divergent quantities chosen such that, when all bare quantities are replaced by renormalized ones, the divergences cancel out, leaving finite results.

It is important to note that the bare quantities are independent of the arbitrary renormalization scale \(\mu\), while the renormalized quantities generally depend on \(\mu\). Within the framework of dimensional regularization, the renormalization constants typically take the following form \cite{Buras:2020xsm}:
\begin{equation}
    Z_i = 1 + \frac{\alpha_s}{4\pi} \left( \frac{a_i}{\varepsilon} + b_i \right) + \mathcal{O}(\alpha_s^2),
    \label{Z_i}
\end{equation}
where \(\alpha_s\) is the strong coupling constant, and \(\varepsilon = \frac{4 - D}{2}\) encodes the deviation from four dimensions. The coefficients \(a_i\) are fixed by the structure of the theory (e.g., here QCD) and are independent of the renormalization scheme. In contrast, the terms \(b_i\) depend on the specific renormalization scheme adopted (e.g. MS or \(\overline{\text{MS}}\)).

It is worth emphasizing that not all quantum field theories are renormalizable. The gauge theory like SM,  however, is a renormalizable theory, meaning its ultraviolet divergences can be absorbed into a finite number of parameters at all orders in perturbation theory. Aside from absorbing infinities, a renormalisation of fields, mass and couplings would still
be necessary even if the loop integrals were finite  \cite{tHooft:1972tcz}.

\begin{itemize}
    \item \textbf{Renormalizable theories:} Only a finite number of divergent subdiagrams appear at each order in perturbation theory.
    \item \textbf{Non-renormalizable theories:} Divergences appear in an infinite number of terms at sufficiently high orders, requiring infinitely many counterterms.
\end{itemize}

\subsection{Counter Term Method}
Having introduced the renormalization of fields and parameters, we now proceed to organize the Lagrangian accordingly. The original Lagrangian, expressed entirely in terms of bare (unrenormalized) quantities, can be rewritten in terms of renormalized quantities and additional counter-terms. Specifically, we write:
\begin{equation}
\mathcal{L}_0 = \mathcal{L} + \mathcal{L}_C,
\end{equation}
where \(\mathcal{L}_0\) is the bare Lagrangian, \(\mathcal{L}\) is the renormalized Lagrangian (in terms of renormalized fields and parameters), and \(\mathcal{L}_C\) is the counter-term Lagrangian. The counter-terms are introduced to cancel the divergences that arise in loop-level functions and decay amplitudes.

The renormalization constants \(Z_i\), introduced in eq.(\ref{Z_i}) through field and parameter redefinitions, determine the structure of \(\mathcal{L}_C\). They are chosen such that the divergences present in the loop contributions are precisely canceled by the corresponding counter-term diagrams. Importantly, this procedure not only cancels the divergent parts (e.g., \(1/\varepsilon\) poles in dimensional regularization) but may also involve subtracting finite parts. However, the subtraction of finite parts is not uniquely defined. Different choices for which finite terms to subtract correspond to different renormalization schemes. As a result, the renormalization constants \(Z_i\), as well as the values of the renormalized parameters and fields, depend on the chosen scheme. These schemes include Minimal Subtraction (MS) and Modified Minimal Subtraction (\(\overline{\text{MS}}\)), where only divergent parts (or divergent parts plus specific constants like \(\gamma_E\) and \(\ln 4\pi\)) are subtracted \cite{tHooft:1972tcz}.

Despite this scheme dependence, physical observables, such as cross sections, decay rates, and scattering amplitudes, must be independent of the renormalization scheme. Thus, the introduction of counter terms through \(\mathcal{L}_C\) plays a central role in the renormalization program. It ensures the finiteness of all physically relevant quantities and organizes divergences in a systematic way that allows for consistent predictions within the framework of quantum field theory.

\subsection{MS and $\overline{\text{MS}}$ Renormalization Schemes}

One of the most widely used renormalization prescriptions is the Minimal Subtraction (MS) scheme, in which only the divergent parts of the loop integrals (typically $1/\epsilon$ poles in dimensional regularization) are subtracted. In this scheme, the renormalization constants $Z_i$ are given by \cite{Buras:2020xsm}
\begin{equation}
    Z_i = \frac{\alpha_s a_{1i}}{4\pi \epsilon} 
    + \left( \frac{\alpha_s}{4\pi} \right)^2 \left( \frac{a_{2i}}{\epsilon} + \frac{b_{2i}}{\epsilon} \right) 
    + \mathcal{O}(\alpha_s^3),
    \label{Eq:Z}
\end{equation}
where $a_{ji}$ and $b_{ji}$ are constants independent of the renormalization scale $\mu$. In the MS scheme, the renormalization constants do not explicitly depend on $\mu$, but indirectly depend on the running coupling constant $g_s(\mu)$. 

An important feature of the MS scheme is that it is mass-independent: the renormalization constants $Z_i$ do not depend on particle masses, which simplifies computations, particularly in gauge theories like QCD \cite{Weinberg:1973xwm}.

A variant of the MS scheme is the Modified Minimal Subtraction ($\overline{\text{MS}}$) scheme, in which the artifacts of dimensional regularization such as $\ln 4\pi$ and the constant $\gamma_E$ are also subtracted \cite{Buras:2020xsm}. The relation between the renormalization scales in the two schemes is given by \cite{Bardeen:1978yd}:
\begin{equation}
    \mu_{\overline{\text{MS}}} = \mu \, e^{\frac{\gamma_E}{2}} (4\pi)^{-\frac{1}{2}}.
    \label{muE}
\end{equation}

Thus, moving from MS to $\overline{\text{MS}}$ is equivalent to:
\begin{align}
    &\text{MS} \rightarrow \overline{\text{MS}} \quad \Leftrightarrow \quad \mu \rightarrow \mu_{\overline{\text{MS}}}, \\
    &Z^{\text{MS}}_i \rightarrow Z^{\overline{\text{MS}}}_i \quad \Leftrightarrow \quad \alpha^{\text{MS}}_s \rightarrow \alpha^{\overline{\text{MS}}}_s.
\end{align}

In the $\overline{\text{MS}}$ scheme, 

\begin{equation}
Z_i = 1 + \sum_{k=1}^{\infty} \frac{1}{\epsilon^k} \, Z_{i,k}(e),
\end{equation}
where, $Z_{i,k}(e)$ are solely dependent on the EM coupling and independent of $\epsilon$. Thus, we
 define $Z_{i,k}$ in QED in the $\overline{\text{MS}}$-scheme such that the divergences are no longer present. The first appearance of a divergence always occurs at one-loop or next-to-leading order (NLO) in a perturbative expansion. For example, if we consider $Z_A$ for the photon field to attain it at NLO, $Z_A$ is defined by the finiteness condition of the photon
propagator,
\begin{equation}
\text{finite} \;=\;
\tikz[baseline=-0.6ex]{
  \begin{feynman}
    \vertex (L)  at (0,0);
    \vertex (v1) at (1.2,0);
    \vertex (v2) at (2.0,0);
    \vertex (R)  at (3.2,0);

    \diagram*{
      (L)  -- [photon] (v1),
      (v2) -- [photon] (R),
    };

    \diagram*{
      (v1) -- [fermion, half left, looseness=1.4] (v2)
           -- [fermion, half left, looseness=1.4] (v1),
    };
  \end{feynman}
}
\;+\;
\tikz[baseline=-0.6ex]{
  \begin{feynman}
    \vertex (L) at (0,0) {};
    \vertex [crossed dot] (C) at (1.6,0) {};
    \vertex (R) at (3.2,0) {};
    \diagram*{
      (L) -- [photon] (C) -- [photon] (R),
    };
  \end{feynman}
}
\end{equation}

The first diagram gives the one-loop vacuum polarisation correction, and second diagram shows its corresponding
counterterm. Requiring that the sum of the two diagrams is finite in the $\epsilon\to0$ limit up to $\mathcal{O}(\alpha_e)$ where $\alpha_e = e^2/(4\pi)$ gives,
\begin{equation}
     Z_A = 1- \frac{4\alpha_e}{3 \pi} \frac{1}{\epsilon}+ \mathcal{O}(\alpha_e^2),
\end{equation}
Similarly, QCD renormalization constants are,
\begin{equation}
    Z_q = 1- \frac{\alpha_s}{4 \pi} C_F \frac{1}{\epsilon}+ \mathcal{O}(\alpha^2_s),
\end{equation}
\begin{equation}
    Z_m = 1- \frac{\alpha_s}{4 \pi}3 C_F \frac{1}{\epsilon}+ \mathcal{O}(\alpha^2_s),
\end{equation}
\begin{equation}
    Z_A = 1- \frac{\alpha_s}{4 \pi} \left[\frac{2}{3}f-\frac{5}{3}N\right] \frac{1}{\epsilon}+ \mathcal{O}(\alpha^2_s),
\end{equation}
\begin{equation}
    Z_{g_s} = 1- \frac{\alpha_s}{4 \pi} \left[\frac{11}{6}N-\frac{2}{6}f\right] \frac{1}{\epsilon}+ \mathcal{O}(\alpha^2_s),
\end{equation}
where $N$ represents number of colors ($N=3$ in QCD), $f$ is the number of quark flavors,and $C_F$(Casimir in fundamental representation) is given as 
\begin{equation}
    C_F = \frac{N^2 -1}{2N}
\end{equation}
$Z_q$ and $Z_A$ are gauge dependent and are given here in the Feynman gauge. However, this gauge dependence is cancelled by other contributions to physical amplitudes.

\subsection{Renormalization Group Equations (RGEs)}

As discussed earlier, the introduction of the arbitrary mass scale $\mu$ in the renormalization procedure leads to a dependence of the renormalized quantities on $\mu$. The evolution of these quantities with respect to $\mu$ is governed by the renormalization group equations (RGEs). For the renormalized coupling constant $g(\mu)$ and the running quark mass $m(\mu)$, the RGEs are given as \cite{Buras:2020xsm},
\begin{align}
    \frac{d g(\mu)}{d\ln \mu} &= \beta(g(\mu), \epsilon), \\
    \frac{d m(\mu)}{d\ln \mu} &= -\gamma_m(g(\mu))m(\mu),
\end{align}
Where,
\begin{equation}
    \beta(g, \epsilon)= -\epsilon g+ \beta(g),
\end{equation}
and
\begin{equation}
    \beta(g)= -g \frac{1}{Z_g}\frac{dZ_g}{dln(\mu)}, 
    \end{equation}
    \begin{equation}
    \gamma(g)= \frac{1}{Z_m}\frac{dZ_m}{dln(\mu)}
\end{equation}
$\beta(g)$ and $\gamma(g)$ are called renormalization group functions. $\beta(g)$  is the beta function that controls the dependence $\mu$ of $g(\mu)$ and $\gamma_m(g)$ is the anomalous dimension of the mass operator $m(\mu)$. In $\overline{\text{MS}}$ scheme there is no explicit $\mu$-dependence, they only depend on g. Including the higher-order contribution we can write $\beta(g)$ as,
\begin{equation}
    \beta(g)= -\beta_0 \frac{g^3}{16\pi^2}-\beta_1\frac{g^5}{(16\pi^2)^2}
\end{equation}
\begin{equation}
\gamma_m(g) = \gamma_m^{(0)} \, \frac{g}{16\pi^2}+ \gamma_m^{(1)} \, (\frac{g}{16\pi^2})^2
\end{equation}
here, $g = g_s$ for higher order terms, whereas $\beta_0$, $\beta_1$, $\gamma^{(0)}_m$), and $\gamma^{(1)}_m$ are gauge independent

\subsection{Running Coupling Constant}

The beta function for the strong coupling constant $\alpha_s = g_s^2 / (4\pi)$ can be expanded perturbatively as
\begin{equation}
    \frac{d\alpha_s}{d \ln \mu} = - \frac{2\beta_0 \alpha_s^2}{4\pi} - \frac{2\beta_1 \alpha_s^3}{(4\pi)^2} + \mathcal{O}(\alpha_s^4),
\end{equation}
where $\beta_0$ and $\beta_1$ are the first two coefficients of the QCD beta function, determined by the gauge group and the fermion content \cite{Buras:2020xsm}.

Solving this differential equation leads to the well-known expression for the running coupling \cite{Bardeen:1978yd}:
\begin{equation}
    \frac{\alpha_s(\mu)}{4\pi} = \frac{1}{\beta_0 \ln \left( \mu^2 / \Lambda_{\overline{\text{MS}}}^2 \right)} 
    - \frac{\beta_1}{\beta_0^3} \frac{\ln \left[ \ln \left( \mu^2 / \Lambda_{\overline{\text{MS}}}^2 \right) \right]}{ \left[ \ln \left( \mu^2 / \Lambda_{\overline{\text{MS}}}^2 \right) \right]^2 },
\end{equation}
where $\Lambda_{\overline{\text{MS}}}$ is a scheme-dependent QCD scale parameter. It can be determined from a known value of $\alpha_s$ at a reference scale, such as $M_Z$.

Relating $\alpha_s$ in the MS and $\overline{\text{MS}}$ schemes, using eq.(\ref{muE}) we have:
\begin{equation}
    \alpha_{s, \text{MS}} = \alpha_{s, \overline{\text{MS}}} \left[ 1 + \frac{\beta_0 (\gamma_E - \ln 4\pi)}{4\pi} \alpha_{s, \overline{\text{MS}}} \right].
\end{equation}
\begin{equation}
    \Lambda^2_{\overline{\text{MS}}}= 4 \pi e^{-\gamma_E}\Lambda^2_{\text{MS}}
\end{equation}
$\Lambda_{\overline{\text{MS}}}$ and $\alpha_s(\mu)$ depend on $f$, the number of effective flavors,
\begin{equation}
f = 
\begin{cases}
6, & \mu \geq m_t \\
5, & m_b \leq \mu \leq m_t \\
4, & m_c \leq \mu \leq m_b \\
3, & \mu \leq m_c
\end{cases}
\end{equation}
To obtain a value of $\alpha_s(\mu)$ on a given scale $\mu$ requires the value of the coupling constant at a set scale. The value of reference that is most commonly used, coming from the Z decays \cite{ParticleDataGroup:2024cfk}, gives an alternative and convenient form for the running coupling that allows a direct comparison with the experimental value is
\begin{equation}
    \alpha_s(\mu) = \frac{\alpha_s(M_Z)}{v(\mu)} \left[ 1 - \frac{\beta_1}{\beta_0} \frac{\alpha_s(M_Z)}{4\pi} \frac{\ln v(\mu)}{v(\mu)} \right],
    \label{alpha}
\end{equation}
where
\begin{equation}
    v(\mu) = 1 - \frac{\beta_0 \alpha_s(M_Z)}{2\pi} \ln \left( \frac{M_Z}{\mu} \right),
    \label{vmu}
\end{equation}
is valid for $f=5$, and the experimentally measured value of $\alpha_s(M_Z)$ is
\begin{equation}
\alpha^5_s(M_Z) = 0.1181 \pm 0.0006.
\end{equation}

\subsection{Running Quark Mass}

The scale dependence of the quark masses arises due to quantum corrections and is governed by the anomalous dimension \(\gamma_m(g)\). Using $\frac{dg}{dln(\mu)}=\beta(g)$, the running mass,
\begin{equation}
        \frac{d m(\mu)}{d\ln \mu} = -\gamma_m(g)m(\mu),
\end{equation}
at scale \(\mu\) is related to its value at a reference scale \(\mu_0\) through:
\begin{equation}
    m(\mu) = m(\mu_0) \exp \left( - \int_{g(\mu_0)}^{g(\mu)}  \frac{\gamma_m(g')}{\beta(g')} \, dg' \right),
\end{equation}
For practical purposes, a simplified leading-order approximation is often sufficient. From~\cite{Buras:2020xsm}, the strong coupling constant at scale \(\mu\) can be approximated by:
\begin{equation}
    \alpha_s (\mu) = \frac{\alpha_s (M_Z)}{1 - \frac{\beta_0 \alpha_s(M_Z)}{2\pi} \ln \left( \frac{M_Z}{\mu} \right)}
\end{equation}

The first two coefficients of the QCD beta function in the $\overline{\text{MS}}$ scheme for five active flavors (\(f = 5\)) are given as
\begin{equation}
\beta_0 = \frac{11N-2f}{3}= \frac{23}{3}, \quad \beta_1 = \frac{34}{3}N^2 -\frac{10}{3}Nf -2 C_F f =\frac{116}{3},
\end{equation}
\begin{equation}
\gamma_m^{(0)} = 6 C_F, \quad \gamma_m^{(1)} = C_F\left(3C_F +\frac{97}{3}N -\frac{10}{3} f\right),
\end{equation}
as reported in~\cite{Buras:2020xsm,Buras:1998raa}. Using these equations, the running top-quark mass at a different scale \(\mu_t\) can be obtained from the following:
\begin{equation}
    m_t(\mu_t) = m_t(m_t) \left[ \frac{\alpha_s(\mu_t)}{\alpha_s(m_t)} \right]^{\frac{4}{\beta_0}}.
\end{equation}
where
\begin{equation}
 \alpha_s(\mu_b) = 0.212, \quad m_t(m_t) = 162.5^{+2.1}_{-1.5}~\text{GeV},
 \quad 
m_b(m_b) = \left( 4.19^{+0.18}_{-0.06} \right) \, \text{GeV}.
\end{equation}
This expression reflects the leading-order scaling behavior of the mass with respect to the strong coupling constant. Higher-order corrections can be included using known perturbative expansions for \(\gamma_m\) and \(\beta(g)\), but the above form provides a reliable estimate for many phenomenological applications.
 
\section{Effective Field Theory}

Effective field theory (EFT) can be interpreted as a framework that one can employ to perform scale separation consistently. The need for such a framework arises in precision studies when attempting to identify deviations from the SM in observables. The main requirement for defining an EFT from a complete theory is the existence of various widely separated scales.

EFT is an explicit tool that manifests scale separation. Let us consider a QFT that has a high energy scale $\Lambda$ (the mass of a heavy field) that we want to describe at a lower energy scale E such that $E<< \Lambda$. One can define a cutoﬀ scale $\mu$, such that $E<<\mu<\Lambda$, which divides the fields into high and low energy modes. Although low energy modes are the relevant external states at the energy scale E, the high energy modes do not propagate on long distances, they only appear as virtual particles and can be “removed” from the theory. The separation of scales is then made explicit at the level of observables. Thus, given an EFT, calculations are greatly simplified, and contributions from different energetic regimes can be calculated.

The basic framework for the theoretical description of weak decays of B hadrons is the EFT, relevant for scales $\mu\approx O(1 - 5) GeV$, which is much smaller than $M_W$, $M_Z$ and $m_t$. It represents a generalization of Fermi theory of weak interactions and allows the (high-energy, or short-distance) electroweak and (low-energy, or long-distance) QCD effects to be handled simultaneously. An eﬀective low-energy theory obtained by integrating out the heavy particles, which, in the SM, are the top quark and the W boson. The standard method of the operator product expansion (OPE) allows for separating the amplitude of a weak meson decay process into two distinct parts: the long-distance contributions contained in the operator matrix elements and the short-distance contributions encoded by the Wilson coeﬃcients. In the case of B decays, the W boson and the top quark with mass bigger than the factorization
scale are integrated out, i.e, removed from the theory as dynamical variables.
The eﬀective Hamiltonian can be written as

\begin{equation}
    \mathcal{H}_{eff}=\frac{G_F}{\sqrt{2}}\sum{i}V_{CKM}^iC_i(\mu)O_i
    \end{equation}
 Here, $G_F$ is the Fermi constant, $V_{CKM}$ is the set of CKM factors relevant to transition. The operators $O_i$ describe the low-energy physics below the energy scale $\mu$. The Wilson coefficients $C_i$ \cite{Wilson 1969, zimmermann1971} describe the strength with which a given operator enters the Hamiltonian and covers the high energy part above the scale $\mu$. The values of these coefficients depend on the scale $\mu$. They can be evaluated at the weak interaction ($M_W$) characteristic scale by matching the effective theory with the complete SM theory. At this scale, QCD corrections are minor and can be calculated perturbatively. Then, the coefficients are transferred to the scale $\mu$ using the renormalization group equations (RGE).
 
 Thus, $\mathcal{H}_{eff}$ is simply a series of effective vertices multiplied by effective coupling constants $C_i$. This series is known under the name of the operator product expansion (OPE) \cite{Wilson1969,zimmermann1971,witten1977}. An amplitude for a decay of a given meson M = K, B,.. into a final state $F = \pi\nu\nu^-,\pi\pi, D, K$ is simply given by
 \begin{equation}
     A(M\rightarrow F)=\left<F|\mathcal{H}_{eff}|M\right>\frac{G_F}{\sqrt{2}}\sum{i}V_{CKM}^iC_i(\mu)\left<F|O_i|M\right>,
 \end{equation}
 where $\left<F|O_i(\mu)|M\right>$ are the hadronic matrix elements of $O_i$ between M and F. They summarize the physics contributions to the amplitude $A(M \rightarrow F )$ from scales lower than $\mu$. The operators $O_i$ describe the low-energy physics below the energy scale $
\mu$. Here is the list of perators of weak decays.
\subsubsection*{\textbf{Current-Current operators:}}
\begin{align}
O_1 &= (\bar{s}_\alpha \gamma^\mu P_L c_\beta)(\bar{c}_\beta \gamma_\mu P_L b_\alpha), \\[2pt]
O_2 &= (\bar{s}_\alpha \gamma^\mu P_L c_\alpha)(\bar{c}_\beta \gamma_\mu P_L b_\beta),
\label{O12}
\end{align}
\subsubsection*{\textbf{QCD Penguins  operators:}}
\begin{align}
O_3 &= \sum_{q=u,d,s,c,b} (\bar{s}_\alpha \gamma^\mu P_L b_\alpha)(\bar{q}_\beta \gamma_\mu P_L q_\beta), \\[2pt]
O_4 &= \sum_{q=u,d,s,c,b} (\bar{s}_\alpha \gamma^\mu P_L b_\beta)(\bar{q}_\beta \gamma_\mu P_L q_\alpha),
\label{O34}
\end{align}

\begin{align}
O_5 &= \sum_{q=u,d,s,c,b} (\bar{s}_\alpha \gamma^\mu P_L b_\alpha)(\bar{q}_\beta \gamma_\mu P_R q_\beta), \\[2pt]
O_6 &= \sum_{q=u,d,s,c,b} (\bar{s}_\alpha \gamma^\mu P_L b_\beta)(\bar{q}_\beta \gamma_\mu P_R q_\alpha),
\label{O56}
\end{align}
\subsubsection*{\textbf{Electromagnetic dipole and chromomagnetic penguin operators:}}
\begin{align}
O_{7\gamma} &= \frac{e}{16\pi^2} m_b \, (\bar{s}_\alpha \sigma^{\mu\nu} P_R b_\alpha) F_{\mu\nu}, \\[2pt]
O_8 &= \frac{g_s}{16\pi^2} m_b \, (\bar{s}_\alpha \frac{\lambda^a}{2} \sigma^{\mu\nu} P_R b_\alpha) G^a_{\mu\nu},
\label{O7}
\end{align}
\subsubsection*{\textbf{Semileptonic operators:}}
\begin{align}
O_9 &= \frac{e^2}{16\pi^2} (\bar{s}_\alpha \gamma_\mu P_L b_\alpha)(\bar{l} \gamma^\mu l), \\[2pt]
O_{10} &= \frac{e^2}{16\pi^2} (\bar{s}_\alpha \gamma_\mu P_L b_\alpha)(\bar{l} \gamma^\mu \gamma^5 l).
\label{O10}
\end{align}

Operators $O_1,.. O_6$ are four-quark operators, classified into current-current tree level W exchange operator ($O_1-O_2$) and QCD penguins mediated by gluons ($O_3-O_6$). Operators $O_7,.. O_{10}$ are electroweak penguin operators, classified as electromagnetic penguin ($O_7$), chromomagnetic operator ($O_8$), and semileptonic operators ($O_9-O_{10}$) and the corresponding Wilson coefficients are $C_i$, where $i= 1,...10$.

Three of them are the most relevant for the description of the $b\rightarrow sl^+l^-$ transition: $O_7$ describing the $b\rightarrow  s\gamma$ transition with an on-shell photon, and $O_9, O_{10}$ are the vector and axial vector operators describing the $b\rightarrow sl^+l^-$ transition. 
Hence, the effective Hamiltonian can be written as
\begin{equation}
    \mathcal{H}_{\text{eff}} = \frac{G_F}{\sqrt{2}}\left( \lambda_u \left( C_1(\mu_b) O_u^1 + C_2(\mu_b) O_u^2 \right) 
+ \lambda_c \left( C_1(\mu_b) O_c^1 + C_2(\mu_b) O_c^2 \right) 
- \lambda_t \sum_{i=3}^6 C_i(\mu) O_i\right)
\end{equation}
where, $\lambda_u = V^*_{us}V_{ub}$ and $\lambda_c = V^*_{cs}V_{cb}$.
\begin{table}[h!]
\centering
\resizebox{\textwidth}{!}{%
\begin{tabular}{|c|c|c|c|c|c|c|c|c|c|}
\hline
\multicolumn{10}{|c|}{\textbf{SM Wilson Coefficients (at $\mu = 4.8$ GeV)}} \\ \hline
$C_1$ & $C_2$ & $C_3$ & $C_4$ & $C_5$ & $C_6$ & $C_{7\gamma}^{\rm eff}$ & $C_8^{\rm eff}$ & $C_9$ & $C_{10}$ \\ \hline
$-0.2632$ & $1.0111$ & $-0.0055$ & $-0.0806$ & $0.0004$ & $0.0009$ & $-0.2923$ & $-0.1663$ & $4.0749$ & $-4.3085$ \\ \hline
\end{tabular}%
}
\caption{Values for the SM Wilson Coefficients at NNLO coming from~\cite{Descotes-Genon:2013vna}}
\label{tab:WilsonCoeffs}
\end{table}

 The OPE systematically separates short-distance physics, allowing complex loop-level processes to be expressed as a sum of operator contributions, each multiplied by its corresponding coefficient. Together, the OPE and RGE methods provide a controlled and systematic approach to incorporate NP contributions and to make precise predictions for rare \( B \)-decay observables.

\subsection{FCNC Processes}

The pattern of flavor violation in the SM is governed by the V-A structure of $W^\pm$ interactions with quarks and leptons, and equally crucial by the natural suppression of FCNC processes with the help of the GIM mechanism. The flavor diagonal structure of the basic vertices involving $\gamma$, gluon $G$ and $Z$ in the SM forbids the appearance of FCNC processes at the tree level. However, with the help of the flavor-changing $W^\pm$ vertex, one can construct one-loop and higher-order diagrams that mediate FCNC processes. The fact that FCNCs occur in the SM only as loop effects makes them particularly useful for testing the quantum structure of the theory and the search for NP beyond the SM. 

\subsubsection{Effective Vertices}
At one loop level, FCNCs are described as basic triple and quadratic effective vertices called penguin and box diagrams, respectively. Using the Feynman rules for elementary vertices and propagators in the SM, we can derive the effective vertices in question.

\subsubsection*{Penguin Vertices}

\begin{figure}[htp]
\centering
\begin{tikzpicture}

\begin{scope}[shift={(0,1.5)}]
\begin{feynman}
\vertex (a) at (-1.2,1.2) {\(b\)};
\vertex (b) at (1.2,1.2) {\(s\)};
\vertex (v1) at (0,1.2);
\vertex (v2) at (0,0.5);
\vertex (out) at (0,-0.5) {\(Z, \gamma\)};
\diagram* {
  (a) -- [fermion] (v1) -- [fermion] (b),
  (v1) -- [boson] (v2) -- [boson] (out),
};
\draw[fill=white, thick] (v1) circle (0.3);
\node at (v1) {\(t\)};
\end{feynman}
\end{scope}

\node at (2.3,2.5) {\Large$=$};

\begin{scope}[shift={(3.5,2.7)}]
\begin{feynman}
\vertex (b1) {\(b\)};
\vertex [right=1.0cm of b1] (b2);
\vertex [right=1.2cm of b2] (b3);
\vertex [right=1.0cm of b3] (b4) {\(s\)};
\vertex at ($(b2)!0.5!(b3)!1cm!-90:(b3)$) (g1);
\vertex [below= of g1] (g2);

\diagram* {
(b1) -- [fermion] (b2) -- [boson, edge label={$W^{\pm}$}] (b3) -- [fermion] (b4),
(b2) -- [fermion, out=90, looseness=1.0, quarter right, edge label' =${t}$] (g1)
     -- [fermion, in=90, looseness=1.0, quarter right, edge label'= ${t}$] (b3),
(g1) -- [boson, edge label' ={\(Z, \gamma\)}] (g2);
};
\end{feynman}
\end{scope}

\node at (7.5,2.3) {\Large$+$};
\begin{scope}[shift={(8.0,2.7)}]
\begin{feynman}
\vertex (b1) {\(b\)};
\vertex [right=1.0cm of b1] (b2);
\vertex [right=1.2cm of b2] (b3);
\vertex [right=1.0cm of b3] (b4) {\(s\)};
\vertex at ($(b2)!0.5!(b3)!1cm!-90:(b3)$) (g1);
\vertex [below= of g1] (g2);

\diagram* {
(b1) -- [fermion] (b2) -- [fermion, edge label={$t$}] (b3) -- [fermion] (b4),
(b2) -- [boson, out=90, looseness=1.0, quarter right, edge label' =${W^\pm}$] (g1)
     -- [boson, in=90, looseness=1.0, quarter right, edge label'= ${W^\pm}$] (b3),
(g1) -- [boson, edge label' ={\(Z, \gamma\)}] (g2);
};
\end{feynman}
\end{scope}

\begin{scope}[shift={(0,-2.5)}]
\begin{feynman}
\vertex (a) at (-1.2,1.2) {\(b\)};
\vertex (b) at (1.2,1.2) {\(s\)};
\vertex (v1) at (0,1.2);
\vertex (v2) at (0,0.5);
\vertex (out) at (0,-0.5) {\(G\)};
\diagram* {
  (a) -- [fermion] (v1) -- [fermion] (b),
  (v1) -- [gluon] (v2) -- [gluon] (out),
};
\draw[fill=white, thick] (v1) circle (0.3);
\node at (v1) {\(t\)};
\end{feynman}
\end{scope}

\node at (2.3,-1.8) {\Large$=$};

\begin{scope}[shift={(3.5,-1.4)}]
\begin{feynman}
\vertex (b1) {\(b\)};
\vertex [right=1.0cm of b1] (b2);
\vertex [right=1.2cm of b2] (b3);
\vertex [right=1.0cm of b3] (b4) {\(s\)};
\vertex at ($(b2)!0.5!(b3)!1cm!-90:(b3)$) (g1);
\vertex [below= of g1] (g2);

\diagram* {
(b1) -- [fermion] (b2) -- [boson, edge label={$W^{\pm}$}] (b3) -- [fermion] (b4),
(b2) -- [fermion, out=90, looseness=1.0, quarter right, edge label' =${t}$] (g1)
     -- [fermion, in=90, looseness=1.0, quarter right, edge label'= ${t}$] (b3),
(g1) -- [gluon, edge label' ={$G$}] (g2);
};
\end{feynman}
\end{scope}
\end{tikzpicture}
\caption{Loop-induced penguin diagrams for FCNC transitions. Left: effective operators with top-quark induced vertices. Right: their one-loop origin via \(W\)–\(t\) loops emitting \(Z\), \(\gamma\), or gluons G.}
\label{fig:FCNC1}
\end{figure}
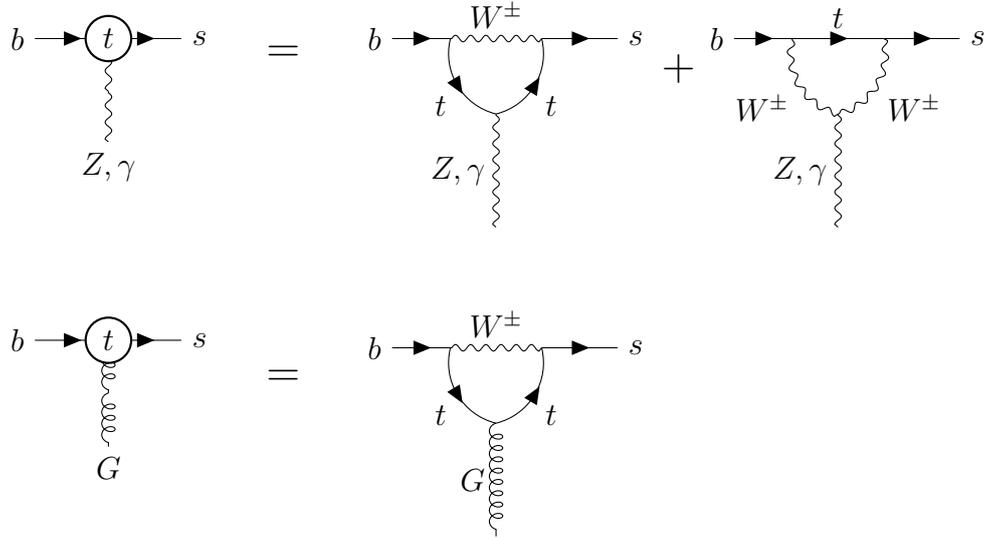

These vertices involve only quarks and are depicted, where $i$ and $j$ in Fig.(\ref{fig:FCNC1}) have the same charge but different flavors, and $t$ is the internal quark. These effective vertices can be calculated using the Feynman rules for elementary vertices and propagators in the SM.
\subsubsection*{Box Vertices}
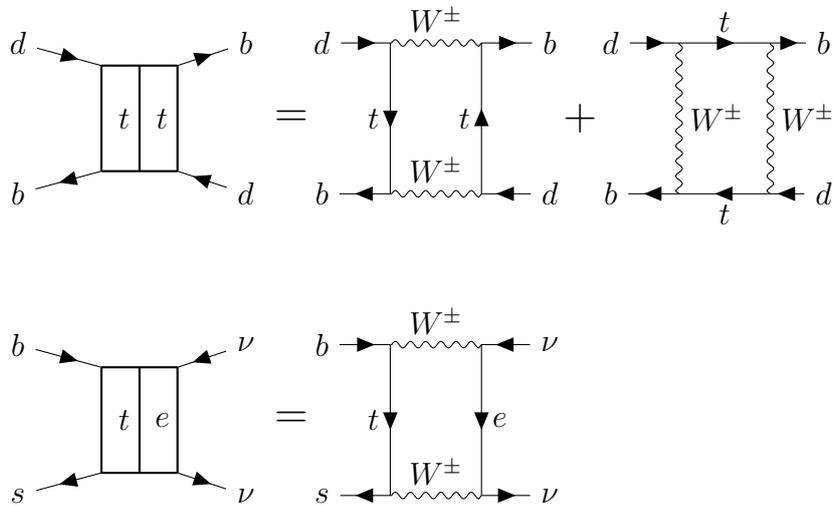
\begin{figure}[htp]
\centering
\begin{tikzpicture}
\begin{scope}[shift={(0,2)}]
\begin{feynman}
\vertex (a1) at (-0.5,1.5) {\(d\)};
\vertex (a2) at (2.5,1.5) {\(b\)};
\vertex (a3) at (-0.5,-0.5) {\(b\)};
\vertex (a4) at (2.5,-0.5) {\(d\)};

\vertex (v1) at (0.6,1.2);    
\vertex (v2) at (0.6,-0.2);   
\vertex (v3) at (1.6,1.2);    
\vertex (v31) at (1.1, 1.2);
\vertex (v4) at (1.6,-0.2);   
\vertex (v41) at (1.1,-0.2);
\diagram* {
  (a1) -- [fermion] (v1),
  (v3) -- [fermion] (a2),
  (v2) -- [fermion] (a3),
  (a4) -- [fermion] (v4)
};

\draw[thick] (v1) -- (v2);  
\draw[thick] (v3) -- (v4);  
\draw[thick] (v1) -- (v3);  
\draw[thick] (v2) -- (v4);  
\draw[thick] (v31) -- (v41);
\node at (0.9, 0.5) {\(t\)};
\node at (1.4, 0.5) {\(t\)};
\end{feynman}
\end{scope}

\node at (3.1,2.5) {\Large$=$};

\begin{scope}[shift={(4,2)}]
\begin{feynman}
\vertex (a1) at (-0.5,1.5) {\(d\)};
\vertex (a2) at (2.5,1.5) {\(b\)};
\vertex (a3) at (-0.5,-0.5) {\(b\)};
\vertex (a4) at (2.5,-0.5) {\(d\)};
\vertex (v1) at (0.4,1.5);
\vertex (v2) at (0.4,-0.5);
\vertex (v3) at (1.6,1.5);
\vertex (v4) at (1.6,-0.5);
\diagram* {
(a1) -- [fermion] (v1) -- [boson, edge label=$W^\pm$] (v3) -- [fermion] (a2),
(v2) -- [fermion] (a3),(v2) -- [boson, edge label=$W^\pm$](v4), (a4) -- [fermion] (v4),
(v1) -- [fermion, edge label'=$t$] (v2),
(v4) -- [fermion, edge label=$t$] (v3),
};
\end{feynman}
\end{scope}

\node at (6.9,2.5) {\Large$+$};

\begin{scope}[shift={(7.8,2.0)}]
\begin{feynman}
\vertex (a1) at (-0.5,1.5) {\(d\)};
\vertex (a2) at (2.3,1.5) {\(b\)};
\vertex (a3) at (-0.5,-0.5) {\(b\)};
\vertex (a4) at (2.3,-0.5) {\(d\)};
\vertex (v1) at (0.4,1.5);
\vertex (v2) at (1.6,1.5);
\vertex (v3) at (0.4,-0.5);
\vertex (v4) at (1.6,-0.5);
\diagram* {
(a1) -- [fermion] (v1) -- [fermion, edge label=$t$] (v2) -- [fermion] (a2),
(v3) -- [fermion](a3), (v4) -- [fermion, edge label=$t$] (v3),(a4) -- [fermion] (v4),
(v1) -- [boson, edge label=$W^\pm$] (v3),
(v2) -- [boson, edge label=$W^\pm$] (v4),
};
\end{feynman}
\end{scope}


\begin{scope}[shift={(0,-2)}]
\begin{feynman}
\vertex (a1) at (-0.5,1.5) {\(b\)};
\vertex (a2) at (2.5,1.5) {\(\nu\)};
\vertex (a3) at (-0.5,-0.5) {\(s\)};
\vertex (a4) at (2.5,-0.5) {\(\nu\)};

\vertex (v1) at (0.6,1.2);    
\vertex (v2) at (0.6,-0.2);   
\vertex (v3) at (1.6,1.2);    
\vertex (v31) at (1.1, 1.2);
\vertex (v4) at (1.6,-0.2);   
\vertex (v41) at (1.1,-0.2);
\diagram* {
  (a1) -- [fermion] (v1),
  (a2) -- [fermion] (v3),
  (v2) -- [fermion] (a3),
  (v4) -- [fermion] (a4)
};

\draw[thick] (v1) -- (v2);  
\draw[thick] (v3) -- (v4);  
\draw[thick] (v1) -- (v3);  
\draw[thick] (v2) -- (v4);  
\draw[thick] (v31) -- (v41);
\node at (0.9, 0.5) {\(t\)};
\node at (1.4, 0.5) {\(e\)};
\end{feynman}
\end{scope}

\node at (3.1,-1.5) {\Large$=$};

\begin{scope}[shift={(4,-2)}]
\begin{feynman}
\vertex (a1) at (-0.5,1.5) {\(b\)};
\vertex (a2) at (2.5,1.5) {\(\nu\)};
\vertex (a3) at (-0.5,-0.5) {\(s\)};
\vertex (a4) at (2.5,-0.5) {\(\nu\)};
\vertex (v1) at (0.4,1.5);
\vertex (v2) at (0.4,-0.5);
\vertex (v3) at (1.6,1.5);
\vertex (v4) at (1.6,-0.5);
\diagram* {
(a1) -- [fermion] (v1) -- [boson, edge label=$W^\pm$] (v3),(a2) -- [fermion] (v3),
(v2) -- [fermion](a3), (v2) -- [boson, edge label=$W^\pm$] (v4) -- [fermion] (a4),
(v1) -- [fermion, edge label'=$t$] (v2),
(v3) -- [fermion, edge label=$e$] (v4),
};
\end{feynman}
\end{scope}
\end{tikzpicture}
\caption{Box diagrams and their decompositions for \( b\bar{b} \to d\bar{d} \) and \( b \to s \nu \bar{\nu} \) transitions.}
\label{fig:box}
\end{figure}
In general, these vertices involve both quarks and leptons given in Fig.(\ref{fig:box}).
\subsubsection{Effective Vertices for FCNC}

The rules for effective vertices are in the ’t Hooft–Feynman gauge are as follows \cite{Buras:1998raa}:

\begin{equation}
    \mathrm{Box}( T_3= \frac{1}{2}) = \lambda_i\frac{ G_F}{\sqrt{2}}\frac{\alpha}{2\pi \sin^2\theta_W} [-4B_0(x_i)] (\bar{s} b)_{V-A} (\bar{\nu}\nu)_{V-A}\label{eq:2}
\end{equation}
\begin{equation}
    \mathrm{Box}( T_3= -\frac{1}{2}) = \lambda_i\frac{ G_F}{\sqrt{2}}\frac{\alpha}{2\pi \sin^2\theta_W} B_0(x_i) (\bar{s} b)_{V-A} (\bar{\mu}\mu)_{V-A}\label{eq:3}
\end{equation}
\begin{equation}
    \bar{s}Z_{\mu}b = i \lambda_i\frac{ G_F}{\sqrt{2}} \frac{e}{2\pi^2} M_Z^2 \frac{\cos\theta_W}{\sin\theta_W} C_0(x_i) \bar{s} \gamma_\mu (1-\gamma_5) b
    \label{eq:4}
\end{equation}
\begin{equation}
    \bar{s}\gamma b = -i \lambda_i\frac{ G_F}{\sqrt{2}} \frac{e}{8\pi^2} D_0(x_i) \bar{s}(q^2 \gamma_\mu-q_\mu \slashed{q}) (1-\gamma_5) b\label{eq:5}
\end{equation}
\begin{equation}
    \bar{s}G^a b = -i \lambda_i\frac{ G_F}{\sqrt{2}} \frac{g_s}{8\pi^2} E_0(x_i) \bar{s_\alpha}(q^2 \gamma_\mu-q_\mu \slashed{q}) (1-\gamma_5)T^a_{\alpha\beta} b_\beta \label{eq:6}
\end{equation}
\begin{equation}
    \bar{s}\gamma'b = -i\bar{\lambda}_i^*\frac{ G_F}{\sqrt{2}} \frac{e}{8\pi^2} D_0'(x_i) \bar{s}(i\sigma_{\mu\lambda}q^\lambda)[m_b(1+\gamma_5)]b \label{eq:7}
\end{equation}
\begin{equation}
    \bar{s}G^{a'}b = -i\bar{\lambda}_i^*\frac{ G_F}{\sqrt{2}} \frac{e}{8\pi^2} E_0'(x_i) \bar{s_\alpha}(i\sigma_{\mu\lambda}q^\lambda)[m_b(1+\gamma_5)]T^a_{\alpha\beta}b_\beta \label{eq:8}
\end{equation}
where, $\lambda_i = \sum_{i=u,c,t}V_{is}^*V_{ib}$ and $V-A= \gamma_\mu-\gamma_\mu \gamma_5$. 

The first rule involves only quarks, and the last two rules involve an on-shell photon and gluon. We have set $m_s = 0$ in these rules. The effective vertex rules, together with the propagator rules for the gauge bosons, provide the framework for calculating the effective Hamiltonians of FCNC processes, though without incorporating QCD corrections. 
The implementation of these rules demands careful attention, as detailed in \cite{Buras:1998raa},
\begin{itemize}
\item Once the mathematical expression associated with a given penguin diagram is determined, its contribution to the effective Hamiltonian is obtained by multiplying the expression by a factor of $i$.
\item For the box diagrams, the structure of the vertices ensures a direct contribution to the effective Hamiltonian.
\item The effective vertices depend on the masses of internal quarks or leptons and are calculable functions of $x_i = \frac{m_i^2}{M_W^2}, i=u,c,t$.
\item The eﬀective vertices depend on elements of the CKM matrix $\lambda_i$. This dependence can be found directly from the diagrams. 
\item The dependences of a given vertex on the CKM factors and the masses of internal
fermions govern the strength of the vertex in question.
\end{itemize}

\subsubsection{Basic Functions}

The basic functions present in eqs. (\ref{eq:2})–(\ref{eq:8}) given by Inami and Lim \cite{inami1981} are,
\begin{equation}
    B_0(x_t) = \frac{1}{4}\left[\frac{x_t}{1-x_t} + \frac{x_t \ln x_t}{(x_t-1)^2}\right]
    \label{eq:B0} 
\end{equation}
\begin{equation}
    C_0(x_t) = \frac{x_t}{4}\left[\frac{x_t-6}{x_t-1} + \frac{(3x_t+2) \ln x_t}{(x_t-1)^2}\right]
\end{equation}
\begin{equation}
    D_0(x_t) = -\frac{4\ln x_t}{9}+\frac{-19x_t^3+25x_t^2}{36(x_t-1)^3} + \frac{x_t^2(5x_t^2-2x_t-6) \ln x_t}{18(x_t-1)^4}
\end{equation}
\begin{equation}
    E_0(x_t) = -\frac{2\ln x_t}{3}+\frac{x_t^2(15-16x_t+4x_t^2)\ln{x_t}}{6(1-x_t)^4} + \frac{x_t(18-11x_t-x_t^2)}{12(1-x_t)^3}
\end{equation}
\begin{equation}
    D_0'(x_t) = -\frac{-8x_t^3+5x_t^2-7x_t}{12(1-x_t)^3} + \frac{x_t^2(2-3x_t) \ln x_t}{2(1-x_t)^4}
\end{equation}
\begin{equation}
    E_0'(x_t) = -\frac{x_t(x_t^2-5x_t-2}{4(1-x_t)^3} + \frac{3x_t^2 \ln x_t}{2(1-x_t)^4}
\end{equation}\label{eq:E0'}

The subscript “$0$” indicates that these functions do not include QCD corrections to the relevant diagrams. The first three functions are gauge dependent and, in an arbitrary gauge $R_\xi$, the functions $B_0(x_t,\xi), C_0(x_t,\xi)$, and $D_0(x_t,\xi)$ are given in \cite{Buchalla:1990qz}. Here we are using ’t Hooft– Feynman gauge with $(\xi = 1)$.
\begin{equation}
    C_0(x_t, \xi) - 4B_0(x_t, \xi, \frac{1}{2}) = C_0(x_t) - 4B_0(x_t) = X_0(x_t)
\end{equation}
\begin{equation}
    C_0(x_t, \xi) - B_0(x_t, \xi, -\frac{1}{2}) = C_0(x_t) - B_0(x_t) = Y_0(x_t)\label{eq:Y0}
\end{equation}
\begin{equation}
    C_0(x_t, \xi) +\frac{1}{4}D_0(x_t, \xi) = C_0(x_t) + \frac{1}{4}D_0(x_t) = Z_0(x_t)\label{eq:Z0}
\end{equation}

Here $X_0(x_t)$ and $Y_0(x_t)$ are linear combinations of the $V-A$ components of $Z$-penguin and box diagrams with final quarks with weak isospin $T_3$ equal to $\frac{1}{2}$ and $-\frac{1}{2}$, respectively. 
\begin{equation}
     X_0(x_t) = \frac{x_t}{8}\left[\frac{x_t+2}{x_t-1} + \frac{(3x_t-6) \ln x_t}{(x_t-1)^2}\right]
\end{equation}
\begin{equation}
     Y_0(x_t) = \frac{x_t}{8}\left[\frac{x_t-4}{x_t-1} + \frac{3x_t \ln x_t}{(x_t-1)^2}\right]
\end{equation}
$Z_0(x_t)$ is a linear combination of the vector component of the $Z^0$-penguin and the $\gamma$-penguin.
\begin{equation}
    Z_0(x_t) = -\frac{1\ln x_t}{9}+\frac{18x_t^4-163x_t^3+259x_t^2-108x_t}{144(x_t-1)^3} + \frac{32x_t^4-38x_t^3-15x_t^2+18x_t}{72(x_t-1)^4}
\end{equation}
Thus, the set of gauge independent basic functions that govern the FCNC processes are $X_0(x_t)$, $Y_0(x_t)$, $Z_0(x_t)$, $E_0(x_t)$, $D'_0(x_t)$ and $E^{'}_0(x_t)$.

\section{Rare B-Decays}
Rare B-decays represent an important class of transitions in which the bottom quark decays through channels other than the dominant $b\rightarrow c$ transition. By definition, these decays exhibit significantly lower branching fractions compared to the Cabibbo-favored modes, typically by several orders of magnitude. The primary categories of rare B-decays include Cabibbo-suppressed $b\rightarrow u$ transitions, loop-mediated $b\rightarrow s$ transitions, or $b\rightarrow d$ processes, and spectator-quark implied mechanisms such as W-exchange or annihilation. 
The theoretical framework underlying these processes is intrinsically connected to the Standard Model's flavor structure. The GIM mechanism is an integral component of the quark flavor mixing that ensures that FCNC transitions do not occur at tree level. FCNC transitions proceed exclusively through higher-order diagrams (penguins and boxes), and their rates get suppressed compared to those of the usually charged-current (CC) induced transitions. 

The study of rare B-decays, particularly those involving FCNCs, provides crucial insights into the fundamental parameters of the SM, including precise determinations of quark masses and CKM matrix elements. These processes serve as sensitive probes of the underlying flavor physics, offering a complementary approach to direct measurements and creating opportunities to test the consistency of the SM framework in the heavy quark sector.

\subsection{Flavor Changing Neutral Current (FCNC) in B-decays}
The FCNC $b\rightarrow sZ$ processes play a crucial role in testing the predictions of the SM and investigating its limits, allowing us to refine our understanding of the known particles and their interactions.

Our research focuses on comprehensive calculations that systematically extend the theoretical understanding of these rare B-decay processes in the presence of vector-like quark models. We establish quantitative relationships between observable decay parameters and the underlying theoretical framework by examining the modifications to FCNC amplitudes arising from these additional degrees of freedom. This approach enables us to identify distinctive signatures that could differentiate between the SM prediction and scenarios involving vector-like quarks, thereby providing a targeted methodology for probing specific BSM physics through rare B-decay phenomenology.

\subsubsection{ Semi-Leptonic B-Decays }
The FCNC transitions $b\rightarrow sl^+l^-$ play an important role in flavor physics. They are responsible for several important decays like $B\rightarrow Kl^+l^-$,$B\rightarrow K^*l^+l^-$,$B\rightarrow X_s\mu^+\mu^-$, $B\rightarrow X_s\gamma$, and $B_s\rightarrow \mu^+\mu^-$. The starting point of $B\rightarrow X_sl^+l^-$ transitions is effective Hamiltonian,
\begin{equation}
\mathcal{H}_{eff}(b\rightarrow sl\bar{l})=\mathcal{H}_{eff}(b\rightarrow s\gamma)-4\frac{G_F}{\sqrt{2}}\frac{\alpha}{4\pi}V_{ts}^*V_{tb}\left[C_9(\mu)O_9+C_{10}(\mu)O_{10}\right],
\end{equation}
 where $B\rightarrow X_s\gamma$ is governed by the operator $O_7$, the diagrams in which the photon couples to $W^\pm$ are analogous to  Z-penguin diagrams in Fig.\ref{fig:z-penguin}. The effective Hamiltonian for $B\rightarrow X_s\gamma$ is
\begin{equation}
\mathcal{H}_{eff}(b\rightarrow s\gamma)=-4\frac{G_F}{\sqrt{2}}\frac{\alpha}{4\pi}V_{ts}^*V_{tb}\left[\sum_{i=1}^{6}C_i(\mu)O_i+C_{7\gamma}(\mu)O_{7\gamma}\right],
\end{equation}
The operators $O_7$, $O_9$ and $O_{10}$ are given in eqs.(\ref{O7}-\ref{O10}). The Wilson coefficients at the scale of $\mu_W= O(M_W)$ are
\begin{equation}
C_7(\mu_W) = - \frac{1}{2} D'_0(x_t),\hspace{0.5cm}
   C_9(\mu_W) = \frac{\left[Y_0(x_t) - 4 \sin^2\theta_W Z_0(x_t)\right]}{ \sin^2 \theta_W }
\end{equation}
\begin{equation}
    \ C_{10} (\mu_W) = -\frac{Y_0 (x_t)}{sin^2 \theta_W}
\end{equation}
with functions $Y_0(x_t)$ and $Z_0(x_t)$ are given in eq.(\ref{eq:Y0}) and eq.(\ref{eq:Z0}). These Wilson coefficients evolved from the $M_W$ scale to the $m_b$ scale using the renormalization group equation,
\begin{equation}
  {C}(\mu_b) = \hat{U}(\mu_b, \mu_W) C(\mu_W)
\end{equation}
where $\hat{U}(\mu_b, \mu_W)$ is the evolution function and $\mu_b = \mathcal{O}(m_b)$ and $\mu_W = \mathcal{O}(M_W)$.At leading order,
\begin{equation}
    C(\mu_b) \approx \left(\frac{\alpha_s(\mu_b)}{\alpha_s(\mu_W)}\right)^{-\frac{\gamma_0}{2\beta_0}} \, C(\mu_W)
\end{equation}
where, $\eta = \frac{\alpha_s(M_W)}{\alpha_s(\mu_b)}$.
\subsubsection{Explicit Calculation of Z-Penguin Diagrams}

\begin{figure}[h!]
    \centering
    \begin{minipage}{0.25\textwidth}
        \centering
        \resizebox{\linewidth}{!}{%
        \begin{tikzpicture}
            \begin{feynman}
                \vertex (b1) {$b$};
                \vertex [right=1.2cm of b1] (b2);
                \vertex [right=1.2cm of b2] (b3);
                \vertex [right=1.2cm of b3] (b4) {$s$};
                \vertex at ($(b2)!0.5!(b3)!0.8cm!-90:(b3)$) (g1);
                \vertex [below= of g1] (g2);
                \diagram* {
                    (b1) -- [fermion] (b2) -- [boson, edge label={$W^{-}$}] (b3) -- [fermion] (b4),
                    (b2) -- [fermion, out=90, looseness=1, quarter right, edge label'=${u,c,t}$] (g1)
                          -- [fermion, in=90, looseness=1, quarter right, edge label'=${u,c,t}$] (b3),
                    (g1) -- [boson, edge label'=$Z$] (g2);
                };
            \end{feynman}
        \end{tikzpicture}}
        (a)
    \end{minipage}
  \hspace{0.20\textwidth}
    \begin{minipage}{0.25\textwidth}
        \centering
        \resizebox{\linewidth}{!}{%
        \begin{tikzpicture}
            \begin{feynman}
                \vertex (b1) {$b$};
                \vertex [right=1.2cm of b1] (b2);
                \vertex [right=1.2cm of b2] (b3);
                \vertex [right=1.2cm of b3] (b4) {$s$};
                \vertex at ($(b2)!0.5!(b3)!0.8cm!-90:(b3)$) (g1);
                \vertex [below= of g1] (g2);
                \diagram* {
                    (b1) -- [fermion] (b2) -- [scalar, edge label={$\chi^{-}$}] (b3) -- [fermion] (b4),
                    (b2) -- [fermion, out=90, looseness=1, quarter right, edge label'=${u,c,t}$] (g1)
                          -- [fermion, in=90, looseness=1, quarter right, edge label'=${u,c,t}$] (b3),
                    (g1) -- [boson, edge label'=$Z$] (g2);
                };
            \end{feynman}
        \end{tikzpicture}}
        (b)
    \end{minipage}

    \vspace{0.5cm} 

    \begin{minipage}{0.25\textwidth}
        \centering
        \resizebox{\linewidth}{!}{%
        \begin{tikzpicture}
            \begin{feynman}
                \vertex (b1) {$b$};
                \vertex [right=1.2cm of b1] (b2);
                \vertex [right=1.2cm of b2] (b3);
                \vertex [right=1.2cm of b3] (b4) {$s$};
                \vertex at ($(b2)!0.5!(b3)!0.8cm!-90:(b3)$) (g1);
                \vertex [below= of g1] (g2);
                \diagram* {
                    (b1) -- [fermion] (b2) -- [fermion, edge label={$u,c,t$}] (b3) -- [fermion] (b4),
                    (b2) -- [boson, out=90, looseness=1, quarter right, edge label'=$W^{-}$] (g1)
                          -- [boson, in=90, looseness=1, quarter right, edge label'=$W^{+}$] (b3),
                    (g1) -- [boson, edge label'=$Z$] (g2);
                };
            \end{feynman}
        \end{tikzpicture}}
        (c)
    \end{minipage}
    \hspace{0.20\textwidth}
    \begin{minipage}{0.25\textwidth}
        \centering
        \resizebox{\linewidth}{!}{%
        \begin{tikzpicture}
            \begin{feynman}
                \vertex (b1) {$b$};
                \vertex [right=1.2cm of b1] (b2);
                \vertex [right=1.2cm of b2] (b3);
                \vertex [right=1.2cm of b3] (b4) {$s$};
                \vertex at ($(b2)!0.5!(b3)!0.8cm!-90:(b3)$) (g1);
                \vertex [below= of g1] (g2);
                \diagram* {
                    (b1) -- [fermion] (b2) -- [fermion, edge label={$u,c,t$}] (b3) -- [fermion] (b4),
                    (b2) -- [scalar, out=90, looseness=1, quarter right, edge label'=$\chi^{-}$] (g1)
                          -- [scalar, in=90, looseness=1, quarter right, edge label'=$\chi^{+}$] (b3),
                    (g1) -- [boson, edge label'=$Z$] (g2);
                };
            \end{feynman}
        \end{tikzpicture}}
        (d)
    \end{minipage}

    \vspace{0.5cm}

    \begin{minipage}{0.25\textwidth}
        \centering
        \resizebox{\linewidth}{!}{%
        \begin{tikzpicture}
            \begin{feynman}
                \vertex (b1) {$b$};
                \vertex [right=1.2cm of b1] (b2);
                \vertex [right=1.2cm of b2] (b3);
                \vertex [right=1.2cm of b3] (b4) {$s$};
                \vertex at ($(b2)!0.5!(b3)!0.8cm!-90:(b3)$) (g1);
                \vertex [below= of g1] (g2);
                \diagram* {
                    (b1) -- [fermion] (b2) -- [fermion, edge label={$u,c,t$}] (b3) -- [fermion] (b4),
                    (b2) -- [boson, out=90, looseness=1, quarter right, edge label'=$W^{-}$] (g1)
                          -- [scalar, in=90, looseness=1, quarter right, edge label'=$\chi^{+}$] (b3),
                    (g1) -- [boson, edge label'=$Z$] (g2);
                };
            \end{feynman}
        \end{tikzpicture}}
        (e)
    \end{minipage}
     \hspace{0.20\textwidth}
    \begin{minipage}{0.25\textwidth}
        \centering
        \resizebox{\linewidth}{!}{%
        \begin{tikzpicture}
            \begin{feynman}
                \vertex (b1) {$b$};
                \vertex [right=1.2cm of b1] (b2);
                \vertex [right=1.2cm of b2] (b3);
                \vertex [right=1.2cm of b3] (b4) {$s$};
                \vertex at ($(b2)!0.5!(b3)!0.8cm!-90:(b3)$) (g1);
                \vertex [below= of g1] (g2);
                \diagram* {
                    (b1) -- [fermion] (b2) -- [fermion, edge label={$u,c,t$}] (b3) -- [fermion] (b4),
                    (b2) -- [scalar, out=90, looseness=1, quarter right, edge label'=$\chi^{-}$] (g1)
                          -- [boson, in=90, looseness=1, quarter right, edge label'=$W^{+}$] (b3),
                    (g1) -- [boson, edge label'=$Z$] (g2);
                };
            \end{feynman}
        \end{tikzpicture}}
        (f)
    \end{minipage}

    \caption{Z-penguin diagrams contributing to $b \to s Z$. Diagrams (a) and (b) show $W$ and $\chi^\pm$ loops, (c) and (d) two-boson loops ($W^+W^-$ or $\chi^+\chi^-$), and (e) and (f) mixed $W$ and $\chi^\pm$ loops. Internal up-type quarks ($u, c, t$) circulate in the loops, adapted from \cite{Buras:1998raa}}
    \label{fig:z-penguin}
\end{figure}
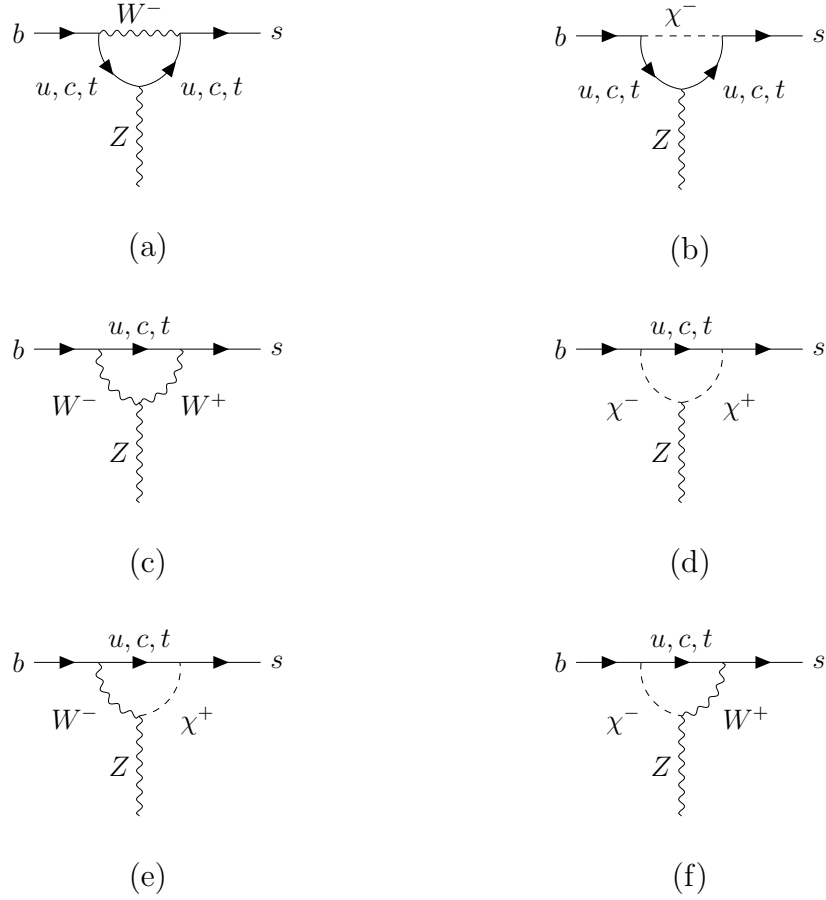

 In evaluating the diagrams shown in Fig.\ref{fig:z-penguin}, one must perform the integration over the internal loop momentum. However, the situation is more subtle: the amplitudes of diagrams in Fig.\ref{fig:z-penguin}(a,c) and Fig.~\ref{fig:z-penguin}(d) turn out to be divergent. These divergences reflect the ultraviolet (UV) behavior of the loop integrals and signal that the calculation must be carried out within a renormalized framework, where the divergences are absorbed into counterterms or matched onto effective operators. Thus, beyond simply performing the loop integration, it is crucial to account for these divergences in order to obtain a finite and physically meaningful effective Hamiltonian.

The induced flavor-violating vertex $\bar{s}Z_\mu b$,
\begin{equation}
    \Gamma_Z^\mu \equiv \frac{i g_2^3}{16\pi^2} \frac{1}{\cos\theta_W} V_{ib} V_{is}^* C_0(x_i) \bar{s} \gamma^\mu (1 - \gamma_5) b
\end{equation}
 with $i=u,c,t$ top quark is dominant in the loop diagram. All diagrams contributing to $\Gamma_{Z}^{\mu}$ given in Fig.(\ref{fig:z-penguin}) and Fig.(\ref{fig:Self energy})are in Feynman gauge $\xi =1$ \cite{Buras:1998raa}. We have to calculate the $C_0(x_t)$ function, given the Z coupling,
\begin{equation}
    i\bar{f}Z_\mu f = i\frac{g_2}{2\cos\theta_W}\gamma_\mu \left[a_f(1+\gamma_5)+b_f(1-\gamma_5)\right],
\end{equation}
From \cite{Buras:1998raa}, $a_f$ and $b_f$ are given as
\begin{equation}
    a_f = -Q_f \sin^2 \theta_W, \quad b_f = T_3(f) - Q_f \sin^2 \theta_W
    \label{afbf}
\end{equation}
with $f$ being the internal quark. Moreover, we will only show external spinors and the CKM factors in the final formula for every diagram.
 Since we have massive internal propagators, we can set all external momenta ($p\ll M$) to zero. Using the Feynman rules, the contribution of the diagram (\ref{fig:z-penguin}a) is given by
\begin{equation}
\begin{aligned}
\Delta_a\Gamma^\mu(Z) &= \int \frac{d^4k}{(2\pi)^4} \left[ \left( \frac{ig_2}{2\sqrt{2}} V_{ts}^* \bar{s} \gamma^\mu(1 - \gamma_5) \right) 
\frac{-ig_{\mu\nu}}{K^2 - M_W^2} \frac{i \slashed{k} + m_t}{k^2 - m_t^2} \right. \\
&\quad \times \left[  \left( \frac{ig_2}{2 \cos \theta_W} \gamma_\rho \left[a_t(1 + \gamma_5) + b_t(1 - \gamma_5)\right] \frac{i \slashed{k} + m_t}{k^2 - m_t^2} \right)
\left( \frac{ig_2}{2\sqrt{2}} V_{tb} \gamma_\nu(1 - \gamma_5) b \right) \right] 
\end{aligned}
\label{eq:Integral1}
\end{equation}
 where $g_2$ is the $SU(2)_L$ coupling, and by simplifying the above equation we have,
\begin{equation}
\Gamma_a^\mu(Z)=\frac{ g_2^3}{16\cos\theta_W}\int \frac{d^4k}{(2\pi)^4} \frac{k_\rho k_\lambda A_1^{\mu\rho\lambda}+m_t^2A_2^\mu}{(k^2 - m_t^2)^2 (k^2 - M_W^2)} V_{tb}V_{ts}^*\bar{s}\gamma^\mu(1- \gamma_5)b,
\label{eq:Integral1}
\end{equation}
where after shifting $(1- \gamma_5)$ to the right of all $\gamma_\mu$ matrices, we have
\begin{equation}
    A_1^{\mu\rho\lambda}= 4b_t\gamma_\nu\gamma^\rho\gamma^\mu\gamma^\lambda\gamma^\nu(1-\gamma_5)
\end{equation}
and
\begin{equation}
    A_2^\mu=4a_t\gamma_\nu\gamma^\mu\gamma^\nu(1-\gamma_5)
\end{equation}
  Using DR, Feynman diagrams are evaluated in $ D = 4 - 2\epsilon$ dimensions to regulate the divergences in the individual diagrams, and the singularities are extracted as poles for $\epsilon\rightarrow 0$.

 The D-dimension integrals in eq.(\ref{eq:Integral1}) are given in Appendix B as,
\begin{equation}
    \int \frac{d^Dk}{(2\pi)^D} \frac{k_\rho k_\lambda}{(k^2 - m_t^2)^2 (k^2 - M_W^2)}=\frac{ig_{\rho\lambda}}{32\pi^2}\left[\frac{1}{\bar{\epsilon}}+\frac{3}{4}+F_1(x_t)]\right],
\end{equation}
and
\begin{equation}
    \int \frac{d^Dk}{(2\pi)^D} \frac{1}{(k^2 - m_t^2)^2 (k^2 - M_W^2)}=\frac{i}{16\pi^2}\frac{1}{M_W^2}R_1(x_t),
\end{equation}
where 
\begin{align}
R_1(x_t) &= \frac{\log(x_t)}{(1-x_t)^2} + \frac{1}{(1-x_t)} \\
F_1(x_t) &= \frac{-1}{2(1-x_t)^2} \Big[x_t^2 \log(x_t) - 2 x_t \log(x_t) - x_t(1-x_t)\Big] 
\\
\frac{1}{\bar{\epsilon}} &= \frac{1}{2\epsilon} + \frac{1}{2} \Big[\log(4\pi) - \gamma_E + \log\Big(\frac{\mu^2}{M^2}\Big)\Big]. 
\label{eq:E}
\end{align}

and \( g^{\mu \nu} g_{\mu \nu} = D \) and $x_t=\frac{m_t^2}{M_W^2}$. Using Dirac matrices in D dimensions, we have
\begin{equation}
    g^{\rho \lambda} A_1^{\mu \rho \lambda} = 16 b_t (1 - 2 \epsilon) \gamma^\mu (1 - \gamma_5)
\end{equation}
\begin{equation}
    A_2^{\mu}= - 8a_t(1-\epsilon)\gamma^\mu(1-\gamma_5)
\end{equation}

Inserting  all expressions in the eq.(\ref{eq:Integral1}), we get
\begin{equation}
\begin{split}
\Delta_a \Gamma^\mu(Z) = \frac{g_2^3}{16 \cos\theta_W} \frac{i}{32 \pi^2} \Big[ 
&16 b_t \, \bar{s}(1 - 2\epsilon) \gamma^\mu (1 - \gamma_5) b 
\Big( \frac{1}{\bar{\epsilon}} - \frac{3}{4} + F_1(x_t) \Big) \\
&- 16 a_t x_t \, \bar{s}(1 - \epsilon) \gamma^\mu (1 - \gamma_5) b \, R_1(x_t) 
\Big] V_{tb} V_{ts}^*.
\end{split}
\end{equation}
 we are keeping $O(\epsilon)$ terms, as they will contribute after the multiplication by $1/\bar{\epsilon}$ defined in eq.(\ref{eq:E}). 
\begin{equation}
\Delta_a\Gamma^\mu(Z)=\frac{i g_2^3}{32\pi^2\cos\theta_W}\left[b_t\left(\frac{1}{\bar{\epsilon}}-\frac{1}{4}+F_1(x_t)\right)-a_tx_tR_1(x_t)\right]V_{tb}V_{ts}^*\bar{s}\gamma^\mu(1- \gamma_5)b,
\label{fig.a}
\end{equation}

 The contribution of the diagram Fig.(\ref{fig:z-penguin}b) is given by
\begin{equation}
\Delta_b\Gamma^\mu(Z)= ix_t\frac{g_2^3}{64\pi^2\cos\theta_W}\left[a_t\left(\frac{1}{\bar{\epsilon}}+\frac{1}{4}+F_1(x_t)\right)+b_tx_tR_t(x_t)\right]V_{tb}V_{ts}^*\bar{s}\gamma^\mu(1- \gamma_5)b,
\label{eq:b}
\end{equation}
The diagrams (c) and (d) of Fig.(\ref{fig:z-penguin}) are calculated as
\begin{equation}
\Delta_c\Gamma^\mu(Z)= - i\frac{g_2^3}{32\pi^2}\cos\theta_W\left[\frac{3}{\bar{\epsilon}}+\frac{5}{4}+ 3F_2(x_t)\right]V_{tb}V_{ts}^*\bar{s}\gamma^\mu(1- \gamma_5)b,
\label{fig.c}
\end{equation}
\begin{equation}
\Delta_d\Gamma^\mu(Z)= i\frac{g_2^3}{128\pi^2}\left[\frac{\sin^2\theta_W-\cos^2\theta_W}{\cos\theta_W}\right]x_i\left[\frac{1}{\bar{\epsilon}}+\frac{3}{4}+ F_2(x_t)\right]V_{tb}V_{ts}^*\bar{s}\gamma^\mu(1- \gamma_5)b,
\label{eq:d}
\end{equation}
The final two diagrams give an identical result and are finite. Including a factor of two, the contribution of these two diagrams combined is given by
\begin{equation}
\Delta_{e+f}\Gamma^\mu(Z)= - i\frac{g_2^3}{32\pi^2}\frac{\sin^2\theta_W}{\cos\theta_W}x_t\left[F_2(x_t)- F_1(x_t)-\frac{1}{2}\right]V_{tb}V_{ts}^*\bar{s}\gamma^\mu(1- \gamma_5)b
\label{fig.ef}
\end{equation}
where,
\begin{equation} 
F_2(x_t)= -\frac{1}{2(1-x_t)^2}\left[x_t^2\log x_t + (1 - x_t)\right]
\end{equation}
 The singularities in eq.(\ref{fig.a}) and eq.(\ref{fig.c}) are canceled separately due to the unitarity of the CKM matrix when summation over the internal quarks is performed ($\lambda_u + \lambda_c +\lambda_t =0$). Also, the contribution in eq.(\ref{fig.ef}) is finite. However, in eq.(\ref{eq:b}) and eq.(\ref{eq:d}), singularities depend on the masses of exchanged quarks; it is evident that these singularities do not cancel each other, so the sum of the five contributions is divergent. These singularities disappear when contributions from the self energy diagrams in Fig.(\ref{fig:Self energy}) are considered.

\subsubsection*{Self-Energy}
 
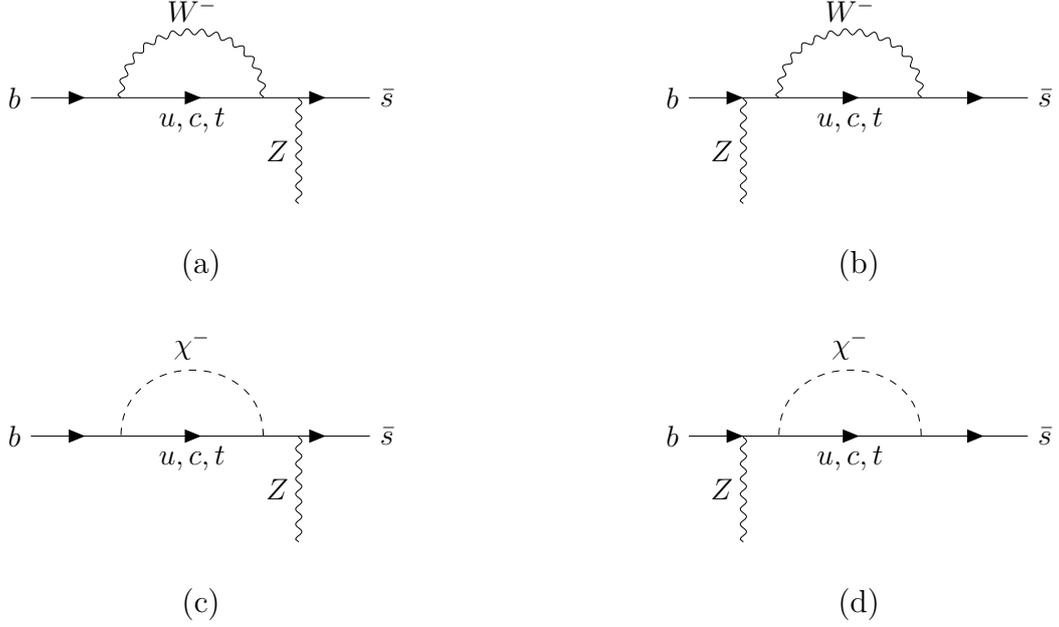
\begin{figure}[h!]
    \centering
    \begin{minipage}{0.35\textwidth}
        \centering
        \resizebox{\linewidth}{!}{%
        \begin{tikzpicture} 
            \begin{feynman}
                \vertex (a1) {$b$};
                \vertex[right=1.5cm of a1] (a2); 
                \vertex[right=1cm of a2] (a3); 
                \vertex[right=1cm of a3] (a4); 
                \vertex[right=0.5cm of a4] (a);
                \vertex[right=1.5cm of a4] (a5) {$\bar{s}$};
                \vertex[below=of a] (c1);

                \diagram* { 
                    (a1) -- [fermion] (a2);
                    (a2) -- [fermion, edge label'={$u,c,t$}] (a4);
                    (a4) -- [fermion] (a5);
                    (a) -- [boson, edge label'=$Z$] (c1),
                    (a4) -- [boson, out=90, in=90, looseness=1.6, edge label'={$W^-$}] (a2);
                };
            \end{feynman} 
        \end{tikzpicture}}
        {(a)}
    \end{minipage}
    \hspace{0.20\textwidth}
    \begin{minipage}{0.35\textwidth}
        \centering
        \resizebox{\linewidth}{!}{%
        \begin{tikzpicture} 
            \begin{feynman}
                \vertex (a1) {$b$};
                \vertex[right=1.5cm of a1] (a2); 
                \vertex[right=1cm of a2] (a3); 
                \vertex[right=1cm of a3] (a4); 
                \vertex[right=1cm of a1] (a);
                \vertex[right=1.5cm of a4] (a5) {$\bar{s}$};
                \vertex[below=of a] (c1);

                \diagram* { 
                    (a1) -- [fermion] (a2);
                    (a2) -- [fermion, edge label'={$u,c,t$}] (a4);
                    (a4) -- [fermion] (a5);
                    (a) -- [boson, edge label'=$Z$] (c1),
                    (a4) -- [boson, out=90, in=90, looseness=1.6, edge label'={$W^-$}] (a2);
                };
            \end{feynman} 
        \end{tikzpicture}}
        {(b)}
    \end{minipage}

    \vspace{0.5cm} 

    \begin{minipage}{0.35\textwidth}
        \centering
        \resizebox{\linewidth}{!}{%
        \begin{tikzpicture} 
            \begin{feynman}
                \vertex (a1) {$b$};
                \vertex[right=1.5cm of a1] (a2); 
                \vertex[right=1cm of a2] (a3); 
                \vertex[right=1cm of a3] (a4); 
                \vertex[right=0.5cm of a4] (a);
                \vertex[right=1.5cm of a4] (a5) {$\bar{s}$};
                \vertex[below=of a] (c1);

                \diagram* { 
                    (a1) -- [fermion] (a2);
                    (a2) -- [fermion, edge label'={$u,c,t$}] (a4);
                    (a4) -- [fermion] (a5);
                    (a) -- [boson, edge label'=$Z$] (c1),
                    (a4) -- [scalar, out=90, in=90, looseness=1.6, edge label'={$\chi^-$}] (a2);
                };
            \end{feynman} 
        \end{tikzpicture}}
        {(c)}
    \end{minipage}
    \hspace{0.20\textwidth}
    \begin{minipage}{0.35\textwidth}
        \centering
        \resizebox{\linewidth}{!}{%
        \begin{tikzpicture} 
            \begin{feynman}
                \vertex (a1) {$b$};
                \vertex[right=1.5cm of a1] (a2); 
                \vertex[right=1cm of a2] (a3); 
                \vertex[right=1cm of a3] (a4); 
                \vertex[right=1cm of a1] (a);
                \vertex[right=1.5cm of a4] (a5) {$\bar{s}$};
                \vertex[below=of a] (c1);

                \diagram* { 
                    (a1) -- [fermion] (a2);
                    (a2) -- [fermion, edge label'={$u,c,t$}] (a4);
                    (a4) -- [fermion] (a5);
                    (a) -- [boson, edge label'=$Z$] (c1),
                    (a4) -- [scalar, out=90, in=90, looseness=1.6, edge label'={$\chi^-$}] (a2);
                };
            \end{feynman} 
        \end{tikzpicture}}
       {(d)}
    \end{minipage}

\caption{Self-energy diagrams contributing to the $b \to s Z$ transition. Diagrams (a) and (b) correspond to loops with $W$ bosons, while (c) and (d) involve charged Goldstone ($\chi^\pm$) loops, and the $Z$ boson couples to the external quark line, adapted from \cite{Buras:1998raa}.}
    \label{fig:Self energy}
\end{figure}

Starting from the self-energy diagram Fig.(\ref{fig:Self energy}a) with external momentum p, we have

\begin{equation}
    A_a =\frac{g_2^2}{2}(1- \epsilon)\int \frac{d^Dk}{(2\pi)^D} \frac{\slashed{p}+\slashed{k}}{(k+p)^2 - m_t^2) (k^2 - M_W^2)}(1- \gamma_5)
\end{equation}
Solving the integral from Appendix B, including the CKM factor and external spinors, we find that
\begin{equation}
A_a = i\frac{g_2^2}{32\pi^2}\left[\frac{1}{\bar{\epsilon}}+\frac{1}{4}+F_2(x_t)\right]V_{tb}V_{ts}^*\bar{s}\slashed{p}(1- \gamma_5)b,
\label{Ka}
\end{equation}
gives a finite contribution after summation over internal quarks, and this is also the case of the self-energy diagram in Fig.(\ref{fig:Self energy}b) that gives the same result $A_b = A_a$. The total amplitude for diagram Fig.(\ref{fig:Self energy}a) is
\begin{align}
    \Delta_a\Gamma^\mu(Z) &=i\frac{g_2}{2\cos{\theta_W}}\gamma_\alpha (a_s(1+\gamma_5)+b_s(1-\gamma_5))\frac{i\slashed{p}+m_s}{p^2-m_s^2}K_a
\end{align}
Considering $m_s\approx0$ and using eq.(\ref{Ka}) we get,
\begin{align}
    \Delta_a\Gamma^\mu(Z) &=  -i \frac{g_3}{32\pi^2} \frac{b_b}{ \cos \theta_W }\left( \frac{1} { \overline{\epsilon}}+\frac{1}{4} + F_2(x_t)\right) V_{td} V^{*}_{ts}{\overline{s}} \gamma_\mu (1 - \gamma_5) b
\end{align}
We get the same result for diagram Fig.(\ref{fig:Self energy}b) as it is independent of external line, $b_s=b_b$. For Fig.(\ref{fig:Self energy}c) we have
\begin{equation}
    A_c = i\frac{g_2^2}{64\pi^2}x_t\left[\frac{1}{\bar{\epsilon}}+\frac{3}{4}+F_2(x_i)\right]V_{tb}V_{ts}^*\bar{s}\slashed{p}(1- \gamma_5)b,
\end{equation}
 also $A_d = A_c$ and is divergent, and cancels the divergence from vertex diagrams. where the internal $s$ and $b$ propagators are massless.

 Hence, we have
 \begin{equation}
     \Delta_{ a + b} \Gamma^\mu (Z) = -i \frac{g_3}{32\pi^2} \frac{b_b}{ \cos \theta_W }\left( \frac{1} { \overline{\epsilon}}+\frac{1}{4} + F_2(x_t)\right) V_{td} V^{*}_{ts}{\overline{s}}  \gamma_\mu (1 - \gamma_5) b
 \end{equation}
 and
 \begin{equation}
     \Delta_{ c + d} \Gamma^\mu (Z) = -i \frac{g_3}{64\pi^2} \frac{b_b}{ \cos \theta_W } x_t\left( \frac{1} { \overline{\epsilon}}+\frac{3}{4} + F_2(x_t)\right) V_{td} V^{*}_{ts}{\overline{s}}  \gamma_\mu (1 - \gamma_5) b
 \end{equation}
 Here, terms of order $O(\epsilon)$ have been neglected. Both contributions introduce finite modifications to the vertex results; however, the second contribution additionally cancels the divergences in eqs.~(\ref{eq:b}) and~(\ref{eq:d}), rendering the resulting $\Gamma_\mu(Z)$ finite.

 Adding all contributions and multiplying the result by i, we find the contribution of the Z-penguin to the effective Hamiltonian for the case of $b\rightarrow sZ$, using
 \begin{align}
\frac{g_2^2}{8 M^2_W} &= \frac{G_F}{\sqrt{2}}, \quad e^2 =g^2_2 \sin^2{\theta_W}, \quad
\label{G_F}
\end{align}
 and 
 \begin{align}
       \frac{M_W}{M_Z} = \cos\theta_W
 \end{align}
is same as given in eq.(\ref{eq:4})
\begin{align}
\mathcal{H}_{eff}= \frac{G_F^2}{\sqrt{2}}\frac{e M^2_Z}{2\pi^2} \frac{\cos{\theta_W}}{\sin{\theta_W}}  \lambda_t C_0(x_t)( \bar{s}b  )_{V-A} 
\end{align}
 Finally, by including all possible diagrams for the decay $b\rightarrow sZ$, we obtain
\begin{equation}
    C_0(x_t) = \frac{x_t}{4}\left[\frac{x_t-6}{x_t-1} + \frac{(3x_t+2) \log x_t}{(x_t-1)^2}\right]
\end{equation}

\subsubsection{Explicit Calculation of Box Diagram}

Let's calculate the box diagram for $b \to s \mu^+ \mu^-$. Setting $m_\nu = m_\mu = 0$, the couplings of Goldstone bosons to the external leptons vanish, so their contributions to the box diagram are zero. Thus, we are left only with the $W^\pm$ exchanges.
\begin{figure}[h!]
\centering
\begin{tikzpicture}
  \begin{feynman}
    \vertex (a) {\(b\)};
    \vertex[right=2cm of a] (b);
    \vertex[below=2cm of a] (c) {\(s\)};
    \vertex[right=2cm of c] (d);
    \vertex[right=2cm of b] (e);
    \vertex[right=2cm of d] (f);
    \vertex[right=2cm of e] (g) {\(\mu\)};
    \vertex[right=2cm of f] (h) {\(\mu\)};

    \diagram* {
      (a) -- [fermion] (b) -- [boson, edge label=\(W^\pm\)](e),
      (g) -- [fermion] (e),
      (d) -- [fermion] (c),(d) -- [boson, edge label'=\(W^\pm\)] (f) -- [fermion] (h),
      (b) -- [fermion, edge label=\(t\)] (d),
      (e) -- [fermion, edge label'=\(\nu_\mu\)] (f),
    };
  \end{feynman}
\end{tikzpicture}
\caption{Box diagram for the $b \to s \mu^+ \mu^-$ transition.}
\end{figure}
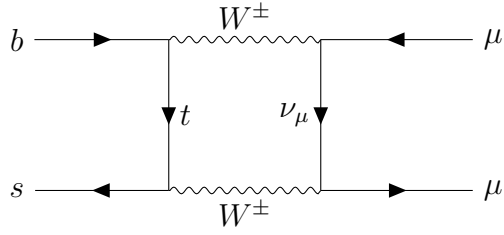

 Concentrating first on the internal top-quark contribution and using the Feynman rules, we have
\begin{equation}
    \mathcal{D}_{\text{box}}= \left(\frac{g_2}{2\sqrt{2}}\right)^4 \lambda_t T_{\sigma \tau}R^{\sigma \tau}
\end{equation}
where
\begin{equation}
R_{\sigma \tau} = \int \frac{d^4k}{(2\pi)^4} \, \frac{k_\sigma k_\tau}{\left[k^2 - m_t^2\right] \left[k^2\right] \left[k^2 - M_W^2\right]^2}
\end{equation}
and
\begin{equation}
    T_{\sigma \tau} = 4\, \bar{s} \gamma^\mu \gamma^\sigma \gamma^\nu (1 - \gamma^5) d \otimes \bar{\mu} \gamma_\nu \gamma_\tau \gamma_\mu (1 - \gamma^5) \mu
\end{equation}
Consequently, using the standard rules for Dirac matrices, we find \cite{Buras:1998raa}
\begin{equation}
g^{\sigma \tau} T_{\sigma \tau} = 16\, (\bar{s} d)_{V-A} \, (\bar{\mu} \mu)_{V-A}
\end{equation}
Each vertex contributes a factor of $\frac{ig}{\sqrt{2}} \gamma^\mu (1 -\gamma^5)$. As the box diagram is finite, we do not have to introduce any regulators. The integral $R_{\sigma \tau}$ can be easily evaluated by using Mathematica package \texttt{FeynCalc}. The resulting amplitude, obtained after performing loop integration and applying the unitarity of the CKM matrix, is

\begin{equation}
\mathcal{M}_{\text{box}} = -i \frac{G_F}{\sqrt{2}} \frac{\alpha}{2\pi \sin^2 \theta_W}  \lambda_t B_0(x_t) 
  (\bar{s} \gamma^\mu (1 - \gamma^5) b)(\bar{\mu} \gamma_\mu (1 - \gamma^5) \mu)
  \label{MB}
\end{equation}
Using eq.(\ref{G_F}) and by multiplying $i$ to eq.(\ref{MB}), we get the effective Hamiltonian
\begin{equation}
\mathcal{H}_{{eff}} = \frac{G_F^2}{2\pi^2}M^2_W  \lambda_t B_0(x_t) (\bar{s} \gamma^\mu (1 - \gamma^5) b)(\bar{\mu} \gamma_\mu (1 - \gamma^5) \mu)
\end{equation}
where,  $\lambda_t = V^*_{ts} V_{tb}$, $x_t = \frac{m^2_{t}}{M_W^2}$ and $B_0(x_t)$ is the loop function given as
  \begin{equation}
     B_0(x_t) = \frac{1}{4} \left[ \frac{x_t}{1 - x_t} + \frac{x_t \ln x_t}{(x_t - 1)^2} \right]  
  \end{equation}
With the effective Hamiltonian for $b \to s \mu^+ \mu^-$ now established, we are equipped to explore its implications in meson systems. In particular, the same loop-induced FCNC structures play a crucial role in $B$ mixing. In the following chapter, we extend the analysis by incorporating VLQs, which modify the standard loop contributions and can induce new effects in the $B$ sector. This sets the stage for a systematic study of their impact on $B$ decays and related observables.
\chapter{Beyond The Standard Model}
\label{chap:BSM}

\section{Vector-Like Quark Model (VQM)}
Throughout the research, several key B meson decays involving $b \to s$ transitions have been central, particularly those sensitive to FCNCs. The decays specified $B_s \to \mu^+\mu^-$, $B \to X_s \mu^+\mu^-$, $B \to X_s \gamma$, and $B \to K \nu\bar{\nu}$ exhibit tensions or constraints that motivate NP interpretations. The only way to confirm potential NP scenarios in the flavor sector, apart from direct detection of new particles, is by observing persistent anomalies in the experimental data.

\subsection*{Experimental tensions in $B$ decays}
Over the past decade, a number of intriguing discrepancies, collectively referred to as $B$ anomalies, have been reported in rare $B$ decays. Notably recent experimental updates show persistent discrepancies in some observables, while others provide stringent bounds.  
\begin{itemize}

  \item \textbf{Branching ratio $B_s\to\mu^+\mu^-$}: Very rare leptonic decays which tightly constrain scalar and some gauge NP; combined LHC measurements (ATLAS, CMS, LHCb) give the latest branching fraction determinations \cite{Czaja:2024the}, $\cal{BR}$ = $(3.34 \pm 0.27) \times 10^{-9}$,~$1.5\sigma$ below SM, $\cal{BR}$=$(3.66 \pm 0.14) \times 10^{-9}$ and should be used for fits. Small tensions or shifts relative to SM predictions have been discussed in combined analysis.

  \item\textbf{ Branching ratio $B\to X_s\mu^+\mu^-$:}
   The updated average branching ratio from Belle, BaBar, LHCb\cite{ParticleDataGroup:2024cfk} is $\cal{BR}$ = $(1.59 \pm 0.11) \times 10^{-6}$,~$1\sigma$ below SM overall, but ~$2-3\sigma$ deficits in low-q² bins (persistent after LHCb 2025 angular updates\cite{Gubernari:2022hxn}). 

 \item\textbf{ Branching ratio $B\to K\nu\bar{\nu}$:} 
 The recent Belle-II measurement is, $\cal{BR}$= $(2.3 \pm 0.7) \times 10^{-5}$,~$2.7\sigma$ above SM expectation $\cal{BR}$ = $(0.45 \pm 0.7)\times 10^{-5}$ \cite{Belle-II:2023esi}.
    
  \item\textbf{Angular observable  in $B\to K^*\mu^+\mu^-$:}
    Long-standing local deviations in one (or several) $q^2$ bins in the $B\to K^*\mu\mu$ angular analysis that motivated global fits to modified Wilson coefficients. Recent amplitude / angular analyses refine the picture but some tensions persist in certain kinematic regions.In $B\to K\mu^+\mu^-$\cite{LHCb:2014cxe}, the dominant uncertainties are hadronic in nature. 

   \item \textbf{Lepton flavor universality (LFU) violation in $b \to s \ell^+\ell^-$:}  
    Ratios such as $R_K$ and $R_{K^*}$, which compare $B \to K^{(*)} \mu^+\mu^-$ to $B \to K^{(*)} e^+e^-$, have consistently been measured below unity~\cite{LHCb:2022qnv}. From the theoretical side, uncertainties are small while experimental sensitivity continues to improve. 
 
\end{itemize}

Although one of these discrepancies alone is conclusive, their collective pattern across different decay modes consistently points towards possible NP in the flavor sector. These results strongly motivate theoretical scenarios that can naturally modify loop-induced processes and introduce new sources of flavor violation.

One of the simplest and most compelling approaches is to consider an extension of the quark sector itself. The SM contains three generations with two quarks each, but there is no fundamental principle requiring the total number of quarks to be limited to six. It is therefore plausible that heavier quarks exist but have not yet been observed at present collider energies. A minimal extension in this direction is achieved by introducing a vector-like isosinglet quark, either of up-type or down-type, into the particle spectrum. Unlike ordinary chiral fermions, these quarks are vector-like, meaning their left- and right-handed components transform identically under the gauge group. As a consequence, they do not introduce gauge anomalies and remain fully consistent within the theoretical framework.  

Vector-like quarks thus provide a natural and economical way to address the observed $B$ anomalies. Their presence modifies loop-induced amplitudes, their mixing with SM quarks, alters the structure of effective operators, and can leave a measurable imprint on processes such as $B\to K \nu \bar{\nu}$. In this sense, they offer a predictive and theoretically robust framework to connect the flavor anomalies discussed above with concrete NP effects.

In this chapter, we explore a specific extension of the SM by introducing an additional down-type vector-like quark (D). We will analyze how the mass term for this new quark appears, ensure the Lagrangian's invariance under the SM gauge group, modify the CKM matrix to a 3×4 structure, and investigate its implications on rare B-meson decays, particularly $b\rightarrow sZ$ and general B-decays. 
Rare radiative decays $B\rightarrow X_s\gamma$ and $  B\rightarrow X_sl^+l^-$  are sensitive probes of new physics. Unlike in the standard model, where FCNC arises only at the loop level, in the vector quark model (VQM), the CKM matrix is non-unitary, leading to $\bar{Z}sb$ interaction at the tree level. Hence, potentially significant NP contributions can be expected. Finally, we compare theoretical predictions with the experimental results from the Large Hadron Collider (LHC) and Belle-II \cite{Belle-II:2023esi}. Current results from Run II of the LHC, with center-of-mass energies of 13 TeV and integrated luminosities of up to 139 fb\(^{-1}\), place the following conservative bound on the down-type VLQ mass ~\cite{Banerjee:2024zvg}:
\begin{equation}
    M_D \geq 1.5\, \text{TeV},
\end{equation}

Consequently, to accurately describe these decays, it becomes necessary to incorporate all factors, including constant factors due to the non-unitarity effects introduced by VLQ into the basic functions. In the VQM, adding an extra isosinglet pair of quarks, U and D, with charges +2/3 and -1/3, respectively to the SM. Yukawa couplings between vector-like and ordinary quarks leads to mixing among the four up- and down-type quarks of the same charge.\cite{Ahmady:2001qh},

The Yukawa Lagrangian in the presence of the additional isosinglet quarks can be written as
\begin{equation}
\mathcal{L}_Y = - \bar{Q}_{Li} (Y_d)_{i\alpha} \, \phi \, d_{0R}^\alpha 
- \bar{Q}_{Li} (Y_u)_{i\beta} \, \tilde{\phi} \, u_{0R}^\beta 
+ \text{h.c.},
\end{equation}
where $i=1,2,3$ labels the SM quark doublets and $\alpha,\beta=1,\dots,4$ run over the three ordinary singlets plus the additional vector-like quark. Bare mass terms for VLQs are gauge invariant and, therefore, are prohibited by the gauge symmetry. Thus, their scale can be significantly larger than the electroweak scale $\mu_W$. The mass terms and mixings between the Standard Model (SM) quarks and Vector-Like Quarks (VLQs) after electroweak symmetry breaking can be written as: 
\begin{equation}
\begin{split}
\mathcal{L}_{\text{mass}} = 
\underbrace{
- \frac{v}{\sqrt{2}} \bar{d}_{L}^{i} (Y_{d})_{i}^{j} d_{R}^{j} 
- \frac{v}{\sqrt{2}} \bar{u}_{L}^{i} (Y_{u})_{i}^{j} u_{R}^{j}
}_{\text{SM quark masses}}
\underbrace{
- \bar{D}_{L}^{a} (M_{D})_{a}^{b} D_{R}^{b} 
- \bar{U}_{L}^{a} (M_{U})_{a}^{b} U_{R}^{b}
}_{\text{VLQ intrinsic masses}}\\
\underbrace{
- \bar{D}_{L}^{a} (\mu_{d})_{a}^{j} d_{R}^{j} 
- \bar{d}_{L}^{i} (\tilde{\mu}_{d})_{i}^{b} D_{R}^{b}
}_{\text{Down-type VLQ-SM mixing}}
\underbrace{
- \bar{U}_{L}^{a} (\mu_{u})_{a}^{j} u_{R}^{j} 
- \bar{u}_{L}^{i} (\tilde{\mu}_{u})_{i}^{b} U_{R}^{b}
}_{\text{Up-type VLQ-SM mixing}}
+ \text{h.c.}
\end{split}
\end{equation}
Here, $d_{L}^{i}, u_{L}^{i}$ are left-handed SM quark doublets with $i = 1,2,3$ generations and $d_{R}^{j}, u_{R}^{j}$ are right-handed SM quark singlets. $Y_{d}, Y_{u}$  are Yukawa couplings for down-type and up-type quarks. $v/\sqrt{2}$ is Higgs vacuum expectation value (vev) generating SM quark masses. 

$D_{L/R}^{a}, U_{L/R}^{a}$ are left- and right-handed vector-like quarks with $a = 1,\dots, n$ for $n$ VLQ generations. $M_{D}, M_{U}$  are gauge-invariant mass terms for VLQs, $\sim$ TeV scale. $\mu_{d}, \mu_{u}$  are mixing between left-handed VLQs and right-handed SM quarks and $\tilde{\mu}_{d}, \tilde{\mu}_{u}$  are mixing between left-handed SM quarks and right-handed VLQs. These terms arise from Yukawa interactions or explicit mass mixings.
   
If VLQs are embedded in $SU(2)_L$ multiplets (e.g., doublets or triplets), additional terms may appear:
\begin{itemize}
    \item Extra Yukawa couplings involving the Higgs field.
    \item Mixing with the SM quark doublets before symmetry breaking.
\end{itemize}
The mass Lagrangian for SM quarks and Vector-Like Quarks (VLQs) can be written compactly in matrix form as:

\begin{equation}
\mathcal{L}_M = - 
\begin{pmatrix}
\bar{d}_0^L & \bar{D}_0^L
\end{pmatrix}
\mathcal{M}_d 
\begin{pmatrix}
d_0^R \\
D_0^R
\end{pmatrix}
-
\begin{pmatrix}
\bar{u}_0^L & \bar{U}_0^L
\end{pmatrix}
\mathcal{M}_u 
\begin{pmatrix}
u_0^R \\
U_0^R
\end{pmatrix}
+ \text{h.c.},
\label{VLQ mass}
\end{equation}
where:
\begin{itemize}
    \item $d_0^L, u_0^L$ are the SM left-handed quark doublets ($3$ generations).
    \item $d_0^R, u_0^R$ are the SM right-handed quark singlets.
    \item $D_0^L, U_0^L$ and $D_0^R, U_0^R$ are the left- and right-handed VLQs.
\end{itemize}
\subsection{ Mass Matrix}
The mass matrices $\mathcal{M}_d$ and $\mathcal{M}_u$  in eq.(\ref{VLQ mass}) are  for down-type and up-type sectors:
\begin{equation}
\mathcal{M}_d = 
\begin{pmatrix}
\frac{v}{\sqrt{2}} Y_d & \tilde{\mu}_d \\
\mu_d & M_D
\end{pmatrix}, \quad
\mathcal{M}_u = 
\begin{pmatrix}
\frac{v}{\sqrt{2}} Y_u & \tilde{\mu}_u \\
\mu_u & M_U
\end{pmatrix}.
\end{equation}

The mass matrix for down-type quarks becomes a $4 \times 4$ matrix:
\begin{equation}
  \mathcal{M}_d =
\begin{pmatrix}
m_{dd} & m_{ds} & m_{db} & y_{dD}  \\
m_{sd}& m_{ss} & m_{sb} & y_{sD}  \\
m_{td} & m_{ts} & m_{tb} & y_{bD}  \\
0 &0 & 0& M_D
\end{pmatrix}, 
\label{md}
\end{equation}
Here, $ Y_{iD}=\lambda_i \frac{v}{\sqrt{2}}$ are Yukawa couplings, $v = \langle H \rangle \approx 246 \, \text{GeV}$ is the vacuum expectation value of the Higgs field, and $M_D$ is the mass of the down-type quark $D$. The zeros in the last row of eq.(\ref{md}) reflect that the left-handed down-type quark does not couple to right-handed SM quarks via Higgs interactions in this minimal model.

The $4\times4$ mass matrices $\mathcal{M}_d$ and $\mathcal{M}_u$ are generally not diagonal; transformations from weak to mass eigenstates make them diagonal.

\subsubsection{Diagonalization of Mass Matrices}

The physical masses and mixing angles of the quarks are obtained by diagonalizing the quark mass matrices, $\mathcal{M}_d$ and $\mathcal{M}_u$. In the presence of a vector-like quark (VLQ), the down-type quark sector is extended to include an additional quark, leading to a $4 \times 4$ mass matrix. These matrices are diagonalized through bi-unitary transformations.
\begin{equation}
\mathcal{M}_d^{\text{diag}} = U_L^{d\dagger} \, \mathcal{M}_d \, U_R^d, \qquad
\mathcal{M}_u^{\text{diag}} = U_L^{u\dagger} \, \mathcal{M}_u \, U_R^u,
\end{equation}
where \(U_L\) and \(U_R\) are unitary $4\times4$ matrices that rotate the quark fields from the gauge eigenstates to the mass eigenstates.
\begin{align}
\begin{pmatrix}
d_L^0 \\
D_L^0
\end{pmatrix}
= U_L^d
\begin{pmatrix}
d_L \\
D_L
\end{pmatrix}, \qquad
\begin{pmatrix}
d_R^0 \\
D_R^0
\end{pmatrix}
= U_R^d
\begin{pmatrix}
d_R \\
D_R
\end{pmatrix}.
\end{align}
To simplify the diagonalization process, first, we diagonalize Standard Model submatrix using $3\times3$ unitary matrices $A_L$ and $A_R$.
\begin{align}
d_L^i = A_L^{im} d_L'^m, \qquad
d_R^i = A_R^{im} d_R'^m,
\end{align}
where \( i = 1,\dots,4 \), and this leads to
\begin{equation}
A_L^\dagger \, \mathcal{M}_d \, A_R = \mathcal{M}_d' = \text{diag}(m_d', m_s', m_b', m_D'),
\end{equation}
where the prime indicates the mass basis of the Standard Model.

Next, we introduce small perturbative unitary martices $V_L$ and $V_R$ of order $\mathcal{O}(v^2/M_D^2)$, by including mixing with VLQ:
\begin{equation}
U_L^d \approx A_L V_L, \qquad U_R^d \approx A_R V_R.
\end{equation}

The SM quarks mix with the VLQ via off-diagonal Yukawa terms $\lambda_i v/\sqrt{2}$. The structure of the left-handed rotation matrix $U_L^d$ can then be written approximately as,
\begin{equation}
U_L^d \approx
\begin{pmatrix}
V_{\text{SM}} & \dfrac{\lambda_i v}{\sqrt{2} M_D} \\
-\left( \dfrac{\lambda_i v}{\sqrt{2} M_D} \right)^\dagger & 1
\end{pmatrix},
\end{equation}
where $V_{\text{SM}}$ is the unitary matrix that diagonalizes the SM submatrix. Since $M_D \gg v$, the mass of VLQ dominates and $\frac{\lambda_i v}{\sqrt{2} M_D}$ is small. In this mass basis, the down-type mass matrix receives VLQ-induced corrections:
\begin{equation}
A_L^{\dagger mj} \left( \delta^{ji} - \frac{v^2}{4 M_D^2} h_d^{ji} \right) \mathcal{M}_d^{ik} A_R^{kn}
= \left( \delta^{mn} - \frac{v^2}{4 M_D^2} h_d'^{mn} \right) \mathcal{M}_d'^n, 
\label{eq:MassMatrixPrime}
\end{equation}
where
\begin{equation}
h_d'^{mn} = (A_L^\dagger h_d A_L)^{mn} = A_L^{\dagger mj} y_d^{j4} y_d^{i4*} A_L^{in}
= y_d'^{m4} y_d'^{n4*}. 
\label{eq:hdPrime}
\end{equation}
The mass matrix in eq.(\ref{eq:MassMatrixPrime}) is not diagonal, to diagonalize this corrected mass matrix up to $\mathcal{O}(v^2/M_D^2)$, we apply an additional unitary transformation \cite{Morozumi:2018cnc},
\begin{align}
d_L'^m = V_L^{mp} d_L''^p, \qquad d_R'^m = V_R^{mp} d_R''^p,
\end{align}
yielding
\begin{equation}
V_L^{\dagger pm} \left( \delta^{mn} - \frac{v^2}{4 M_D^2} h_d'^{mn} \right) m_d'^n V_R^{nq}
= m_d''^p \delta^{pq},
\end{equation}
where double primes denote the final physical mass eigenbasis including VLQ, and $m_d''^p = (m_d'', m_s'', m_b'', m_D'')$. The mixing angles in $V_L$ and $V_R$ are of the order $\mathcal{O}(v^2/M_D^2)$. Hereafter, we omit the double primes, and denote $h_d \equiv h_d'$.

In this basis, the effective $3\times3$ CKM matrix is defined as:
\begin{equation}
V_{\text{CKM}} = A_L \left( 1 - \frac{v^2}{4 M_D^2} h_d \right) V_L.
\label{eq:CKM}
\end{equation}

FCNCs arise in the $Z$, $h$, and $\chi_0$ interactions through the non-diagonal structure of the matrix $Z_{\text{NC}}$, given by,
\begin{equation}
Z_{\text{NC}} = V_L^\dagger \left( 1 - \frac{v^2}{2 M_D^2} h_d \right) V_L
\simeq 1 - \frac{v^2}{2 M_D^2} h_d + \mathcal{O}\left( \frac{v^4}{M_D^4} \right). 
\label{eq:ZNC}
\end{equation}
Using Eqs.~\eqref{eq:CKM} and \eqref{eq:ZNC}, the unitarity relation becomes,
\begin{equation}
\sum_{i=u,c,t} V_{\text{CKM}}^{ip*} V_{\text{CKM}}^{iq} =
Z_{\text{NC}}^{pq} = \delta^{pq} - \frac{v^2}{2 M_D^2} h_d^{pq} + \mathcal{O}\left( \frac{v^4}{M_D^4} \right).
\label{eq:unitarity}
\end{equation}
where, $p, q$ are the mixing quarks. This result shows that the CKM matrix is no longer unitary due to mixing with the vector-like quark. The deviation from unitarity is encoded in the $Z_{\text{NC}}$ matrix. The unitarity is restored in the decoupling limit $M_D \to \infty $\cite{Morozumi:2018cnc}.

\section{Extended CKM Matrix}

In the presence of a down-type vector-like quark (VLQ), the left- and right-handed quark fields are rotated by unitary matrices to go from the weak to the mass eigenbasis:
\begin{align}
    d'_{L,R} &= U^{d}_{L,R} \, d_{L,R}, 
    & u'_{L,R} &= U^{u}_{L,R} \, u_{L,R}.
\end{align}

where primes denote mass eigenstates, and unprimed fields are in the weak basis. These rotations can be represented explicitly as:
\begin{align}
\begin{pmatrix}
d'_L \\
s'_L \\
b'_L \\
D'_L
\end{pmatrix}
=
U_L^d
\begin{pmatrix}
d_L \\
s_L \\
b_L \\
D_L
\end{pmatrix}, \quad
\begin{pmatrix}
u'_L \\
c'_L \\
t'_L
\end{pmatrix}
=
U_L^u
\begin{pmatrix}
u_L \\
c_L \\
t_L
\end{pmatrix}.
\end{align}
\( U_{L,R}^d \) are the full unitary matrices that diagonalize the $4 \times 4$ mass matrix $\mathcal{M}_d$:
\begin{equation}
    \mathcal{M}_d^{\rm diag} = U_L^{d\dagger} \, \mathcal{M}_d \, U_R^d.
    \label{eq:md_diag}
\end{equation}
and
\begin{equation}
U_L^d = A_L V_L, \quad U_R^d = A_R V_R.
\label{eq:UL_UR_factor}
\end{equation}

\subsection*{Charged Current}

The charged current interaction in the weak basis is:
\begin{equation}
  \mathcal{L}_W = -\frac{g}{\sqrt{2}} \bar{u}_L \gamma^\mu d_L W^+_\mu + \text{h.c.}  
\end{equation}
After rotating to the mass basis, the generalized CKM matrix appears:
\begin{equation}
\mathcal{L}_W = -\frac{g}{\sqrt{2}} \bar{u}_L' \gamma^\mu V_{\text{CKM}} d_L'' W^+_\mu + \text{h.c.},
\end{equation}
The generalized CKM matrix arises from the mismatch between up-type and down-type field rotations.
\begin{equation}
V_{\text{CKM}} = U_L^{u\dagger} U_L^d.
\end{equation}
Since the up-sector has no VLQ, its field rotation remains \( U_L^u = A_L^u \). Therefore
\begin{equation}
V_{\text{CKM}} = A_L^{u\dagger} A_L^d V_L.
\label{eq:CKMcombined}
\end{equation}
Hence,
\begin{equation}
V_{\text{CKM}} =
\begin{pmatrix}
V_{ud} & V_{us} & V_{ub} & V_{uD} \\
V_{cd} & V_{cs} & V_{cb} & V_{cD} \\
V_{td} & V_{ts} & V_{tb} & V_{tD}
\end{pmatrix},
\end{equation}
The CKM matrix in this framework becomes a \(3 \times 4\) matrix because of the additional down-type VLQ. The inclusion of the fourth (VLQ) column makes the matrix non-unitary.  

\subsection*{Neutral Currents }
Neutral current interactions are modified as 
\begin{equation}
\mathcal{L}_Z = \frac{g}{\cos \theta_W} Z_\mu \left( \bar{u}_L \gamma^\mu U^u u_L - \bar{d}_L \gamma^\mu U^d d_L \right),
\end{equation}
with neutral mixing matrices,
\begin{equation}
    U^u  = V_{\text{CKM}} V_{\text{CKM}}^\dagger, \qquad
    U^d = V_{\text{CKM}}^\dagger V_{\text{CKM}}. 
\end{equation}
   
Using eq.(\ref{eq:CKMcombined}),we get
\begin{align}
U^d &= V_{\text{CKM}}^\dagger V_{\text{CKM}} 
= V_L^\dagger A_L^{d\dagger} A_L^u A_L^{u\dagger} A_L^d V_L.
\end{align}
The up-type rotation matrix \( A_L^u \) is unitary (since no VLQ is introduced in the up-sector),
\begin{equation}
U^d = V_L^\dagger A_L^{d\dagger} A_L^d V_L.
\label{eq:Ud_simplified}
\end{equation}
Since \( A_L^d \) is a \( 4\times4 \) unitary matrix, its projection onto the SM \( 3\times3 \) subspace is no longer unitary. 
\begin{equation}
\left( A_L^{d\dagger} A_L^d \right)^{\alpha\beta} = \delta^{\alpha\beta} - A_L^{4\alpha*} A_L^{4\beta},
\end{equation}
where \( \alpha, \beta = d, s, b \). Substituting into eq.(\ref{eq:Ud_simplified}), we obtain:
\begin{equation}
U^d = \delta^{\alpha\beta} - V_L^\dagger A_L^{4\alpha*} A_L^{4\beta} V_L + \cdots.
\end{equation}

This clearly shows that \( U^d \neq \mathbb{I} \), indicating that the CKM matrix is not unitary. The deviation from unitarity arises from mixing with the fourth-generation down-type VLQ through the elements \( A_L^{4\alpha} \), and results in FCNC at tree level. These FCNCs can also be enhanced by loop effects, such as Higgs-mediated contributions at loop level also mentioned in~\cite{Morozumi:2018cnc}.
\begin{equation}
\sum_{i=u,c,t} V_{\text{CKM}}^{ip*} V_{\text{CKM}}^{iq} =
Z_{\text{NC}}^{pq} =U^{pq}.
\end{equation}

The presence of a vector-like quark, whose left- and right-handed components are both $SU(2)$ singlets, leads to the violation of unitarity in the effective CKM and neutral current mixing matrices.

\begin{align}
U^{\alpha\beta} &\equiv \sum_{i=1}^{3} U_L^{qi\alpha *} U_L^{qi\beta}
= \delta^{\alpha\beta} - U_L^{q\alpha 4} U_L^{q\beta 4*}, \\
&= 
\begin{cases}
(V_{\text{CKM}}^\dagger V_{\text{CKM}})^{\alpha\beta}, & \text{down-type}, \\
(V_{\text{CKM}} V_{\text{CKM}}^\dagger)^{\alpha\beta}, & \text{up-type}.
\end{cases}
\end{align}
These matrices are non-unitary due to the mixing with the heavy VLQ, and their off-diagonal elements lead to flavor-changing neutral currents:
\begin{equation}
U^{\alpha\beta} = \sum_{i=u,c,t} V_{i\alpha}^* V_{i\beta} \neq 0, \quad \text{for } \alpha \neq \beta.
\end{equation}
Since the various $U^{\alpha\beta}$ are non-vanishing they would signal NP and the presence of FCNC at the tree level and Higgs loop can substantially modify the predictions of the SM for the FCNC processes.

Another effect of the extended mixing involving both types of quark is that the Z-mediated neutral currents may no longer be diagonal in flavor. Therefore, models with VLQs generically predict Z-mediated FCNCs.

In the limit \( M_D \rightarrow \infty \), the VLQ decouples and the mixing with the VLQ vanishes.
\begin{equation}
  A_L^{\alpha 4} \rightarrow 0, \quad V_L \rightarrow I \quad \Rightarrow \quad Z_{\text{NC}}^{pq} \rightarrow \delta^{pq},  
\end{equation}
and we recover SM unitarity,
\begin{equation}
    \sum_{i=u,c,t} V_{\text{CKM}}^{ip*} V_{\text{CKM}}^{iq} \rightarrow \delta^{pq}.
\end{equation}
or
\begin{equation}
\lim_{M_D \to \infty} U^d = \mathbb{I}.
\end{equation}

\section{Vector-Like Quark Contribution in B-Decays}

After spontaneous symmetry breaking, the Lagrangian gives rise to the physical quark masses, which together with the couplings to the SM-like Higgs can be denoted as,
\begin{equation}
\begin{aligned}
\mathcal{L}_H &= - \bar{d}_L \mathcal{M}_d^{\rm diag} d_R - \bar{u}_L \mathcal{M}_u^{\rm diag} u_R \frac{\sqrt{2} \chi^+}{v} \\
&\quad - \left[ \bar{u}_L V_{\text{CKM}} \mathcal{M}_d^{\rm diag} d_R + \text{h.c.} \right]
- i \frac{\chi^0}{v} \left[ \bar{d}_L \mathcal{M}_d^{\rm diag} d_R - \bar{u}_L \mathcal{M}_u^{\rm diag} u_R \right] \\
&\quad - \frac{h}{v} \left[ \bar{d}_L \mathcal{M}_d^{\rm diag} d_R + \bar{u}_L \mathcal{M}_u^{\rm diag} u_R \right] + \text{h.c.}
\end{aligned}
\end{equation}
Here, \( v \) is the vacuum expectation value, and \( h, \chi^+, \chi^0 \) are the physical Higgs and Goldstone fields.Finally, we obtain the following Lagrangian,
\begin{align}
  \mathcal{L}_\mathrm{SM} + \mathcal{L}_\mathrm{Eff}^\mathrm{tree}
  =
  \mathcal{L}_0 + \mathcal{L}_{A} + \mathcal{L}_{W} + \mathcal{L}_{Z} + \mathcal{L}_{\chi^\pm} + \mathcal{L}_{h} + \mathcal{L}_{\chi_0} + \cdots ~,\label{Eq:LVLQ}
\end{align}
where the ellipsis represents the terms that contain more than four fields. Each part of the Lagrangian is given in \cite{Morozumi:2018cnc}:
\begin{align}
\mathcal{L}_0&= \overline{u^i}i\slashed{\partial}u^i + \overline{d^p}i\slashed{\partial}d^p  -\left[m_u^{i}\overline{u^i}u^i + m_d^p\overline{d^p}d^p\right]~,
\label{Eq:L0}
 \end{align}
 \begin{align}
  \mathcal{L}_{A}&= -e\left[Q_u\overline{u^i}\gamma^\mu u^i+ Q_d\overline{d^p}\gamma^\mu d^p\right]A_\mu ~,\label{Eq:LA}
\end{align}
\begin{align}
  \mathcal{L}_{W}&= -\frac{g}{\sqrt{2}}\overline{u^i} \gamma^\mu V_\mathrm{CKM}^{iq} P_L d^q W_\mu^+ +\mathrm{h.c},
  \label{Eq:LW}
\end{align}
\begin{align}
  \mathcal{L}_{Z}&=  -\frac{g}{c_w}\left[\overline{u^i}\gamma^\mu
  \left(\frac{1}{2}P_L-Q_us_w^2\right) u^i
  -\overline{d^p}\gamma^\mu
  \left(\frac{1}{2}Z_\mathrm{NC}^{pq}P_L+Q_ds_w^2\delta^{pq}\right) d^q
  \right]Z_\mu ~,
  \label{Eq:LZ}
\end{align}
\begin{align}
  \mathcal{L}_{\chi^\pm} &=  \frac{g}{\sqrt{2}M_W}
  \overline{u^i}V_\mathrm{CKM}^{ip}
  \left(
  m_u^{i}P_L-m_d^{p}P_R
  \right)d^p\chi^+ + \mathrm{h.c.}~,
  \label{Eq:Lchipm}
\end{align}
\begin{align}
  \mathcal{L}_{h}&=
  -\frac{g}{2M_W}\overline{d^p}Z_\mathrm{NC}^{pq}\left(m_d^{q}P_R+m_d^{p}P_L\right)d^qh~,
  \label{Eq:Lh}
\end{align}
\begin{align}
  \mathcal{L}_{\chi^0} &=
  -\frac{ig}{2M_W}
  \overline{d^p}Z_\mathrm{NC}^{pq}\left(m_d^{q}P_R-m_d^{p}P_L\right)d^q\chi_0 ~.
  \label{Eq:Lchi0}
\end{align}
In eqs.~\eqref{Eq:L0}-\eqref{Eq:Lchi0}, $P_L$ and $P_R$ denote the chiral projection operators, $P_L\equiv\frac{1-\gamma_5}{2}, P_R\equiv\frac{1+\gamma_5}{2}$. $Q_u$ and $Q_d$ are the electromagnetic charge of up-type and down-type quarks, respectively.
After diagonalizing the extended $4\times4$ down-type quark mass matrix, the $Z$-boson couplings to down-type quarks are modified. The relevant interaction Lagrangian given in \cite{Alok:2022pjb}:
\begin{equation}
\mathcal{L}_Z = \frac{g}{2\cos\theta_W} \sum_{i,j} \bar{d}_i \gamma^\mu \left[ (g_L)_{ij} P_L + (g_R)_{ij} P_R \right] d_j Z_\mu,
\end{equation}
where $d_i = (d, s, b, D)$ and the off-diagonal couplings $(g_L)_{ij}$ are generated due to VLQ mixing. Specifically for $b\rightarrow s$ decay,
\begin{equation}
(g_L)_{sb} \simeq -\frac{1}{2} \left( V_{L}^{4\times4} \right)_{s4} \left( V_{L}^{4\times4} \right)_{b4}^*,
\end{equation}
where $V_L^{4\times4}$ is the left-handed mixing matrix. This leads to a tree-level $b \to sZ$ transition, enhancing processes such as $B \to K^{(*)} \ell^+ \ell^-$ and $B_s \to \mu^+\mu^-$.

\subsection{$b \to sZ$ Transitions with VLQs}

The neutral current transition with changing flavor $b \to sZ$ deserves particular attention in VLQ models. Unlike in the SM, where such processes are forbidden at tree level by the GIM mechanism \cite{GlashowIliopoulosMaiani1970}, the introduction of VLQs creates a fundamentally different scenario. Following the diagonalization of the extended quark mass matrices that incorporate both Standard Model and vector-like states, the $Z$ boson couplings undergo significant modification \cite{Ishiwata:2015cga}. 

Specifically, the left-handed $Z$ couplings acquire off-diagonal elements in flavor space, enabling tree-level $b \to sZ$ transitions \cite{Bobeth2017}. This phenomenon represents a distinctive feature of Vector-like quark models and constitutes a marked departure from the loop-suppressed nature of such processes within the Standard Model framework. The generation of these flavor-violating neutral current interactions stems directly from the non-unitary nature of the mixing matrices that relate the weak and mass eigenstates in the presence of vector-like fermions \cite{Branco:2022gja}.

These modified $Z$ couplings can substantially influence the Wilson coefficients $C_9$ and $C_{10}$, which parameterize the semileptonic operators governing rare B-decay processes \cite{Buchalla:1998ba, Buras:2014fpa}. Consequently, experimental observables in transitions such as $B^+ \to K^+ \ell^+ \ell^-$ and $B \to K^*\ell^+ \ell^-$ exhibit significant deviations from their SM predictions \cite{CMS:2014xfa, Descotes-Genon:2012isb,Belle-II:2023esi}. The potential enhancement of these decay channels provides a promising avenue for detecting indirect signatures of vector-like quarks, potentially at scales beyond the direct reach of current collider experiments \cite{Altmannshofer:2017wqy}.

The following section details the emergence of tree-level and additional Higgs loop-level FCNCs in VLQ models. We focus on the specific mechanisms by which these processes arise and their implications for rare B-decay phenomenology.

\subsection{Tree-Level Digram of $b\rightarrow s\mu^+\mu^-$}

Z-mediated FCNCs appear at tree level in the left-handed sector due to the addition of VLQs, whereas it is forbidden in the SM due to GIM suppression. In particular, a $Zbs$ coupling \cite{Alok:2012xm} can be generated as,
\begin{equation}
    \mathcal{L}_Z=-\frac{g}{2\cos{\theta_W}} U_{sb} \bar{s}\gamma_\mu P_L b Z_\mu
\end{equation}
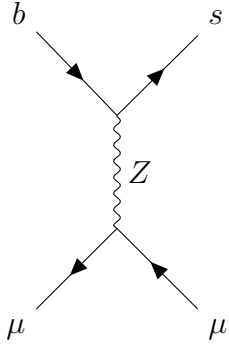
\begin{figure}
    \centering
    \begin{tikzpicture}
\begin{feynman}
    \vertex (a);
    \vertex [above left=of a] (b) {$b$};
      \vertex [above right=of a] (c) {$s$};
      \vertex [below=of a] (d);
       \vertex [below right=of d] (e) {${\mu}$};
        \vertex [below left=of d] (f) {$\mu$};
        
      \diagram{
      (b) -- [fermion] (a) -- [fermion] (c);
      (a) -- [boson, edge label=$Z$] (d);
      (e) -- [fermion](d) -- [fermion](f);
      };
      
\end{feynman}  
\end{tikzpicture}
    \caption{Tree-Level FCNC of $b\rightarrow s\mu^+\mu^-$ with VLQ}
    \label{fig:TreeFCNC}
\end{figure}
Here, $U_{sb}$ represents the VLQ contribution at the tree level. Using the Feynman rules, the amplitude of tree-level FCNCs in Fig.(\ref{fig:TreeFCNC}) is,  
   \begin{equation}
   \mathcal{M}_{tree}= -\frac{ig^2}{4M^2_Z\cos^2{\theta_{W}}} U_{sb}\bar{s}\gamma_\mu P_{L}b \bar{\mu}(\gamma^\mu(a_f(1
   +\gamma_5)+b_f(1-\gamma_5)))\mu
   \label{Amp-1}
   \end{equation}
  From eq(\ref{afbf}), $a_f$ and $b_f$ are,
   \begin{equation}
       a_f= -Q_f\sin^2{\theta_W}, \hspace{0.5cm}
       b_f= T_3 (f)-Q_f \sin^2{\theta_W}
        \label{af}
    \end{equation}
 where $f=i$ represents the internal quark $u,c,t$, and the charge and isospin of $\mu$ are 
\begin{equation}
    Q_f = -1,\quad T_f =-\frac{1}{2}
    \label{Q_f}
\end{equation}
By plugging eq(\ref{af}) and eq(\ref{Q_f}) into eq(\ref{Amp-1}), we have
\begin{equation}
   \mathcal{M}_{tree}= -\frac{ig^2}{4M^2_W} U_{sb}\bar{s}\gamma_\mu P_{L}b \bar{\mu}(\gamma^\mu(\sin^2{\theta_W}(1
   +\gamma_5)+(-\frac{1}{2}+\sin^2{\theta_W})(1-\gamma_5)))\mu
   \label{Amp-2}
   \end{equation}
  Using eq(\ref{G_F}) and considering the operator, $O^\mu_L=\frac{\alpha}{4\pi}\Bar{s}\gamma_\mu P_{L}b \Bar{\mu}(\gamma^\mu(1-\gamma_5))\mu$ the tree-level amplitude is,
\begin{equation}
   \mathcal{M}_{tree}= - \frac{iG_F}{\sqrt{2}} U_{sb}\bar{s}\gamma_\mu P_{L}b \bar{\mu}(\gamma^\mu(-1+2\sin^2{\theta_W})(1-\gamma_5)))\mu
   \end{equation}
   Now we set $\mathcal{H}=i\mathcal{M}$ to get the effective Hamiltonian.
   \begin{equation}
   \mathcal{H}_{eff}=  \frac{G_F}{\sqrt{2}} U_{sb}\bar{s}\gamma_\mu P_{L}b \bar{\mu}(\gamma^\mu(-1+2\sin^2{\theta_W})(1-\gamma_5)))\mu
   \end{equation}
   Comparing the result with the Standard Model effective Hamiltonian, 
    \begin{equation}
   \mathcal{H}_{e f f}=-\frac{4 G_F}{\sqrt{2}} V_{t s}^* V_{t b}  C_L(\mu) O_L(\mu)
   \end{equation}
   where $C^\mu_{L,SM}= \frac{Y_0}{\sin^2{\theta_W}}$, we can write
   \begin{equation}
  C^\mu_L= C^\mu_{L,SM} - \frac{\pi}{\alpha}\frac{U_{sb}}{V_{ts}^*V_{tb}}(-1+2\sin^2{\theta_W})
  \label{Clt}
   \end{equation}
The second part in eq.(\ref{Clt}) is tree level VLQ's contribution.

\subsubsection*{Wilson Coefficients ($C_9$ and $C_{10}$)}
To obtain the NP contribution in the Wilson coefficients $C_9$ and $C_{10}$, considering the amplitude in eq.(\ref{Amp-2}) of tree-level $b\rightarrow s \mu^+ \mu^-$,
  and by simplifying we can write,  
   \begin{equation}
   \mathcal{M}_{tree}= -\frac{ig^2}{8M^2_W} U_{sb}\bar{s}\gamma_\mu P_{L}b \bar{\mu}(\gamma^\mu(-1+4\sin^2{\theta_W}+\gamma_5))\mu
   \end{equation}
 The effective Hamiltonian is then obtained by multiplying $i$ with amplitude,
   \begin{equation}
   \mathcal{H}_{eff}=  \frac{G_f}{\sqrt{2}} U_{sb}\bar{s}\gamma_\mu P_{L}b \bar{\mu}(\gamma^\mu(-1+4\sin^2{\theta_W}+\gamma_5))\mu
   \label{Htree}
   \end{equation}
 The effective Hamiltonian of Standard Model for $b\rightarrow sl\bar{l}$ is,
\begin{equation}
\mathcal{H}_{eff}(b\rightarrow sl\bar{l})=\mathcal{H}_{eff}(b\rightarrow s\gamma)-4\frac{G_F}{\sqrt{2}}\frac{\alpha}{4\pi}V_{ts}^*V_{tb}\left[C_9(\mu)Q_9+C_{10}(\mu)Q_{10}\right],
\label{HSM}
\end{equation}
The  corresponding operators $O_9$ and $O_{10}$ are given as;

\begin{equation}
   O_9 = \frac{e^2}{16\pi^2}(\bar{s} \gamma_\mu P_L b)(\bar{l} \gamma^\mu l)=\frac{\alpha}{4\pi}(\bar{s} \gamma_\mu P_L b)(\bar{l} \gamma^\mu l)=\frac{\alpha}{4\pi}Q_9
\end{equation}
\begin{equation}
      O_{10} =\frac{e^2}{16\pi^2} (\bar{s} \gamma_\mu P_L b)(\bar{l} \gamma^\mu \gamma^5 l)= \frac{\alpha}{4\pi} Q_{10}
\end{equation}
Hence, comparing eq.(\ref{Htree}) with eq.(\ref{HSM}),the Wilson coefficients $C_9 $ and $C_{10}$ for tree-level FCNCs are 
\begin{equation}
    C_{9,VLQ}=-\frac{U_{sb}}{\lambda_t}\frac{\pi}{\alpha}(4\sin^2{\theta_W}-1)
    \label{C_9T}
    \end{equation}
    \begin{equation}
    C_{10,VLQ}=-\frac{U_{sb}}{\lambda_t}\frac{\pi}{\alpha}
    \label{C_10T}
    \end{equation}
These Wilson coefficients are the result of a vector-like quark contribution the same as that given in \cite{Alok:2012xm}.

\subsection{Loop Diagrams of $b\rightarrow s\mu^+\mu^-$}
  In the Standard Model, $ \ b\to sZ$ decay is a loop-induced process, dominated by the $W$-boson and top-quark loop. However, new contributions arise from loops involving the heavy $D$ and its mixing with SM quarks in the VQM. We have a Higgs loop (physical and virtual) with vector-like contributions coming into the loop. Here we calculate the amplitude of each additional Feynman diagrams due to the VLQ contribution in Fig.(\ref{VH}).
\subsubsection{Higgs Loop}

From \cite{Handoko:1994xw}, the Lagrangian of the quark-Higgs (physicaland unphysical) couplings are given as
\begin{align}
\mathcal{L}_H & =\frac{-g}{2 M_W}\left[z_u^{\alpha \beta} \bar{u}^\alpha\left(m_{u^\alpha} P_L+m_{u^\beta} P_R\right) u^\beta+z_d^{\alpha \beta} \bar{d}^\alpha\left(m_{d^\alpha} P_L+m_{d^\beta} P_R\right) d^\beta\right] H \\
\mathcal{L}_{\chi^0} & =\frac{-i g}{2 M_W}\left[z_u{ }^{\alpha \beta} \bar{u}^\alpha\left(m_{u^\alpha} P_L-m_{u^\beta} P_R\right) u^\beta-z_d^{\alpha \beta} \bar{d}^\alpha\left(m_{d^\alpha} P_L-m_{d^\beta} P_R\right) d^\beta\right] \chi^0
\end{align}
The appearance of the terms proportional to $\frac{m_\alpha}{M_W}$  in the $\mathcal{L}_H$ and $\mathcal{L}_{\chi_0}$ may predict a significant contribution of the vector-like quarks in the internal line, which have a large mass $ m_\alpha$.

\begin{figure}[h!]
 \begin{center}
 \begin{tikzpicture} 
\begin{feynman}
\vertex (a1) {\(b\)};
\vertex[right=1.5cm of a1] (a2);
\vertex[right=1.2cm of a2] (a3);
\vertex[right=1.5cm of a3] (a5) {\(s\)};

\vertex at ($(a2)!0.5!(a3)+(0,-1.2)$) (c1);
\vertex[below=1.2cm of c1] (c2);
\vertex[below right=of c2] (c3) {\(\mu\)};
\vertex[below left=of c2] (c4) {\({\mu}\)};

\diagram* {
  (a1) -- [fermion] (a2) -- [scalar, edge label={$H$}] (a3) --  [fermion] (a5),
  
  (a2) -- [fermion, out=90, in=180, looseness=1.0, edge label'={$D$}] (c1) -- [fermion, in=90, out=0, looseness=1.0, edge label'={$D$}] (a3),
  (c1) -- [boson, edge label'={\(Z\)}] (c2),
  (c3) -- [fermion] (c2) -- [fermion] (c4),
};
\end{feynman}
\end{tikzpicture} 
\quad
 \begin{tikzpicture} 
\begin{feynman}
\vertex (a1) {\(b\)};
\vertex[right=1.5cm of a1] (a2);
\vertex[right=1.2cm of a2] (a3);
\vertex[right=1.5cm of a3] (a5) {\(s\)};

\coordinate (c1) at ($(a2)!0.5!(a3)+(0,-1.2)$);
\vertex[below=1.2cm of c1] (c2);
\vertex[below right=of c2] (c3) {\(\mu\)};
\vertex[below left=of c2] (c4) {\({\mu}\)};

\diagram* {
  (a1) -- [fermion] (a2) -- [scalar, edge label={$\chi^0$}] (a3) -- [fermion] (a5) ,
  
  (a2) -- [fermion, out=90, in=180, looseness=1.0, edge label'={$D$}] (c1) -- [fermion, in=90, out=0, looseness=1.0, edge label'={$D$}] (a3),
  (c1) -- [boson, edge label'={\(Z\)}] (c2),
  (c3) -- [fermion] (c2) -- [fermion] (c4),
};
\end{feynman} 
\end{tikzpicture}  
 \end{center}
 \caption{Vertex diagram $b\rightarrow s \mu^+\mu^-$ with Higgs in the loop}
 \label{VH}
 \end{figure}
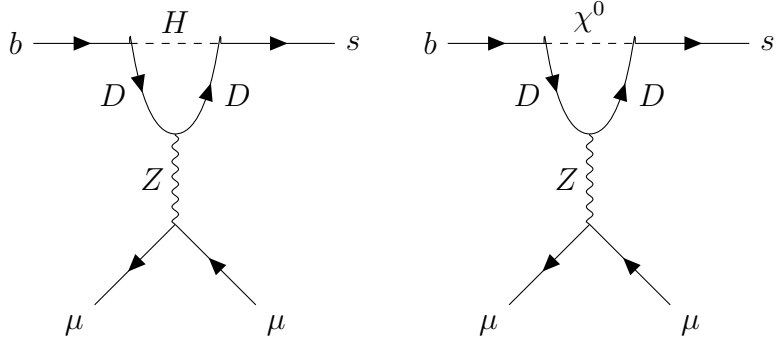

Starting from the Higgs loop in $b\rightarrow s\mu^+\mu^-$ given in Fig.(\ref{VH}).
  First, we calculate the amplitude of the Higgs loop in $b\rightarrow sZ$ with down-type VLQ given in \cite{Ahmady:2001qh}. Using Feynman rules for vector-like quarks from \cite{Morozumi:2018cnc}, the amplitude is
\begin{equation}
\begin{aligned}
      \mathcal{M_H}&=\int \frac{d^4k}{(2\pi)^4}\bar{s}(\frac{ -g}{2M_W})A^{42{*}}A^{44}(m_s(\frac{1-\gamma_5}{2})+M_D(\frac{1+\gamma_5}{2}) )\frac{i}{(k^2 - m_H^2)} (\frac{i\slashed{k}+M_D}{k^2 - M_D^2})\\
&\quad\times((-\frac{ig}{\cos{\theta_W}})\gamma_\mu(-\frac{1}{3} \sin^2{\theta_W}))(\frac{i(\slashed{k}+M_D)}{k^2-M^2_D})(\frac{ -g}{2M_W})A^{44{*}}A^{43}(M_D(\frac{1-\gamma_5}{2})\\
&\quad+m_b(\frac{1+\gamma_5}{2})))b  
\end{aligned}
\end{equation}

Since the vector-like quark $D$ in the loop is heavy compared to the masses of $s$ and $b$ quarks, we ignore $m_s$ and $m_b$.
\begin{equation}
\begin{aligned}
    \mathcal{M_H}&=\int \frac{d^4k}{(2\pi)^4}\bar{s}(\frac{ -g}{2M_W})A^{42{*}}A^{44}A^{44{*}}A^{43} (M_D(\frac{1+\gamma_5}{2}) )\frac{i}{(k^2 - m_H^2)} (\frac{i\slashed{k}+M_D}{k^2 - M_D^2})
\\
&\quad\times((-\frac{ig}{\cos{\theta_W}})\gamma_\mu(-\frac{1}{3} \sin^2{\theta_W}))(\frac{i(\slashed{k}+M_D)}{k^2-M^2_D})(\frac{ -g}{2M_W})(M_D(\frac{1-\gamma_5}{2})))b
\end{aligned}
\end{equation}
where $A^{44{*}}A^{44} \apeq 1$ and $A^{42{*}}A^{43}= U_{sb}$, we have
\begin{equation}
\begin{aligned}
  \mathcal{M_H}&=(\frac{g^3}{16M_W^2\cos{\theta_W}})U_{sb}M^2_D(\frac{1}{3}\sin^2{\theta_W})
    \int \frac{d^4k}{(2\pi)^4}\frac{1}{k^2 - m_H^2}\frac{1}{(k^2 - M_D^2)^2}\\
    &\quad\times\bar{s}\left[(1+\gamma_5) (\slashed{k}+M_D)\gamma_\mu(\slashed{k}+M_D)(1-\gamma_5)\right]b  
\end{aligned}
\end{equation}

where the numerator is simplified by shifiting $(1-\gamma_5)$ to the right of all $\gamma_\mu$ matrices, we have
\begin{equation}
\begin{aligned}
    \bar{s}[(1+\gamma_5) (\slashed{k}+M_D)\gamma_\mu(\slashed{k}+M_D)(1-\gamma_5)]b
   & =2\bar{s}[ \slashed{k}\gamma_\mu\slashed{k}(1-\gamma_5)+M^2_D(1-\gamma_5)]b\\
   & =\bar{s}[ \gamma_\alpha \gamma_\mu \gamma_\beta {k^\alpha k^\beta}(1-\gamma_5)+M^2_D(1-\gamma_5)]b
\end{aligned}
\end{equation}

Hence the amplitude,
\begin{equation}
    \mathcal{M_H}=(\frac{g^3}{16M_W^2\cos{\theta_W}})U_{sb}M^2_D(\frac{2}{3}\sin^2{\theta_W})
    \int \frac{d^4k}{(2\pi)^4}\frac{\bar{s}[ \gamma_\alpha \gamma_\mu \gamma_\beta {k^\alpha k^\beta}(1-\gamma_5)+M^2_D(1-\gamma_5)]b}{(k^2 - m_H^2)(k^2 - M_D^2)^2}
    \label{amp-1}
\end{equation}
whereas the integrals from Appendix B are given as
\begin{equation}
    \int \frac{d^Dk}{(2\pi)^D} \frac{k^\alpha k^\beta}{(k^2 - m_D^2)^2 (k^2 - M_H^2)}=\frac{ig^{\alpha\beta}}{32\pi^2}\left[\frac{1}{\bar{\epsilon}}+\frac{3}{4}+F_1(x_1)\right],
\end{equation}
and
\begin{equation}
    \int \frac{d^Dk}{(2\pi)^D} \frac{1}{(k^2 - m_D^2)^2 (k^2 - M_H^2)}=\frac{i}{16\pi^2}\frac{1}{M_H^2}R_1(x_1),
\end{equation}
where $R_1(x_1)$ and $F_1(x_1)$ are given as,
\begin{equation}
    R_1(x_1)= \frac{\log{x_1}}{(1-x_1)^2}+\frac{1}{(1-x_1)}
    \label{R_1}
\end{equation}
\begin{equation}
    F_1(x_1)= -\frac{1}{2(1-x_1)^2}\left[x_1^2\log x_1- 2x_1\log x_1 - x_1(1 - x_1)\right],
    \label{F_1}
\end{equation}
and $x_1=\frac{M_D^2}{M_H^2}$. Inserting integrals in equation(\ref{amp-1}), we get
\begin{equation}
    \begin{aligned}
        \mathcal{M_H}&=(\frac{g^3}{16M_W^2\cos{\theta_W}})U_{sb}M^2_D(\frac{2}{3}\sin^2{\theta_W})\left[\frac{ig^{\alpha\beta}}{32\pi^2}(\frac{1}{\bar{\epsilon}}+\frac{3}{4}+F_1(x_1))\right]
    \bar{s}( \gamma_\alpha \gamma_\mu \gamma_\beta (1-\gamma_5))b\\
    &+\left[M^2_D\frac{i}{16\pi^2}\frac{1}{M_H^2}R_1(x_1)\right]\bar{s}\gamma_\mu(1-\gamma_5)b) ,  
    \end{aligned}
\end{equation}
   using eq.(\ref{R_1}) and
  \begin{equation}
         g^{\alpha\beta}\bar{s}( \gamma_\alpha \gamma_\mu \gamma_\beta (1-\gamma_5))b= -2(1-\epsilon)\bar{s}( \gamma_\mu (1-\gamma_5))b 
  \end{equation} 
where we kept $\mathcal{O}(\epsilon)$ terms, as they will contribute after the multiplication by $\frac{1}{\bar{\epsilon}}$, 
\begin{equation}
\begin{aligned}
    \mathcal{M_H}&=(\frac{ig^3}{32\pi^2\cos{\theta_W}})U_{sb}\frac{M^2_D}{M_W^2}(\frac{1}{24}\sin^2{\theta_W})\left[ -2(1-\epsilon)(\frac{1}{\bar{\epsilon}}+\frac{3}{4}+F_1(x_1))\right]
    \bar{s}\gamma_\mu (1-\gamma_5)b\\
    &+2\frac{M^2_D}{M_H^2}\left[\frac{\log{x_1}}{(1-x_1)^2}+\frac{1}{(1-x_1)}\right]\bar{s}\gamma_\mu(1-\gamma_5)b) ,
\end{aligned}  
\end{equation}
From Appendix B, given $\frac{1}{\bar{\epsilon}}=\frac{1}{2\epsilon}+\frac{1}{2}\left[\log{4\pi}-\gamma_E+\log{\frac{\mu^2}{M^2}}\right]$ , we get
\begin{equation}
\begin{aligned}
\mathcal{M_H} &= \frac{i g^{3} M_D^{2} U_{sb}}{32 \pi^{2} M_W^{2} \cos \theta_W} 
\left( \frac{\sin^2 \theta_W}{24} \right) 
\Biggl\{ 
    -2 \Bigl( \frac{1}{\bar{\epsilon}} + \frac{1}{4} + F_1(x_1) \Bigr) \\
    &\quad + \frac{2 x_1 \log x_1}{(1 - x_1)^2} + \frac{2 x_1}{1 - x_1} 
\Biggr\} 
\bar{s} \gamma_\mu (1 - \gamma_5) b
\end{aligned}
\end{equation}
From equation (\ref{F_1}), using $F_1(x_1)$, 
the amplitude of Higgs loop $b\rightarrow sZ$ is,
\begin{align}
  \mathcal{M_H}=\frac{i g^{3} U_{s b}}{32 \pi^{2}\cos \theta_{W}}\left(\frac{1}{24} \sin ^{2} \theta_{W}\right)\frac{M_{D}^{2}}{M_{W}^{2}}\left(\frac{-2}{\epsilon}-\frac{1}{2}+\frac{x_1^{2} \log x_1}{(1-x_1)^{2}}+\frac{x_{1}}{\left(1-x_{1}\right)}\right) \bar{s}\gamma_\mu \left(1-\gamma_{5}\right) b
  \label{H}
\end{align}
Similarly, the amplitude for unphysical Higgs $\chi^0$ loop is,
\begin{align}
    \mathcal{M}_{\chi^{0}}=\frac{i g^{3} U_{s b}}{32 \pi^{2} \cos \theta_{W}}\left(\frac{1}{24} \sin ^{2} \theta_{W}\right) \frac{M_{D}^{2}}{M_{W}^{2}}\left(\frac{-2}{\epsilon}-\frac{1}{2}+\frac{x_2^{2} \log x_2}{(1-x_2)^{2}}+\frac{x_{2}}{\left(1-x_{2}\right)}\right) \bar{s}\gamma_\mu \left(1-\gamma_{5}\right) b
\label{chi}
\end{align}
where $x_2=\frac{M_D^2}{M_Z^2}$.
\subsubsection*{Wilson Coefficients $C_{9}$ and $ C_{10}$}
To get NP contribution in  Wilson coefficients $C_{9}$ and $C_{10}$ due to VLQ in the Higgs loop, we need to calculate the amplitude for  $b\rightarrow s\mu^+ \mu^-$. Using eq.(\ref{H}) we have,
\begin{equation}
\begin{aligned}
    {\mathcal{M_H}}&=\frac{i g^{3}U_{s b}}{32 \pi^{2} \cos \theta_{W}} \left(\frac{1}{24} \sin ^{2} \theta_{W}\right) \frac{M_{D}^{2}}{M_{W}^{2}}\left(\frac{-2}{\bar{\epsilon}}-\frac{1}{2}+\frac{x_1^{2} \log x_1}{(1-x_1)^{2}}+\frac{x_{1}}{\left(1-x_{1}\right)}\right) \bar{s}\gamma_\mu \left(1-\gamma_{5}\right) b\\
    &\quad \times\frac{i}{M_Z^2}\left(\frac{ig}{2\cos{\theta_W}}\bar{\mu}\gamma^{\mu}(\frac{-1}{2}+2\sin^2{\theta_W}+\frac{1}{2}\gamma_5)\mu \right)
\end{aligned}
\end{equation}

    
and from equation (\ref{G_F}), and using $\alpha= \frac{e^2}{4\pi}$ the amplitude is
\begin{equation}
\begin{aligned}
     {\mathcal{M_H}}&=\frac{-i G_FU_{s b}}{\sqrt{2}}\frac{\alpha}{\pi} \left(\frac{1}{96} \right) \frac{M_{D}^{2}}{M_{W}^{2}}\left(\frac{-2}{\epsilon}-\frac{1}{2}+\frac{x_1^{2} \log x_1}{(1-x_1)^{2}}+\frac{x_{1}}{\left(1-x_{1}\right)}\right) \\
    &\times\bar{s}\gamma_\mu \left(1-\gamma_{5}\right) b
 \bar{\mu}\gamma^{\mu}(-1+4\sin^2{\theta_W}+\gamma_5)\mu
\end{aligned}
\end{equation}
   
Hence the effective Hamiltonian of given Higgs loop is,
\begin{equation}
    \begin{aligned}
       \mathcal{H}&=\frac{ G_F U_{s b}}{\sqrt{2}}\frac{\alpha}{\pi} \left(\frac{1}{96} \right) \frac{M_{D}^{2}}{M_{W}^{2}}\left(\frac{-2}{\epsilon}-\frac{1}{2}+\frac{x_1^{2} \log x_1}{(1-x_1)^{2}}+\frac{x_{1}}{\left(1-x_{1}\right)}\right) \\
    &\quad\times\bar{s}\gamma_\mu \left(1-\gamma_{5}\right) b
    \left(\bar{\mu}\gamma^{\mu}(-1+4\sin^2{\theta_W}+\gamma_5)\mu \right) 
    \label{HH}
    \end{aligned}
\end{equation}
whereas, the Standard Model effective Hamiltonian is
\begin{equation}
    \mathcal{H}_{eff}=-4 \frac{G_F}{\sqrt{2}}\lambda_t
\frac{\alpha}{4\pi}(C_{9}O_{9}+C_{10}O_{10})
\label{HHSM}
\end{equation}

Hence, by comparing eq.(\ref{HH}) and eq.(\ref{HHSM}), the Wilson Coefficient $C_{9}$ and $C_{10}$  corresponding to operators $O_{9}$ and $O_{10}$ are,

\begin{align}
    C_{9}= -\frac{U_{sb}}{\lambda_t}\frac{M_D^2}{M_W^2}\frac{1}{48}\left(\frac{-2}{\epsilon}-\frac{1}{2}+\frac{x_1^{2} \log x_1}{(1-x_1)^{2}}+\frac{x_{1}}{\left(1-x_{1}\right)}\right)(4\sin^2{\theta_W}-1)
    \end{align}
\begin{align}
    C_{10}= -\frac{U_{sb}}{\lambda_t}\frac{M_D^2}{M_W^2}\frac{1}{48}\left(\frac{-2}{\epsilon}-\frac{1}{2}+\frac{x_1^{2} \log x_1}{(1-x_1)^{2}}+\frac{x_{1}}{\left(1-x_{1}\right)}\right)
\end{align}
    Similarly for the $\chi^0$ loop, the Wilson coefficients are
\begin{align}
    C_{9}=- \frac{U_{sb}}{\lambda_t}\frac{M_D^2}{M_W^2}\frac{1}{48}\left(\frac{-2}{\epsilon}-\frac{1}{2}+\frac{x_2^{2} \log x_2}{(1-x_2)^{2}}+\frac{x_{2}}{\left(1-x_{2}\right)}\right)(4\sin^2{\theta_W}-1)
\label{C_9H}
\end{align}
\begin{align}
    C_{10}= -\frac{U_{sb}}{\lambda_t}\frac{M_D^2}{M_W^2}\frac{1}{48}\left(\frac{-2}{\epsilon}-\frac{1}{2}+\frac{x_2^{2} \log x_2}{(1-x_2)^{2}}+\frac{x_{2}}{\left(1-x_{2}\right)}\right)
    \label{C_10H}
\end{align}    
\subsubsection{Self Energy Diagrams}

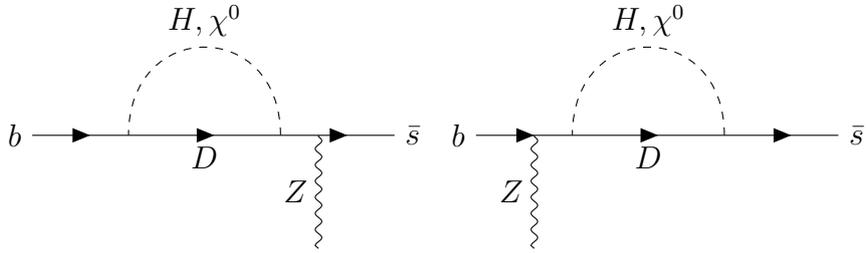
\begin{figure}[h!]
\begin{center}
\begin{tikzpicture} 
\begin{feynman}
\vertex (a1) {\( b\)};
\vertex[right=1.5cm of a1] (a2); 
\vertex[right=1cm of a2] (a3); 
\vertex[right=1cm of a3] (a4); 
\vertex[right=0.5cm of a4](a);
\vertex[right=1.5cm of a4] (a5) {\( \bar{s}\)};
\vertex [below=of a](c1);

\diagram* { 
(a1)--[fermion](a2);
(a2)--[fermion,edge label' ={$D$}] (a4);
(a4)--[fermion](a5);
 (a) -- [boson, edge label' = \(Z\)] (c1),
(a4) -- [scalar, out=90, in=90, looseness=2.0, edge label'={$H,\chi^0$}] (a2);
};
\end{feynman} 
\end{tikzpicture}
\begin{tikzpicture} 
\begin{feynman}
\vertex (a1) {\( b\)};
\vertex[right=1.5cm of a1] (a2); 
\vertex[right=1cm of a2] (a3); 
\vertex[right=1cm of a3] (a4); 
\vertex[right=1cm of a1](a);
\vertex[right=1.5cm of a4] (a5) {\( \bar{s}\)};
\vertex [below=of a](c1);

\diagram* { 
(a1)--[fermion](a2);
(a2)--[fermion,edge label' ={$D$}] (a4);
(a4)--[fermion](a5);
 (a) -- [boson, edge label' = \(Z\)] (c1),
(a4) -- [scalar, out=90, in=90, looseness=2.0, edge label'={$H,\chi^0$}] (a2);
};
\end{feynman} 
\end{tikzpicture}
\end{center}
\caption{ Higgs Self-Energy of $b\rightarrow s Z$}
\label{HSF}
\end{figure}

In the Standard Model, ultraviolet divergences in loop-level processes such as $b \rightarrow s Z$ are cancelled by a combination of self-energy diagrams and the unitarity of the CKM matrix as we have seen in previous chapter. However, in the presence of a vector-like quark (VLQ), the CKM matrix becomes non-unitary due to mixing with the fourth-generation quark. As a result, the cancellation is no longer complete. Nevertheless, some of the singularities are still cancelled by self-energy contributions. The remaining divergences are absorbed into renormalization counterterms. The self-energy contributions relevant for the Higgs loop in the $b \rightarrow s Z$ transition are shown in Fig.~(\ref{HSF}), and the Higgs coupling with vector-like quark $D$ and SM quarks $b, s$  is given in Fig.(\ref{fig:Higgs-D-couplings}).

\begin{figure}[h!]
\centering
\begin{minipage}{0.45\textwidth}
\centering
\begin{tikzpicture}
\begin{feynman}
\vertex (a1) {\( H\)};
\vertex[right=1.5cm of a1] (a2); 
\vertex[above right=1cm of a2] (a3); 
\vertex[below right=1cm of a2] (a4); 
\diagram* { 
 (a1) -- [scalar] (a2),
 (a3) -- [fermion, edge label'={$D$}] (a2),
 (a2) -- [fermion, edge label'={$s$}] (a4),
};
\end{feynman}
\end{tikzpicture}

\vspace{0.2cm}
\begin{equation*}
= z^{s D} \bar{s}\left(m_{s} L+M_{D} R\right) D H
\end{equation*}
\end{minipage}
\hspace{1cm}
\begin{minipage}{0.45\textwidth}
\centering
\begin{tikzpicture}
\begin{feynman}
\vertex (a1) {\( H\)};
\vertex[right=1.5cm of a1] (a2); 
\vertex[above right=1cm of a2] (a3); 
\vertex[below right=1cm of a2] (a4); 
\diagram* { 
 (a1) -- [scalar] (a2),
 (a3) -- [fermion, edge label'={$b$}] (a2),
 (a2) -- [fermion, edge label'={$D$}] (a4),
};
\end{feynman}
\end{tikzpicture}

\vspace{0.1cm}
\begin{equation*}
= z^{Db} \bar{D}\left(M_{D} L+m_{b} R\right) b H
\end{equation*}
\end{minipage}
\caption{Higgs couplings involving the down-type vector-like quark \(D\) and SM quarks \(b\), \(s\).}
\label{fig:Higgs-D-couplings}
\end{figure}
We first calculate the self-energy diagram present in Fig.(\ref{HSF}) with external momentum p and given as
\begin{equation}
    \begin{aligned}
     \mathcal{A_H}&=\int \frac{d^4k}{(2\pi)^4}\bar{s}\left(\frac{ -g}{2M_W}A^{42{*}}A^{44}M_D(\frac{1+\gamma_5}{2})\right)\frac{i}{(k^2 - m_H^2)} \left(\frac{i(\slashed{k}+\slashed{p}+M_D)}{(k+p)^2 - M_D^2}\right)\\
&\quad\times\left(\frac{ -g}{2M_W}A^{44{*}}A^{43}M_D(\frac{1-\gamma_5}{2})\right)b   
    \end{aligned}
\end{equation}
where $A^{42{*}}A^{43}= U_{sb}$,
\begin{equation}
\begin{aligned}
 \mathcal{A_H}&=\frac{-g^{2}M_{D}^{2} U_{s b}}{16 M_{W}^{2}}  \int \frac{d^{4} K}{(2 \pi)^{4}} \frac{1}{\left.\left(k^{2}-M_{W}^{2}\right)((k+P)^{2}-M_{D}^{2})\right)}\\
 &\quad\times\bar{s}\left(1+\gamma_{5}\right)\left(\slashed{k}+\slashed{p}+M_{D}\right)\left(1-\gamma_{5}\right) b 
\end{aligned}
\end{equation}
and
\begin{equation}
    \begin{split}
    \mathcal{A_H}=\frac{-g^{2}}{16} \frac{M_{D}^{2}}{M_{W}^{2}} U_{s b} \int \frac{d^{4} K}{(2 \pi)^{4}} \frac{2 \bar{s}(\slashed{k}+\slashed{P})\left(1-\gamma_{5}\right) b}{\left.\left(k^{2}-M_{W}^{2}\right)((k+P)^{2}-M_{D}^{2})\right)} \\
 =\frac{-g^{2}}{8} \frac{M_D^{2}}{M_{W}^{2}} U_{s b} \int \frac{d^{4} K}{(2 \pi)^{4}} \frac{(k+p)^{\beta}}{\left.\left(k^{2}-M_{W}^{2}\right)((k+P)^{2}-M_{D}^{2})\right)} \bar{s} \gamma_{\beta}\left(1-\gamma_{5}\right) b
    \end{split}
\end{equation}

using the integral from Appendix B,
\begin{align}
     \int \frac{d^{D} k}{(2 \pi)^{D}} \frac{(k+P)^{\beta}}{\left.\left(k^{2}-M_{H}^{2}\right)(k+P)^{2}-M_{D}^{2}\right)}=P^{\beta} \frac{i}{16 \pi^{2}}\left(\frac{1}{\epsilon}+\frac{3}{4}+F_{2}(x_1)\right)
\end{align}
The self-energy amplitude including the external spinor is, 
\begin{align}
    \mathcal{A_H}=\frac{-i g^{2}U_{s b}}{8 \times 16 \pi^{2}} \frac{M_{D}^{2}}{m_{W}^{2}} \left(\frac{1}{\epsilon}+\frac{3}{4}+F_{2}(x_1)\right) \bar{s} \slashed{p}\left(1-\gamma_{5}\right) b
    \label{A_H}
\end{align}
where $F_2(x_1)$ is given as,
\begin{equation} 
F_2(x_1)= -\frac{1}{2(1-x_1)^2}\left[x_1^2\log x_1 + (1 - x_1)\right]
\end{equation}
So, the total amplitude of the self-energy diagram of $b\rightarrow sZ$ in Fig.(\ref{HSF}) is written as,

\begin{equation}
  \mathcal{M_H}=\frac{i g_{2} }{2 \cos \theta_{W}} \bar{s}\gamma_{\alpha}\left(a_{f}\left(1+\gamma_{s}\right)+b_{f}\left(1-\gamma_{s}\right)\right) i \frac{\slashed{p}+m_{s}}{p^{2}-m_{s}^{2}} (\mathcal{A_H}) b  
\end{equation}
Using $ \mathcal{A_H}$ from equation(\ref{A_H}) and ignoring the mass of $s$ quark, the amplitude is,
\begin{equation}
\begin{aligned}
  \mathcal{M_H} &=\frac{i g_{2}}{2 \cos \theta_{W}} \bar{s} \gamma_{\alpha}\left(a_{f}\left(1+\gamma_{5}\right)+b_{f}\left(1-\gamma_{5}\right)\right) \frac{i \slashed{p}}{P^{2}}\\
  &\times\left(-\frac{i g_2^{2}}{8 \times 16 \pi^{2}} \frac{M_D^{2}}{M_{W}^{2}} U_{s b}\left(\frac{1}{\bar{\epsilon}}+\frac{3}{4}+F_{2}(x_1)\right) \slashed{p}\left(1-\gamma_{5}\right)\right) b  
\end{aligned}
\end{equation}
using $b_f$ from eq.(\ref{af}) and simplifying above equation, we get
\begin{equation}
 \mathcal{M_H}=\frac{i g^{3}}{4 \times 32 \pi^{2}}\frac{M_D^2}{M_W^2} U_{ sb }\frac{\left(T_{3 f}-Q_{f} \sin ^{2} \theta_{W}\right)}{\cos \theta_{W}}\left(\frac{1}{\bar{\epsilon}}+\frac{3}{4}+F_{2}\left(x_{1}\right)\right) \bar{s} \gamma_{\alpha}\left(1-\gamma_{5}\right) b
\end{equation}

Using $b_f=\frac{-1}{2}+\frac{1}{3}\sin^2{\theta_W}$, we get the following result for the amplitude,

\begin{align}
   \mathcal{M_H}=\frac{i g^{3}}{32 \pi^{2}} \frac{M_D^{2}}{M_W^{2}} U^{s b}\left(\frac{-\frac{1}{2}-\frac{1}{3} \sin ^{2} \theta_{W}}{4 \cos \theta_{W}}\right)\left(\frac{1}{\epsilon}+\frac{3}{4}-\frac{x_1^{2} \log x_1}{2(1-x_1)^{2}}-\frac{1}{2(1-x_1)}\right) \bar{s} \gamma_\mu\left(1-\gamma_{5}\right) b
   \label{HS1}
\end{align}

Similarly, for the amplitude of $\chi^0$ we have

\begin{align}
   \mathcal{M}_{\chi^{0}}=\frac{i g^{3}}{32 \pi^{2}} \frac{M_D^{2}}{M_W^{2}} U_{s b}\left(\frac{-\frac{1}{2}-\frac{1}{3} \sin ^{2}\theta_{W}}{4 \cos \theta_{W}}\right)\left(\frac{1}{\epsilon}+\frac{3}{4}-\frac{x_2^{2} \log x_2}{2(1-x_2)^{2}}-\frac{1}{2(1-x_2)}\right) \bar{s} \gamma_\mu\left(1-\gamma_{5}\right) b
\end{align}
\subsubsection*{Wilson Coefficients $C_{9}$ and $ C_{10}$}
We consider the self-energy diagrams of $b\rightarrow \ s\mu^+\mu^-$ to obtain the Wilson coefficients $C_9$ and $C_{10}$.
\begin{figure}[h!]
\begin{center}
    \bigskip
\begin{tikzpicture} 
\begin{feynman}
\vertex (a1) {\( b\)};
\vertex[right=1.5cm of a1] (a2); 
\vertex[right=1cm of a2] (a3); 
\vertex[right=1cm of a3] (a4); 
\vertex[right=0.5cm of a4](a);
\vertex[right=1.5cm of a4] (a5) {\(s\)};
\vertex [below=of a](c1);
\vertex [below right=of c1] (e) {$\mu$};
\vertex [below left=of c1] (f) {$\mu$};
\diagram* { 
(a1)--[fermion](a2);
(a2)--[fermion,edge label' ={$D$}] (a4);
(a4)--[fermion](a5);
 (a) -- [boson, edge label' = \(Z\)] (c1),
(a4) -- [scalar, out=90, in=90, looseness=2.0, edge label'={$H,\chi^0$}] (a2);
(e) -- [fermion](c1) -- [fermion](f);
};
\end{feynman} 
\end{tikzpicture}
\hspace{1em}
\begin{tikzpicture} 
\begin{feynman}
\vertex (a1) {\( b\)};
\vertex[right=1.5cm of a1] (a2); 
\vertex[right=1cm of a2] (a3); 
\vertex[right=1cm of a3] (a4); 
\vertex[right=1cm of a1](a);
\vertex[right=1.5cm of a4] (a5) {\( s\)};
\vertex [below=of a](c1);
\vertex [below right=of c1] (e) {$\mu$};
\vertex [below left=of c1] (f) {$\mu$};
\diagram* { 
(a1)--[fermion](a2);
(a2)--[fermion,edge label' ={$D$}] (a4);
(a4)--[fermion](a5);
 (a) -- [boson, edge label' = \(Z\)] (c1),
(a4) -- [scalar, out=90, in=90, looseness=2.0, edge label'={$H,\chi^0$}] (a2);
(e) -- [fermion](c1) -- [fermion](f);
};
\end{feynman} 
\end{tikzpicture}
\end{center}
\caption{Self Energy  diagrams $b\rightarrow s \mu^+\mu^-$}
\label{SEM}
\end{figure}
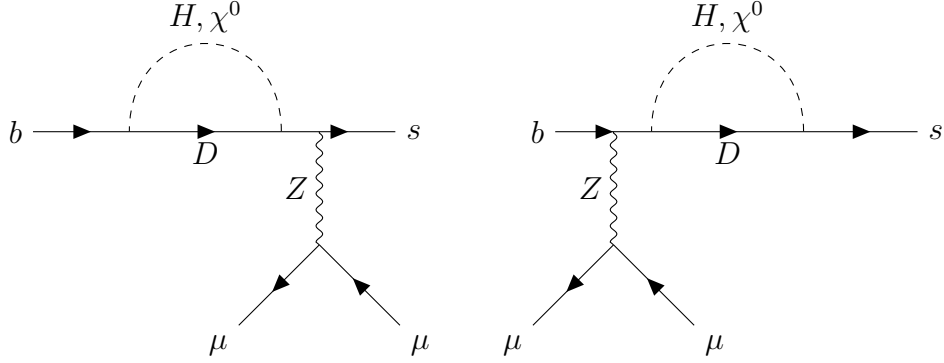

Using the eq.(\ref{HS1}),amplitude for the self-energy  of $b\rightarrow s \mu^+\mu^-$ in Fig.(\ref{SEM}) is written as,
\begin{equation}
    \begin{aligned}
      \mathcal{M_H}&=\frac{i g^{3}}{32 \pi^{2}} \frac{M_D^{2}}{M_W^{2}} U_{s b}\left(\frac{\frac{-1}{2}+\frac{1}{3} \sin ^{2} \theta_{W}}{16 M_Z^2 \cos^2\theta_{W}}\right)\left(\frac{1}{\epsilon}+\frac{3}{4}-\frac{x_1^{2} \log x_1}{2(1-x_1)^{2}}-\frac{1}{2(1-x_1)}\right) \bar{s} \gamma_\mu\left(1-\gamma_{5}\right) b \\
  &\times \frac{i}{M_Z^2}\left(\frac{ig}{2\cos{\theta_W}}\bar{\mu}(\frac{-1}{2}+2\sin^2{\theta_W}+\frac{1}{2}\gamma_5)\mu \right)  
    \end{aligned}
\end{equation}
Using $G_F$ from eq.(\ref{G_F}) we get,
\begin{equation}
    \begin{aligned}
     \mathcal{M_H}&=\frac{-i G_F}{\sqrt{2}} \frac{M_D^{2}}{M_W^{2}} \frac{\alpha}{\pi \sin^2{\theta_W}}U_{s b}\left(\frac{\frac{-1}{2}+\frac{1}{3} \sin ^{2} \theta_{W}}{16}\right)\left(\frac{1}{\epsilon}+\frac{3}{4}-\frac{x_1^{2} \log x_1}{2(1-x_1)^{2}}-\frac{1}{2(1-x_1)}\right)\\& \times \bar{s} \gamma_\mu\left(1-\gamma_{5}\right) b 
   \left(\bar{\mu}(-1+4\sin^2{\theta_W}+\gamma_5)\mu \right)   
    \end{aligned}
\end{equation}
Hence, the effective Hamiltonian $\mathcal{H}=i\mathcal{M}$ is,
\begin{equation}
    \begin{aligned}
        \mathcal{H}_{eff}&=\frac{G_F}{\sqrt{2}} \frac{M_D^{2}}{M_W^{2}} \frac{\alpha}{\pi \sin^2{\theta_W}}U_{s b}\left(\frac{\frac{-1}{2}+\frac{1}{3} \sin ^{2} \theta_{W}}{16}\right)\left(\frac{1}{\epsilon}+\frac{3}{4}-\frac{x_1^{2} \log x_1}{2(1-x_1)^{2}}-\frac{1}{2(1-x_1)}\right) \\
        & \times\bar{s} \gamma_\mu\left(1-\gamma_{5}\right) b 
   \left(\bar{\mu}(-1+4\sin^2{\theta_W}+\gamma_5)\mu \right)   
    \end{aligned}
\end{equation}
By comparing this with eq.(\ref{HHSM}), the Wilson Coefficients $C_{9}$ and $C_{10}$ are 
\begin{equation}
    C_{9}= -\frac{U_{sb}}{\lambda_t}\frac{M_D^2}{M_W^2}\left(\frac{\frac{-1}{2}+\frac{1}{3}\sin^2{\theta_W}}{16\sin^2{\theta_W}}\right)\left(\frac{1}{\epsilon}+\frac{3}{4}-\frac{x_1^{2} \log x_1}{2(1-x_1)^{2}}-\frac{1}{2(1-x_1)}\right)(4\sin^2{\theta_W}-1)
\end{equation}
\begin{equation}
    C_{10}= -\frac{U_{sb}}{\lambda_t}\frac{M_D^2}{M_W^2}\left(\frac{\frac{-1}{2}+\frac{1}{3}\sin^2{\theta_W}}{16\sin^2{\theta_W}}\right)\left(\frac{1}{\epsilon}+\frac{3}{4}-\frac{x_1^{2} \log x_1}{2(1-x_1)^{2}}-\frac{1}{2(1-x_1)}\right)
    \end{equation}
   Similarly for self-energy with $\chi^0$, we have 
\begin{equation}
    C_{9}= -\frac{U_{sb}}{\lambda_t}\frac{M_D^2}{M_W^2}\left(\frac{\frac{-1}{2}+\frac{1}{3}\sin^2{\theta_W}}{16\sin^2{\theta_W}}\right)\left(\frac{1}{\epsilon}+\frac{3}{4}-\frac{x_2^{2} \log x_2}{2(1-x_2)^{2}}-\frac{1}{2(1-x_2)}\right)(4\sin^2{\theta_W}-1)
    \label{C_9X}
 \end{equation}
 \begin{equation}
    C_{10}= -\frac{U_{sb}}{\lambda_t}\frac{M_D^2}{M_W^2}\left(\frac{\frac{-1}{2}+\frac{1}{3}\sin^2{\theta_W}}{16\sin^2{\theta_W}}\right)\left(\frac{1}{\epsilon}+\frac{3}{4}-\frac{x_2^{2} \log x_2}{2(1-x_2)^{2}}-\frac{1}{2(1-x_2)}\right)
    \label{C_10X}
\end{equation}

\subsubsection*{Total Wilson Coefficients}

The total Wilson coefficients ($C_9, C_{10}$)  combined from the tree-level, vertex, and self-energy loop diagrams of $b\rightarrow s\mu^+\mu^-$ in Fig.(\ref{fig:TreeFCNC}), Fig.(\ref{VH}) and Fig.(\ref{SEM}) is
\begin{align}
        C_{9}^{'}= C_{9, SM}+C_{9,VLQ}
 \end{align}
 \begin{align}
        C_{10}^{'}= C_{10, SM}+C_{10,VLQ}
 \end{align}
 from eqs.(\ref{C_9T}),(\ref{C_9H}) and (\ref{C_9X}) the total Wilson coefficient $C_9$ is,
\begin{equation}
     \begin{aligned}
        C_{9}^{'} &=C_{9, SM}-\frac{U_{sb}}{\lambda_t}\frac{\pi}{\alpha}-\frac{U_{sb}}{\lambda_t}\frac{M_D^2}{M_W^2}(4\sin^2{\theta_W}-1) \Bigg\{\frac{1}{48}\Bigg[\left(\frac{-2}{\epsilon}-\frac{1}{2}+\frac{x_1^{2} \log x_1}{(1-x_1)^{2}}+\frac{x_{1}}{\left(1-x_{1}\right)}\right)
    &+\left(\frac{-2}{\epsilon}-\frac{1}{2}+\frac{x_2^{2} \log x_2}{(1-x_2)^{2}}+\frac{x_{2}}{\left(1-x_{2}\right)}\right)\Bigg]
    +\left(\frac{\frac{-1}{2}+\frac{1}{3}\sin^2{\theta_W}}{16\sin^2{\theta_W}}\right)\\&\times\Bigg[\left(\frac{1}{\epsilon}+\frac{3}{4}-\frac{x_1^{2} \log x_1}{2(1-x_1)^{2}}-\frac{1}{2(1-x_1)}\right)+\left(\frac{1}{\epsilon}+\frac{3}{4}-\frac{x_2^{2} \log x_2}{2(1-x_2)^{2}}-\frac{1}{2(1-x_2)}\right)\Bigg]\Bigg\} (4\sin^2{\theta_W}-1) 
     \end{aligned}
\end{equation}
Similarly, using the eqs. (\ref{C_10T}) (\ref{C_10H}) and (\ref{C_10X}) we have combined Wilson coefficient $C_{10}$,
\begin{equation}
    \begin{aligned}
        C_{10}^{'} &=C_{10, SM}-\frac{U_{sb}}{\lambda_t}\frac{\pi}{\alpha} -\frac{U_{sb}}{\lambda_t}\frac{M_D^2}{M_W^2}\Bigg\{\frac{1}{48}\Bigg[\left(\frac{-2}{\epsilon}-\frac{1}{2}+\frac{x_1^{2} \log x_1}{(1-x_1)^{2}}+\frac{x_{1}}{\left(1-x_{1}\right)}\right)\\
   & +\left(\frac{-2}{\epsilon}-\frac{1}{2}+\frac{x_2^{2} \log x_2}{(1-x_2)^{2}}+\frac{x_{2}}{\left(1-x_{2}\right)}\right)\Bigg]+\left(\frac{\frac{-1}{2}+\frac{1}{3}\sin^2{\theta_W}}{16\sin^2{\theta_W}}\right)\\&\times\Bigg[\left(\frac{1}{\epsilon}+\frac{3}{4}-\frac{x_1^{2} \log x_1}{2(1-x_1)^{2}}-\frac{1}{2(1-x_1)}\right)+\left(\frac{1}{\epsilon}+\frac{3}{4}-\frac{x_2^{2} \log x_2}{2(1-x_2)^{2}}-\frac{1}{2(1-x_2)}\right)\Bigg]\Bigg\} 
    \end{aligned}
\end{equation}
 Hence, the simplified form of total Wilson coefficient is, 
\begin{equation}
    \begin{aligned}
      C_{9}^{'} & =C_{9, SM}-\frac{U_{sb}}{\lambda_t}\frac{\pi}{\alpha} (4\sin^2{\theta_W}-1)-\frac{U^{sb}}{\lambda_t}\frac{M_D^2}{M_W^2}\\
  & \times \left[\frac{1}{48}(F(x_1)+F(x_2))
    +\left(\frac{\frac{-1}{2}+\frac{1}{3}\sin^2{\theta_W}}{16\sin^2{\theta_W}}\right)\left(F^{'}(x_1)+F^{'}(x_2)\right)\right](4\sin^2{\theta_W}-1)   
    \end{aligned}
\end{equation}
and 
\begin{equation}
    \begin{aligned}
    C_{10}^{'} &=C_{10, SM}-\frac{U_{sb}}{\lambda_t}\frac{\pi}{\alpha} -\frac{U^{sb}}{\lambda_t}\frac{M_D^2}{M_W^2}\Bigg[\frac{1}{48}(F(x_1)+F(x_2))\\
    & +\left(\frac{\frac{-1}{2}+\frac{1}{3}\sin^2{\theta_W}}{16\sin^2{\theta_W}}\right)\left(F^{'}(x_1)+F^{'}(x_2)\right)\Bigg]
\end{aligned}
\end{equation}

Where, from Appendix B, $F(x)$ and $F'(x)$ are given as
\begin{equation}
    F(x)=\left(\frac{-2}{\epsilon}-\frac{1}{2}+\frac{x^{2} \log x}{(1-x)^{2}}+\frac{x}{\left(1-x)\right)}\right)
\end{equation}

\begin{equation}
    F^{'}(x)= \left(\frac{1}{\epsilon}+\frac{3}{4}-\frac{x^{2} \log x}{2(1-x)^{2}}-\frac{1}{2(1-x)}\right)
\end{equation}
After summing all relevant one-loop contributions to the $b \rightarrow s Z$ vertex including those from the extended quark sector, we find that the resulting amplitude remains ultraviolet divergent. This residual divergence arises due to the non-unitary structure induced by the VLQ mixing, which prevents the full cancellation of divergences among vertex and self-energy diagrams. While part of the divergence in the amplitude is cancelled by self-energy diagrams, the non-unitarity leaves behind residual divergences. These remaining divergent terms propagate into the loop-induced Wilson coefficients and are absorbed into renormalized counterterms for the effective operators. This ensures the finiteness of physical observables, though we do not explicitly display the counterterms here, as our focus lies on the finite contributions arising from VLQ-induced flavor mixing.

\subsection{$b \to s\gamma$ Transition with VLQs}
 VLQs contribute via new loop diagrams involving the heavy $D$ quark and its mixing with SM quarks. The process remains loop-suppressed, with modifications to the Wilson coefficient $C_7$. The branching ratio can shift, but not as dramatically as $ b\to sZ$.

The process $ b \rightarrow s \gamma$ is a neutral current transition with changing flavor (FCNC) at the Standard Model's loop level. Flavor-changing processes such as $ b\rightarrow s \gamma$ are forbidden at the tree level due to the GIM mechanism but arise at the loop level via electroweak corrections. The decay involves an initial b-quark transitioning to the s-quark and emission of a photon.
The dominant contributions to $ b \to s \gamma $ come from Fig.(\ref{feyngamma}).
\begin{figure}[h]
    \centering
    \begin{minipage}{0.25\textwidth}
        \centering
        \resizebox{\linewidth}{!}{%
        \begin{tikzpicture}
            \begin{feynman}
                \vertex (b1) {$b$};
                \vertex [right=1.2cm of b1] (b2);
                \vertex [right=1.2cm of b2] (b3);
                \vertex [right=1.2cm of b3] (b4) {$s$};
                \vertex at ($(b2)!0.5!(b3)!0.8cm!-90:(b3)$) (g1);
                \vertex [below= of g1] (g2);
                \diagram* {
                    (b1) -- [fermion] (b2) -- [boson, edge label={$W^{-}$}] (b3) -- [fermion] (b4),
                    (b2) -- [fermion, out=90, looseness=1, quarter right, edge label'=${u,c,t}$] (g1)
                          -- [fermion, in=90, looseness=1, quarter right, edge label'=${u,c,t}$] (b3),
                    (g1) -- [boson, edge label'=$\gamma$] (g2);
                };
            \end{feynman}
        \end{tikzpicture}}
        (a)
    \end{minipage}
  \hspace{0.25\textwidth}
    \begin{minipage}{0.25\textwidth}
        \centering
        \resizebox{\linewidth}{!}{%
        \begin{tikzpicture}
            \begin{feynman}
                \vertex (b1) {$b$};
                \vertex [right=1.2cm of b1] (b2);
                \vertex [right=1.2cm of b2] (b3);
                \vertex [right=1.2cm of b3] (b4) {$s$};
                \vertex at ($(b2)!0.5!(b3)!0.8cm!-90:(b3)$) (g1);
                \vertex [below= of g1] (g2);
                \diagram* {
                    (b1) -- [fermion] (b2) -- [scalar, edge label={$\chi^{-}$}] (b3) -- [fermion] (b4),
                    (b2) -- [fermion, out=90, looseness=1, quarter right, edge label'=${u,c,t}$] (g1)
                          -- [fermion, in=90, looseness=1, quarter right, edge label'=${u,c,t}$] (b3),
                    (g1) -- [boson, edge label'=$\gamma$] (g2);
                };
            \end{feynman}
        \end{tikzpicture}}
        (b)
    \end{minipage}
    \vspace{0.5cm}
    \begin{minipage}{0.25\textwidth}
        \centering
        \resizebox{\linewidth}{!}{%
        \begin{tikzpicture}
            \begin{feynman}
                \vertex (b1) {$b$};
                \vertex [right=1.2cm of b1] (b2);
                \vertex [right=1.2cm of b2] (b3);
                \vertex [right=1.2cm of b3] (b4) {$s$};
                \vertex at ($(b2)!0.5!(b3)!0.8cm!-90:(b3)$) (g1);
                \vertex [below= of g1] (g2);
                \diagram* {
                    (b1) -- [fermion] (b2) -- [fermion, edge label={$u,c,t$}] (b3) -- [fermion] (b4),
                    (b2) -- [boson, out=90, looseness=1, quarter right, edge label'=$W^{-}$] (g1)
                          -- [boson, in=90, looseness=1, quarter right, edge label'=$W^{+}$] (b3),
                    (g1) -- [boson, edge label'=$\gamma$] (g2);
                };
            \end{feynman}
        \end{tikzpicture}}
        (c)
    \end{minipage}
    \hspace{0.25\textwidth}
    \begin{minipage}{0.25\textwidth}
        \centering
        \resizebox{\linewidth}{!}{%
        \begin{tikzpicture}
            \begin{feynman}
                \vertex (b1) {$b$};
                \vertex [right=1.2cm of b1] (b2);
                \vertex [right=1.2cm of b2] (b3);
                \vertex [right=1.2cm of b3] (b4) {$s$};
                \vertex at ($(b2)!0.5!(b3)!0.8cm!-90:(b3)$) (g1);
                \vertex [below= of g1] (g2);
                \diagram* {
                    (b1) -- [fermion] (b2) -- [fermion, edge label={$u,c,t$}] (b3) -- [fermion] (b4),
                    (b2) -- [scalar, out=90, looseness=1, quarter right, edge label'=$\chi^{-}$] (g1)
                          -- [scalar, in=90, looseness=1, quarter right, edge label'=$\chi^{+}$] (b3),
                    (g1) -- [boson, edge label'=$\gamma$] (g2);
                };
            \end{feynman}
        \end{tikzpicture}}
        (d)
    \end{minipage}

    \vspace{0.5cm}

    \begin{minipage}{0.25\textwidth}
        \centering
        \resizebox{\linewidth}{!}{%
        \begin{tikzpicture}
            \begin{feynman}
                \vertex (b1) {$b$};
                \vertex [right=1.2cm of b1] (b2);
                \vertex [right=1.2cm of b2] (b3);
                \vertex [right=1.2cm of b3] (b4) {$s$};
                \vertex at ($(b2)!0.5!(b3)!0.8cm!-90:(b3)$) (g1);
                \vertex [below= of g1] (g2);
                \diagram* {
                    (b1) -- [fermion] (b2) -- [fermion, edge label={$u,c,t$}] (b3) -- [fermion] (b4),
                    (b2) -- [boson, out=90, looseness=1, quarter right, edge label'=$W^{-}$] (g1)
                          -- [scalar, in=90, looseness=1, quarter right, edge label'=$\chi^{+}$] (b3),
                    (g1) -- [boson, edge label'=$\gamma$] (g2);
                };
            \end{feynman}
        \end{tikzpicture}}
        (e)
    \end{minipage}
     \hspace{0.25\textwidth}
    \begin{minipage}{0.25\textwidth}
        \centering
        \resizebox{\linewidth}{!}{%
        \begin{tikzpicture}
            \begin{feynman}
                \vertex (b1) {$b$};
                \vertex [right=1.2cm of b1] (b2);
                \vertex [right=1.2cm of b2] (b3);
                \vertex [right=1.2cm of b3] (b4) {$s$};
                \vertex at ($(b2)!0.5!(b3)!0.8cm!-90:(b3)$) (g1);
                \vertex [below= of g1] (g2);
                \diagram* {
                    (b1) -- [fermion] (b2) -- [fermion, edge label={$u,c,t$}] (b3) -- [fermion] (b4),
                    (b2) -- [scalar, out=90, looseness=1, quarter right, edge label'=$\chi^{-}$] (g1)
                          -- [boson, in=90, looseness=1, quarter right, edge label'=$W^{+}$] (b3),
                    (g1) -- [boson, edge label'=$\gamma$] (g2);
                };
            \end{feynman}
        \end{tikzpicture}}
        (f)
    \end{minipage}
    
    \caption{Photon penguin diagrams contributing to $b \to s \gamma$ via $W$ and $\chi^\pm$ loops}
    \label{feyngamma}
\end{figure}

From \cite{Buras:1998raa}, the Standard Model amplitude for $ b \to s \gamma $ is,
\begin{equation}
    \bar{s}\gamma b = i\lambda_i\frac{ G_F}{\sqrt{2}} \frac{e}{8\pi^2} D_0'(x_i) \bar{s}(i\sigma_{\mu\nu}q^\nu)[m_b(1+\gamma_5)]b 
    \label{eq:7}
\end{equation}
where $x_t = m_t^2/m_W^2$ and $D'_0(x_t)$is given as
\begin{equation}
D'_0(x_t) = \frac{x_t(7-5x_t-8x_t^2)}{24(x_t-1)^3} + \frac{x_t^2(3x_t-2)}{4(x_t-1)^4}\ln x_t
\end{equation}
Hence, the SM amplitude of $ b \to s \gamma $ is:
\begin{equation}
     \mathcal{M}_{\gamma}=i\lambda_t\frac{e}{8\pi^2} m_b D_0'(x_t) \bar{s}(i \sigma_{\mu\nu} q^\nu (1 + \gamma^5)) b \epsilon^\mu(q),
     \label{gamma1}
\end{equation}
We used Mathematica with BSM Package(Peng4BSM@LO)\cite{Bednyakov:2015penguin} in order to get the amplitude with non-unitary CKM due to  the VLQ contribution.
\begin{multline}
    \mathcal{M}_{\gamma}=\frac{e^3 m_b}{1152M^2_W \pi^2 \sin^2{\theta_W}}\frac{1}{(x_1^2-1)^4}(V_{ub}V^*_{us}+V_{cb}V^*_{cs}+V_{tb}V^*_{ts})\\
    \left[46-205x_1^2+24(13-3\log{x_1})x_1^4+(-175+108\log{x_1})x_1^6+22x_1^6\right]\bar{s} \sigma_{\mu\nu} q^\nu (\frac{1 + \gamma^5}{2}) b 
\end{multline}
From the above equation, the non-unitary contribution is 
\begin{equation}
    \mathcal{M}_{\gamma}=-i\frac{e g^2 \sin^2{\theta_W} m_b}{1152M^2_W \pi^2 \sin^2{\theta_W}}(U_{sb})
    (46)\bar{s} (i\sigma_{\mu\nu} q^\nu (\frac{1 + \gamma^5}{2})) b 
\end{equation}
Using equation(\ref{G_F}) and $i \, \sigma_{\mu\nu} \, q^\nu \, \varepsilon^\mu = \frac{1}{2} \, \sigma^{\mu\nu} F_{\mu\nu}$,
\begin{equation}
    \mathcal{M}_{\gamma}=i\frac{G_F}{\sqrt{2}}\frac{e }{8\pi^2 }U_{sb}
    \frac{23}{18} m_b\bar{s} (\frac{1}{2}\sigma^{\mu\nu} F_{\mu\nu} (1 +\gamma^5) b 
    \label{gamma2}
\end{equation}
Adding equations (\ref{gamma1}) and (\ref{gamma2}), we get the total amplitude,
\begin{equation}
    \mathcal{M}_{\gamma}=i\frac{G_F}{\sqrt{2}}\frac{e }{8\pi^2 }\lambda_t(D'_0(x_i)-\frac{U_{sb}}{\lambda_t}
    \frac{23}{18} )m_b\bar{s} (\frac{-1}{2}\sigma^{\mu\nu} F_{\mu\nu} (1 +\gamma^5)) b 
    \label{gamma2}
\end{equation}
The effective Hamiltonian of $b\rightarrow s\gamma$ is 
\begin{equation}
    \mathcal{H}_{eff}=\frac{G_F}{\sqrt{2}}\frac{e }{8\pi^2 }\lambda_t\frac{1}{2}(D'_0(x_t)-\frac{U_{sb}}{\lambda_t}
    \frac{23}{18} )m_b\bar{s} \sigma^{\mu\nu} F_{\mu\nu} (1 +\gamma^5) b 
    \label{gamma2}
\end{equation}
The result is matched to the Standard Model effective Hamiltonian:
\begin{equation}
\mathcal{H}_{\text{eff}} = -4\frac{G_F}{\sqrt{2}}\lambda_t (C_7 O_7)
\label{HG}
\end{equation}
where $O_7$ operator;
\[
O_7 = \frac{e}{16\pi^2} m_b (\bar{s} \sigma^{\mu\nu} P_R b) F_{\mu\nu}
\]
And $ C_7$  is the corresponding Wilson coefficient. Hence, by comparing eq.(\ref{gamma2}) and eq.(\ref{HG}) the Wilson coefficient $C_7$ is,
\begin{equation}
    C_7 = -\frac{1}{2}( D'_0(x_t)-\frac{U_{sb}}{\lambda_i}
    \frac{23}{18})
\end{equation}
 
where $C_7 = -\frac{1}{2}( D'_0(x_t)$ is the standard model Wilson coefficient and $\frac{1}{2}\frac{23}{18}\frac{U_{sb}}{\lambda_i}$ is the non-unitarity contribution due to VLQ.
In Table(\ref{Comp}) below is the comparison between $b\rightarrow sZ$ and $b\rightarrow s \gamma$ in the context of vector-like quark Model, where $B\rightarrow s \gamma$ has small effects on SM whereas, Z channel $b\rightarrow s Z$ gives significant contribution.
\begin{table}[h!]
\centering
\resizebox{\textwidth}{!}{%
\begin{tabular}{|c|c|c|}
\hline
\textbf{Feature} & \textbf{$b \to sZ$ (with VLQ)} & \textbf{$b \to s\gamma$ (with VLQ)} \\
\hline
SM Mechanism & Loop-level only & Loop-level only \\
\hline
VLQ Effect & Tree-level FCNC possible & Only modifies loop contributions \\
\hline
Main Impact & Large enhancement possible & Moderate shift in branching ratio \\
\hline
Affected Wilson Coeff. & $C_9$, $C_{10}$ & $C_7$ \\
\hline
Experimental Sensitivity & $B \to K^{(*)} \ell^+ \ell^-$, $B_s \to \mu^+\mu^-$ & $b \to s\gamma$ inclusive/exclusive \\
\hline
New Physics Signature & Deviations in angular observables,
branching ratios, lepton universality & Small deviations in $b \to s\gamma$ branching ratio \\
\hline

\end{tabular}%
}
\caption{Comparison of $b \to sZ$ and $b \to s\gamma$ in the context of vector-like quarks.}
\label{Comp}
\end{table}

\section{ Exclusive B-Decay $B\rightarrow K \nu \bar{\nu}$}

Vector-like quark models significantly alter the decay \( B \rightarrow K \nu \bar{\nu} \) through non unitarity of CKM matrix and  additional contribution from penguin diagrams , modifying the effective Wilson coefficient \( C_L^{\text{eff}} \). These models predict correlated deviations in the branching ratios \( \text{BR}(B \rightarrow K^+ \nu \bar{\nu}) \) and \( \text{BR}(B \rightarrow K^{(*)} \ell^+ \ell^-) \), offering a way to distinguish VLQs from other scenarios of  NP\cite{Buras:2014fpa}, \cite{Hou:2024vyw}.
The SM prediction for the branching ratio (see e.g.\cite{Buras:2014fpa}, \cite{Altmannshofer:2009ma}) is given by:
\begin{equation}
\mathcal{B}(B^+ \rightarrow K^+ \nu \bar{\nu})_{\text{SM}} \approx (4.0 \pm 0.5) \times 10^{-6}
\end{equation}
Recent  Belle II measurement \cite{Belle-II:2023esi},  
\begin{equation}
    \text{BR}(B^+ \rightarrow K^+ \nu \bar{\nu}) = (2.3 \pm 0.7) \times 10^{-5} 
\end{equation}
resulted a branching ratio for \( B^+ \rightarrow K^+ \nu \bar{\nu} \) which is \( 2.7\sigma \) above the Standard Model prediction \cite{Belle-II:2023esi}, \cite{Altmannshofer:2023hkn}, suggesting the possibility of VLQ-induced effects. 

Compared to inclusive processes such as \( B \rightarrow X_s \nu \bar{\nu} \), exclusive decays are experimentally more accessible, as they involve a fully reconstructible meson in the final state. While inclusive decays are theoretically cleaner due to parton-hadron duality, the exclusive channel \( B\rightarrow K \nu\bar {\nu} \) is a more practical observable at $ B$-factories like Belle II.  A VLQ explanation of such an anomaly would require TeV-scale VLQ masses and specific flavor structures, which can be further tested through polarization measurements in \( B \rightarrow K^* \nu \bar{\nu} \) and complementary direct searches at high-energy colliders.

Furthermore, when compared to rare kaon decays such as \( K^+ \rightarrow \pi^+ \nu \bar{\nu} \) or \( K_L \rightarrow \pi^0 \nu \bar{\nu} \), which are also FCNCs driven by \( s \rightarrow d \nu \bar{\nu} \), the $B$-decay probes different elements of the quark flavor structure namely, the \( b \rightarrow s \) transition governed by \( V_{tb}V_{ts}^* \), rather than \( V_{ts}V_{td}^* \). This makes $ B$-decays complementary to kaon decays in studying the flavor sector. Moreover, the energy scale involved in $ B$-decays allows for better sensitivity to heavy New Physics states.

In models with Vector-Like Quarks,  new heavy fermions mix with SM quarks without violating gauge invariance, thereby inducing tree-level FCNCs through modified $Z$-boson couplings. In particular, the $Z$-penguin contribution to \( b \rightarrow s \nu \bar{\nu} \) is modified, leading to corrections to the effective Wilson coefficient \( C_L^\nu \), and potentially the appearance of a right-handed operator with a new coefficient \( C_R^\nu \), which is absent in the SM.

This can significantly affect both the decay rate and the angular distributions (in the case of \( B \rightarrow K^* \nu \bar{\nu} \)), and leads to deviations from the SM prediction of the branching ratio. The new contributions can be parametrized by defining \cite{Altmannshofer:2009ma}

\begin{equation}
\mathcal{H}_{\text{eff}} = -\frac{4 G_F}{\sqrt{2}} V_{tb} V_{ts}^* \left( C_L^\nu \mathcal{O}_L^\nu + C_R^\nu \mathcal{O}_R^\nu \right),
\label{Hvv}
\end{equation}

With the operators,
\begin{equation}
\mathcal{O}_L^\nu = \frac{e^2}{16 \pi^2} (\bar{s} \gamma_\mu P_L b)(\bar{\nu} \gamma^\mu (1 - \gamma_5) \nu), \quad
\mathcal{O}_R^\nu = \frac{e^2}{16 \pi^2} (\bar{s} \gamma_\mu P_R b)(\bar{\nu} \gamma^\mu (1 - \gamma_5) \nu)
\end{equation}
In the SM, $C^\nu_R$ is negligible and
\begin{align}
    C^\nu_L= \frac{-X(x_t)}{\sin^2{\theta_W}}
    \label{CL}
\end{align}
 where $x_t=m_t^2/M_W^2$ and the function $X(x_t)$ at the next-to-leading order in QCD \cite{Buchalla:1998ba}.
 \begin{equation}
X(x_t) = \eta_Y X_0(x_t) 
\end{equation}
where $\eta_Y= 0.89$ is the QCD contribution, and $X_0(x_t) $is
\begin{equation}
X_0(x_t) = \frac{x_t}{8}\Big[\frac{2 + x_t}{x_t -1}+\frac{3x_t-6}{(x_t-1)^2}\ln x_t\Big]\;.
\label{X0}
\end{equation}

In the Standard Model \cite{Buras:2014fpa},
\begin{equation}
(C^\nu_L)^\text{SM} = - 6.38 \pm 0.06~,
\label{eq:CLSM}
\end{equation}
where the uncertainty of the top quark mass dominates the error, the corresponding operator is not renormalized by QCD, so the only dependence of the renormalization scale enters $X(x_t)$ through the running top quark mass, which is, however, canceled mainly through NLO QCD corrections. The residual scale dependence is considered in the error in eq.(\ref{eq:CLSM}).

\subsection{Vector Like Quark Contribution}

Vector-like quark contribution significantly affects the \( b \rightarrow s \nu \bar{\nu} \) transition, as observed deviation in branching ratios. It can also uncover right-handed current contributions through angular observables in \( B \rightarrow K \nu \bar{\nu} \), making these modes key targets in the search for New Physics.

In VLQ scenarios, heavy fermions mix with SM quarks without breaking gauge invariance. This induces tree-level FCNCs via modified \( Z \)-boson couplings, leading to both left- and right-handed operators in the effective Hamiltonian given in eq.(\ref{Hvv}), where, \( C_L^\nu \) and \( C_R^\nu \) receive contributions proportional to the VLQ couplings and mixing angles. This can lead to enhancements or suppressions in the branching ratio depending on the VLQ representation and coupling structure.




It is essential to compute quark-level box Fig.(\ref{Boxvv}) and penguin diagrams involving VLQs Fig.(\ref{fig:higgs-bsvv}), as these contribute to the short-distance physics encoded in the Wilson coefficients \( C_L^\nu \) and \( C_R^\nu \). These diagrams encapsulate how VLQs modify FCNCs, directly influencing both decay rates and angular observables. Accurate evaluation of these loop-level contributions allows one to connect deviations in experimental data with the structure and couplings of new heavy states, thereby testing specific VLQ scenarios and distinguishing them from other models of New Physics.

\subsubsection {Box Diagram  $ b \to s \nu \bar{\nu} $}

\begin{figure}[h!]
\centering
\begin{tikzpicture}
  \begin{feynman}
    \vertex (a) {\(b\)};
    \vertex[right=2cm of a] (b);
    \vertex[below=2cm of a] (c) {\(s\)};
    \vertex[right=2cm of c] (d);
    \vertex[right=2cm of b] (e);
    \vertex[right=2cm of d] (f);
    \vertex[right=2cm of e] (g) {\(\nu\)};
    \vertex[right=2cm of f] (h) {\(\nu\)};

    \diagram* {
      (a) -- [fermion] (b) -- [boson, edge label=\(W^\pm\)](e),
      (g) -- [fermion] (e),
      (d) -- [fermion] (c),(d) -- [boson, edge label'=\(W^\pm\)] (f) -- [fermion] (h),
      (b) -- [fermion, edge label=\(t\)] (d),
      (e) -- [fermion, edge label'=\(e\)] (f),
    };
  \end{feynman}
\end{tikzpicture}
\caption{Box diagram with internal $W$ bosons of $b\rightarrow s \nu \bar{\nu}$}
\label{Boxvv}
\end{figure}
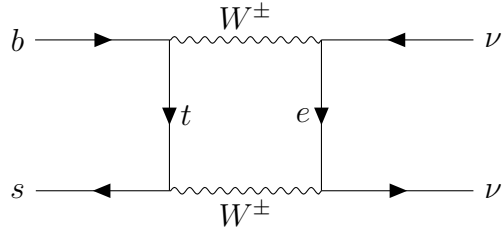

The process $ b \to s \nu \bar{\nu} $ involves a box diagram mediated by two W-bosons and a top quark t in the loop. The Feynman diagram in Fig.(\ref{Boxvv}) shows an incoming b-quark decaying into an s-quark and two neutrinos ($ \nu $ and $ \bar{\nu} $) emitted as final-state particles.

The full amplitude from the Feynman rules is,
\begin{equation}
\begin{aligned}
\mathcal{M}_{\text{box}} &= \int \frac{d^4k}{(2\pi)^4} \left[ \bar{s} \left( \frac{-ig}{\sqrt{2}} \gamma^\mu (1 - \gamma_5) V^*_{ts} \right) 
\frac{i(\slashed{k} + m_t)}{k^2 - m_t^2} 
\left( \frac{-ig}{\sqrt{2}} \gamma^\nu (1 - \gamma_5) V_{tb} \right) b \right] \\
&\quad \times \left[ \bar{\nu} \left( \frac{-ig}{\sqrt{2}} \gamma^\rho (1 - \gamma_5) \right)
\frac{i \slashed{k}}{k^2} 
\left( \frac{-ig}{\sqrt{2}} \gamma^\sigma (1 - \gamma_5) \right) \nu \right]  \cdot \frac{-i g_{\mu\rho}}{k^2 - M_W^2} \cdot \frac{-i g_{\nu\sigma}}{k^2 - M_W^2}
\end{aligned}
\end{equation}
We have ignored the mass of electron $m_e \approx0$ and  $\lambda_t = V^*_{ts} V_{tb}$. Factorizing the spinor and loop structure, we have
\begin{equation}
\mathcal{M}_{\text{box}} = \left( \frac{g^4}{4} \right) \lambda_t  T^{\sigma\tau} R_{\sigma\tau}
\label{mbox}
\end{equation}
where the Dirac (spinor) structure is
\begin{equation}
T^{\sigma\tau} = -4\, \bar{s} \gamma^\mu \gamma^\sigma (1 - \gamma_5) b \otimes \bar{\nu} \gamma_\mu \gamma^\tau (1 - \gamma_5) \nu
\end{equation}
and the loop integral is
\begin{equation}
R^{\sigma\tau} = \int \frac{d^4k}{(2\pi)^4} \frac{k^\sigma k^\tau}{(k^2 - m_t^2)(k^2 - M_W^2)^2}
\end{equation}
Using Lorentz symmetry,
\begin{equation}
R^{\sigma\tau} = g^{\sigma\tau}  \int \frac{d^4k}{(2\pi)^4} \frac{k^2}{4(k^2 - m_t^2)(k^2 - M_W^2)^2}
\end{equation}

Evaluating the integral, we obtain
\begin{equation}
R^{\sigma\tau} = \frac{i}{16\pi^2 M_W^2} \left[ 4 B_0(x_t) + 1 \right] g^{\sigma\tau}, \quad x_t = \frac{m_t^2}{M_W^2}
\end{equation}
Substituting in eq.(\ref{mbox}) gives the amplitude
\begin{equation}
\begin{aligned}
\mathcal{M}_{\text{box}} &= \frac{g^4}{4} \lambda_t \left( -4\, \bar{s} \gamma^\mu \gamma^\sigma (1 - \gamma_5) b  \cdot\bar{\nu} \gamma_\mu \gamma^\tau (1 - \gamma_5) \nu \right) 
 \times \left( \frac{i}{16\pi^2 M_W^2} \left[ 4 B_0(x_t) + 1 \right] g_{\sigma\tau} \right)
\end{aligned}
\end{equation}
Using the identity $\gamma^\mu \gamma^\sigma \gamma_\mu = -2 \gamma^\sigma$, we get
\begin{equation}
\bar{s} \gamma^\mu \gamma^\sigma (1 - \gamma_5) b \cdot \bar{\nu} \gamma_\mu \gamma_\sigma (1 - \gamma_5) \nu \to -8 \bar{s} \gamma^\mu (1 - \gamma_5) b \cdot \bar{\nu} \gamma_\mu (1 - \gamma_5) \nu  
\end{equation}
Then the amplitude simplifies to
\begin{equation}
\mathcal{M}_{\text{box}} = \lambda_t  \frac{g^4}{64\pi^2 M_W^2} \left[ 4 B_0(x_t) + 1 \right]  \left( \bar{s} \gamma^\mu (1 - \gamma_5) b \cdot \bar{\nu} \gamma_\mu (1 - \gamma_5) \nu \right)
\end{equation}
$ B_0(x_t) $ is the scalar one-loop function given in eq.(\ref{eq:B}). Using
\begin{equation}
    \frac{g^4}{64\pi^2 M_W^2} = \frac{G_F}{\sqrt{2}}  \frac{\alpha}{2\pi \sin^2\theta_W}
\end{equation}
we obtain the effective Hamiltonian,
\begin{equation}
\mathcal{H}_{\text{eff}}^{\text{box}} = \lambda_t \frac{G_F}{\sqrt{2}} \frac{\alpha}{2\pi \sin^2\theta_W} \left[ 4 B_0(x_t) + 1 \right] \cdot (\bar{s} \gamma^\mu (1 - \gamma^5) b)(\bar{\nu} \gamma_\mu (1 - \gamma^5) \nu)
\end{equation}

In the Standard Model, the unitarity of the CKM matrix implies:
\begin{equation}
  \lambda_u + \lambda_c + \lambda_t = 0  
\end{equation}
which causes the constant term \( +1 \) to cancel. However, in extensions with VLQs, the CKM matrix is no longer unitary, so this term remains, providing a window for new physics.
\subsubsection{Tree-Level Diagram of $b\rightarrow s \nu \bar{\nu}$}
\begin{figure}[h!]
        \centering
        \begin{tikzpicture}
        \begin{feynman}
            \vertex (a) at (0,0);
            \vertex (b) at (-2,1) {\(b\)};
            \vertex (c) at (2,1) {\(s\)};
            \vertex (d) at (0,-1.5);
            \vertex (e) at (1.5,-2.5) {\(\nu\)};
            \vertex (f) at (-1.5,-2.5) {\(\bar{\nu}\)};
            \diagram* {
                (b) -- [fermion] (a) -- [fermion] (c),
                (a) -- [boson, edge label'=\(Z\)] (d),
                (f) -- [fermion] (d) -- [fermion] (e),
            };
        \end{feynman}
        \end{tikzpicture}
        \caption{Tree-level diagram for \( b \rightarrow s \nu \bar{\nu} \).}
        \label{fig:tree-bsvv} 
\end{figure}
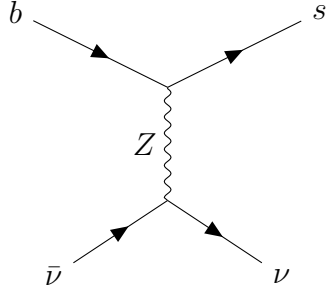
Now considering Tree level process $b\rightarrow s \nu \bar{\nu}$  in Fig.(\ref{fig:tree-bsvv}).The process is same as in section(4.2) and using eq.(\ref{CL}), we get
\begin{equation}
    X = \frac{U_{sb}}{\lambda_t}\frac{\pi}{\alpha}\sin^2{\theta_W}
\end{equation}
and
\begin{equation}
     C_L = \frac{U_{sb}}{\lambda_t}\frac{\pi}{\alpha}
     \label{C_LT}
\end{equation}
\subsubsection{Loop Diagrams of $b\rightarrow s \nu \bar{\nu}$}
\begin{figure}
        \centering
        \begin{tikzpicture} 
        \begin{feynman}
            \vertex (a1) at (0,0) {\( b\)};
            \vertex (a2) at (1.5,0);
            \vertex (a3) at (2.5,0);
            \vertex (a4) at (3.5,0);
            \vertex (a5) at (5,0) {\(s\)};
            
            \vertex (c2) at (2.5,-2);  
            \vertex (c1) at (3.5,-3) {\(\nu\)};
            \vertex (c3) at (1.5,-3) {\(\bar{\nu}\)};
        
            \diagram* { 
                (a1) -- [fermion] (a2) -- [fermion, edge label=$D$] (a3) -- [fermion, edge label=$D$] (a4) -- [fermion] (a5),
                (c3) -- [fermion] (c2) -- [fermion] (c1),
                (a3) -- [boson, edge label'=$Z$] (c2),
                (a4) -- [scalar, out=90, in=90, looseness=2.0, edge label'=\({H,\chi^{0}}\)] (a2)
            };
        \end{feynman} 
        \end{tikzpicture}
        \caption{VLQ-induced Higgs loop diagram for \( b \rightarrow s \nu \bar{\nu} \).}
        \label{fig:higgs-bsvv}
    \end{figure}
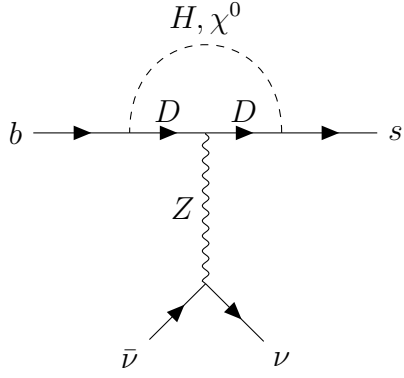
From the Higgs loop shown in Fig.(\ref{fig:higgs-bsvv}), using the Fyenman rules we have the amplitude,
\begin{equation}
    \begin{aligned}
           {\mathcal{M_H}}&=\frac{i g^{3}}{32 \pi^{2} \cos \theta_{W}} U^{s b}\left(\frac{1}{12} \sin ^{2} \theta_{W}\right) \frac{m_{D}^{2}}{M_{W}^{2}}\left(\frac{-2}{\epsilon}-\frac{1}{2}+\frac{x_1^{2} \log x_1}{(1-x_1)^{2}}+\frac{x_{1}}{\left(1-x_{1}\right)}\right)\\ 
           &\times\bar{s}\gamma_\mu \left(1-\gamma_{5}\right) b\frac{i}{M_Z^2}\left(\frac{ig}{4\cos{\theta_W}}\bar{\nu}\gamma^\mu(1-\gamma_5)\nu \right)
    \end{aligned}
\end{equation}
Using eq.(\ref{G_F}) and simplifying above equation we have the effective Hamiltonian for $b\rightarrow s \nu \bar{\nu}$

\begin{align}
    \mathcal{H}=\frac{ G_F}{\sqrt{2}}\frac{\alpha}{\pi} U^{s b}\left(\frac{1}{48} \right) \frac{m_{D}^{2}}{M_{W}^{2}}\left(\frac{-2}{\epsilon}-\frac{1}{2}+\frac{x_1^{2} \log x_1}{(1-x_1)^{2}}+\frac{x_{1}}{\left(1-x_{1}\right)}\right) \bar{s}\gamma_\mu \left(1-\gamma_{5}\right) b \bar{\nu}\gamma^\mu(1-\gamma_5)\nu 
\end{align}
Hence, by comparing to SM effective Hamiltonian in eq.(\ref{Hvv}), we get the Wilson coefficient $C_L$,
\begin{align}
    C_{L}= \frac{U^{sb}}{\lambda_t}\frac{m_D^2}{M_W^2}\frac{1}{24}\left(\frac{-2}{\epsilon}-\frac{1}{2}+\frac{x_1^{2} \log x_1}{(1-x_1)^{2}}+\frac{x_{1}}{\left(1-x_{1}\right)}\right)
    \end{align}
Similarly for $\chi^0$, we can write the $C_L$,
\begin{align}
    C_{L}= \frac{U^{sb}}{\lambda_t}\frac{m_D^2}{M_W^2}\frac{1}{24}\left(\frac{-2}{\epsilon}-\frac{1}{2}+\frac{x_2^{2} \log x_2}{(1-x_2)^{2}}+\frac{x_{2}}{\left(1-x_{2}\right)}\right)
    \end{align} 
\begin{figure}[h!]
    \centering
    \begin{minipage}{0.45\textwidth}
    \centering
    \begin{tikzpicture} 
    \begin{feynman}
        \vertex (a1) at (0,0) {\( b \)};
        \vertex [right=1.5cm of a1] (a2); 
        \vertex [right=1.5cm of a2] (a3);
        \vertex [right=0.75cm of a3] (a); 
        \vertex [right=1.5cm of a3] (a4) {\( s \)};
        
        \vertex [below=1.7cm of a] (c1);
        \vertex [below left=1.2cm of c1] (f) {\( \nu \)};
        \vertex [below right=1.2cm of c1] (e) {\( \bar{\nu} \)};
    
        \diagram* { 
            (a1) -- [fermion] (a2) -- [fermion, edge label' = {\( D \)}] (a3) -- [fermion] (a4),
            (a2) -- [scalar, out=90, in=90, looseness=2.0, edge label' = {\( H,\chi^0 \)}] (a3),
            (a) -- [boson, edge label' = {\( Z \)}] (c1),
            (f) -- [fermion] (c1) -- [fermion] (e),
        };
    \end{feynman} 
    \end{tikzpicture}
    \end{minipage}
    \hfill
    \begin{minipage}{0.45\textwidth}
    \centering
    \begin{tikzpicture} 
    \begin{feynman}
        \vertex (a1) at (0,0) {\( b \)};
        \vertex [right=1.5cm of a1] (a2); 
        \vertex [right=1.5cm of a2] (a3);
        \vertex [right=0.75cm of a1] (a); 
        \vertex [right=1.5cm of a3] (a4) {\( s \)};
        
        \vertex [below=1.7cm of a] (c1);
        \vertex [below left=1.2cm of c1] (f) {\( \nu \)};
        \vertex [below right=1.2cm of c1] (e) {\( \bar{\nu} \)};
    
        \diagram* { 
            (a1) -- [fermion] (a2) -- [fermion, edge label' = {\( D \)}] (a3) -- [fermion] (a4),
            (a2) -- [scalar, out=90, in=90, looseness=2.0, edge label' = {\( H,\chi^0 \)}] (a3),
            (a) -- [boson, edge label' = {\( Z \)}] (c1),
            (f) -- [fermion] (c1) -- [fermion] (e),
        };
    \end{feynman} 
    \end{tikzpicture}
    \end{minipage}

    \caption{Self-energy type diagrams for \( b\rightarrow s \nu \bar{\nu} \) via VLQ and scalar loop contributions.}
    \label{fig:self-energy-bsvv}
\end{figure}
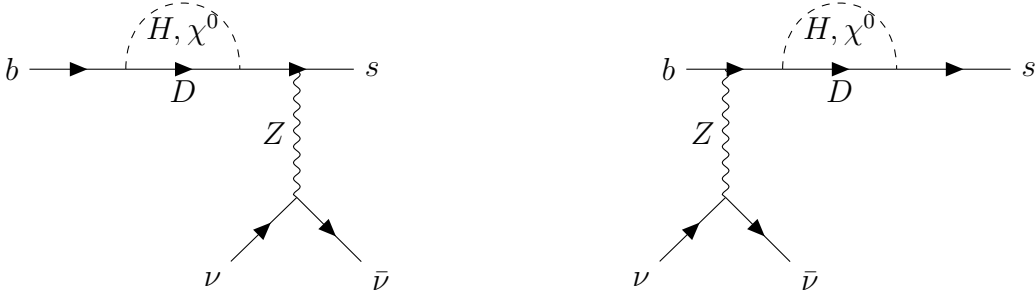
And from the self-energy given in Fig.(\ref{fig:self-energy-bsvv}), we have
\begin{align}
    C_{L}= -\frac{U^{sb}}{\lambda_t}\frac{m_D^2}{M_W^2}\left(\frac{\frac{-1}{2}+\frac{1}{3}\sin^2{\theta_W}}{8\sin^2{\theta_W}}\right)\left(\frac{1}{\epsilon}+\frac{3}{4}-\frac{x_1^{2} \log x_1}{2(1-x_1)^{2}}-\frac{1}{2(1-x_1)}\right)
    \label{CLH}
    \end{align}
Similarly for $\chi^0$, we have,
   \begin{align}
    C_{L}= -\frac{U^{sb}}{\lambda_t}\frac{m_D^2}{M_W^2}\left(\frac{\frac{-1}{2}+\frac{1}{3}\sin^2{\theta_W}}{8\sin^2{\theta_W}}\right)\left(\frac{1}{\epsilon}+\frac{3}{4}-\frac{x_2^{2} \log x_2}{2(1-x_2)^{2}}-\frac{1}{2(1-x_2)}\right)
    \label{CLX}
    \end{align}

\subsection{Branching Ratios of B-Decays with Vector-Like Quarks}
In the following section, we analytically compute the branching ratios for both inclusive and exclusive rare \( B \) decays in the presence of a singlet down-type VLQ. The inclusion of VLQs modifies the flavor structure of the Standard Model by inducing tree-level FCNCs and altering the loop-induced penguin and box diagrams. These effects enter through modified Wilson coefficients, particularly \( C_L^\nu \) and potentially a new right-handed contribution \( C_R^\nu \), thus impacting decay rates such as \( \mathcal{BR}(B \rightarrow K \nu \bar{\nu}) \) and \( \mathcal{BR}(B \rightarrow X_s \nu \bar{\nu}) \). Exclusive decays are especially sensitive to the underlying hadronic form factors, while inclusive modes benefit from parton-level cleanliness. By first obtaining analytical expressions in terms of the effective operators and VLQ parameters (masses and mixings), we lay the foundation for a detailed numerical analysis to follow. This enables us to compare theoretical predictions with current experimental bounds and probe the parameter space of the VLQ model.

\subsubsection{Inclusive B-Decay $B\rightarrow X_s \nu \bar{\nu}$}

The decay \( \bar{B} \to X_s \nu \bar{\nu} \) proceeds via a loop-induced FCNC transition \( b \to s \nu \bar{\nu} \), 
described by an effective Hamiltonian with a single operator weighted by the Inami--Lim function \( X_0(x_t) \)  in SM and CKM factor \( V_{ts}^*V_{tb} \). The effective Hamiltonian for the decay $\bar{B} \to X_s \nu \bar{\nu}$ is given by
\begin{equation}
   H_{eff}= \frac{G_F}{\sqrt{2}} \frac{\alpha}{2 \pi \sin^2 \theta_W}
V^*_{ts}V_{tb}X_0(x_t) (\bar{s} b)_{V-A} (\bar{\nu} \nu)_{V-A} + {\rm h.c.}\;, 
\end{equation} 
Using the effective Hamiltonian, the differential and total branching ratios for the inclusive decay \( \bar{B} \to X_s \nu \bar{\nu} \) can be computed \cite{Alok:2014yua}. The leading-order expression of branching ratio is given by
\begin{align}
 \mathcal{BR}(B\rightarrow X_s \nu \bar{\nu})=\frac{\alpha^2}{2\pi^2\sin^4{\theta_W}}BR((B\rightarrow X_c e \bar{\nu})\left(\frac{\bar{\eta}(|V_{ts}^* V_{tb}| X^{'})^2}{(V_{cb})^2 f(\hat{m_c})\kappa(\hat{m_c})}\right)
 \label{Xvv}
 \end{align}
  The factor $\bar{\eta}= 0.87$ represents the QCD correction to the matrix element of the $b\rightarrow s \nu \bar{\nu}$ transition due to the contributions of virtual and bremsstrahlung gluons, $f(\hat{m_c})$ is the phase-space factor in $BR((B\rightarrow X_c e \bar{\nu})$ and $\kappa(\hat{m_c})$ is the QCD correction of one-loop given in \cite{Kim:1989ac},
 \begin{equation}
     f(\hat{m_c})=1-8(\frac{m_c}{m_b})^2+8(\frac{m_c}{m_b})^6-(\frac{m_c}{m_b})^8-24(\frac{m_c}{m_b})^4\ln{\frac{m_c}{m_b}}
     \label{fmc}
\end{equation}
\begin{equation}
    \kappa(\hat{m_c})=1-\frac{2\alpha_s(m_b)}{3\pi}\left((\pi^2-\frac{31}{4})(1-\frac{m_c}{m_b})^2+\frac{3}{2}\right)
    \label{kmc}
 \end{equation}
where $\alpha_s(m_b)$ is the QCD coupling constant at the energy scale $\mu=m_b$. The presence of a tree-level $Z{\bar b}s$ coupling changes the value of the structure function $X_0(x_t)$ given in equation (\ref{X0}).  The  structure function within the VLQ  model can be written as
 \begin{equation}
X'(x_t) = X_0(x_t) + \Big(\frac{\pi \sin^2 \theta_W}{\alpha V^*_{ts}V_{tb}}\Big) U_{sb}\;.
\end{equation}
 Now adding the contribution of all the diagrams (non-unitarity, tree, vertex, and self-energy) from eqs.( \ref{C_LT}),(\ref{CLH}) and (\ref{CLX}) we have 
\begin{equation}
\begin{aligned}
X' &= X_0 + \frac{U^{sb}}{V_{ts}^* V_{tb}} \Bigg\{ 
\frac{1}{4}+\frac{\pi}{\alpha} \sin^2{\theta_W} 
+\Bigg[ \frac{1}{48} \frac{M_D^2}{M_W^2} \sin^2{\theta_W} \frac{1}{48}(F(x_1)+F(x_2))\\
    & +\left(\frac{\frac{-1}{2}+\frac{1}{3}\sin^2{\theta_W}}{16\sin^2{\theta_W}}\right)\frac{M_D^2}{M_W^2} \left(F^{'}(x_1)+F^{'}(x_2)\right)\Bigg]
\Bigg\}
\end{aligned}
\label{X00}
\end{equation}

The SM value of $\mathcal{BR}(B\rightarrow X_s \nu \bar{\nu})=(2.14\pm 0.23)\times10^{-5}$, whereas, experimental upperbound is $\mathcal{BR}(B \to X_s \bar{\nu} \nu) < 64 \times 10^{-5}$ at $90\%$ CL \cite{ALEPH:2000vvi}, SM value is well with the experimental bound.

\subsubsection{Exclusive B-Decay $B\rightarrow K \nu \bar{\nu}$}

The exclusive decay \( B \rightarrow K \nu \bar{\nu} \) is particularly important for this study, as it constitutes the main focus of our analytical calculation. It involves hadronic matrix elements parameterized by form factors, with the branching ratio sensitively depending on short-distance Wilson coefficients and long-distance QCD effects. 

The matrix element for the decay amplitude of $B \rightarrow K \nu \bar{\nu}$ can be written as
\begin{equation}
\mathcal{M} = \frac{G_F \alpha}{\sqrt{2} \pi} V_{tb} V_{ts}^* \, C_L^{\nu} 
\langle K | \bar{s} \gamma^\mu P_L b | B \rangle (\bar{\nu} \gamma_\mu (1 - \gamma_5) \nu),
\end{equation}
where \( C_L^{\nu} \) is the Wilson coefficient encoding short-distance contributions from $Z$-penguin and box diagrams. The hadronic matrix element $\langle K | \bar{s} \gamma^\mu P_L b | B \rangle$ is parametrized in terms of the $B \to K$ vector form factor $f_+^{B\to K}(q^2)$ as
\begin{equation}
\langle K(p_K) | \bar{s} \gamma^\mu b | B(p_B) \rangle = 
f_+^{K}(q^2) \left[ (p_B + p_K)^\mu - \frac{m_B^2 - m_K^2}{q^2} q^\mu \right] 
+ f_0^{K}(q^2) \frac{m_B^2 - m_K^2}{q^2} q^\mu,
\end{equation}
where \( q^\mu = p_B^\mu - p_K^\mu \) is the momentum transfer, and physical range of $q^2$ is $0 \leq q^2 \leq (m_B - m_K)^2$.

The form factors \( f_+^{K}(q^2) \) and \( f_0^{K}(q^2) \) are determined using a combination of light-cone sum rules (LCSR) at low \( q^2 \) and lattice QCD at high \( q^2 \), with a combined fit provided by \cite{Bailey:2015dka}. These inputs are essential for a precise Standard Model prediction of the differential and total branching ratio for the $B \to K \nu \bar{\nu}$ decay. The branching ratio of exclusive $B\rightarrow K \nu \bar{\nu}$\cite{Parrott:2022zte}.
 \begin{align}
     \frac{d\mathcal{BR}(B\rightarrow K \nu \bar{\nu})}{dq^2}=\frac{G_F^2\alpha^2_{EW}(M_Z)}{32\pi^2 \sin^4{\theta_W}}X^{'2}\tau_B |V_{ts}^* V_{tb}|^2 |p_k|^3 f^{2}_+(q^2)
     \end{align}
From \cite{Altmannshofer:2009ma,Ali:1999mm}, we now express the differential branching ratio in terms of the dimensionless variable \( s_B = \frac{q^2}{M_B^2} \), which rescales the invariant mass of the neutrino pair. This transformation simplifies the kinematic limits and makes the expression more suitable for numerical analysis:
\begin{align}
     \frac{d\mathcal{BR}(B\rightarrow K \nu \bar{\nu})}{ds_B}=\frac{G_F^2\alpha^2 M_B^5}{256\pi^5 \sin^4{\theta_W}}X^{'2} |V_{ts}^* V_{tb}|^2 \lambda^{3/2}(s_B, \tilde{m}_K^2, 1) (f_+^K(s_B))^2,
 \end{align}
 or,
 \begin{align}
     \frac{d\mathcal{BR}(B\rightarrow K \nu \bar{\nu})}{ds_B}=\frac{G_F^2\alpha^2 M_B^5}{256\pi^5 \sin^4{\theta_W}}X^{'2}\tau_B |V_{ts}^* V_{tb}|^2 \lambda^{\frac{3}{2}} (f_+^K(s_B))^2,
 \end{align}
 From \cite{Ball:2004ye}, the form factor \( f^K_+(q^2) \) characterizes the hadronic matrix element involved in semileptonic or rare decays of mesons, encoding the non-perturbative QCD effects. It depends on the squared momentum transfer \( q^2 \) between the initial and final states. A commonly used parametrization of \( f^K_+(q^2) \) is given by a double-pole form:
\begin{align}
    f_+^K(q^2) = \frac{r_1}{1 - \left(\frac{q}{m_1}\right)^2} + \frac{r_2}{\left(1 - \left(\frac{q}{m_1}\right)^2\right)^2},
\end{align}
where \( q^2 \) is the momentum transfer squared, \( m_1 \) corresponds to the mass of the dominant vector resonance that couples to the hadronic current, often identified with the \( B^* \) or \( K^* \) meson depending on the process, and $m_1 = m_{{B^*}, {B^*_s}}$  is fixed. Whereas, \( r_1 \) and \( r_2 \) are phenomenological parameters that are determined by fits to experimental data or lattice QCD calculations. 

This form factor is crucial in predicting decay rates and distributions in processes such as \( B \to K \ell^+ \ell^- \) or \( B \to K \nu \bar{\nu} \), which probe the flavor structure of the Standard Model and possible new physics effects. Accurate knowledge of \( f^K_+(q^2) \) reduces theoretical uncertainties in these rare decays, enabling stringent tests of the Standard Model and constraints on BSM theories. For $f_0$, one can write a decomposition \cite{Ball:2004ye},
      \begin{align}
         f_0(q^2)=\frac{r_2}{1-(\frac{q}{m_{fit}})^2}
     \end{align}
\begin{table}
    \centering
    \begin{tabular}{|c|c|c|c|c|}
     \hline
        F & $r_1$ & $r_2$ & $m_1^2$ &$m_{fit}^2$ \\ \hline
        f(K) & 0.162 & 0.173 & $m_1^2 (K)$&-- \\
         $f_0$ & 0 & 0.330 & - & 37.46\\ \hline
    \end{tabular}
    \caption{Fit parameters}
    \label{tab:my_label}
\end{table}
The accuracy of the fits of the LCSR results to the above parametrizations is generally very high and best for sets 1 to 3 of Table \ref{tab:my_label} with $m_b = (4.80 \pm 0.05)$ GeV , with a maximum $1.2\%$  deviation given in \cite{Ball:2004ye}. Then we have
     \begin{align}
         f^K_+(s_B)=\frac{0.162}{1-s_B(\frac{M_B}{5.41})^2}+\frac{0.173}{(1-s_B(\frac{M_B}{5.41})^2)^2}  
     \end{align}
\begin{equation}
     \lambda=s_B^2+(\frac{m_K}{M_B})^4+1-2\left(s_B(\frac{m_K}{M_B})^2+(\frac{m_K}{M_B})^2+s_B\right)
\end{equation}
In the calculation of the total branching ratio for the decay process, the differential branching ratio is integrated over the kinematic variable \( s_B \), which typically represents the normalized squared momentum transfer or a related invariant quantity. In our analysis, \( s_B \) varies within the range $0 \leq s_B \leq (1 - \tilde{m_K})^2 \approx 0.82$ \cite{Altmannshofer:2009ma}, and $\tilde{m_i} = m_i/M_B$, which corresponds to the physically allowed phase-space region excluding the range dominated by resonance contributions or thresholds.

The total branching ratio is then obtained by integrating the differential branching ratio over SB:
\begin{equation}
    \mathcal{BR}_\text{total} = \int_0^{0.82} \frac{d\mathcal{BR}}{ds_B} \, ds_B.
\end{equation}
and 
\begin{equation}
    \mathcal{BR}_\text{total}(B\rightarrow K \nu \bar{\nu})=\frac{G_F^2\alpha^2 M_B^5}{256\pi^5 \sin^4{\theta_W}}X^{'2}\tau_B |V_{ts}^* V_{tb}|^2 \int_0^{0.82}\lambda^{\frac{3}{2}} (f_+^K(s_B))^2ds_B,
    \label{BRK}
\end{equation}
where
\begin{align}
         \int_0^{0.82}{\lambda^{\frac{3}{2}} (f^K(s_B))^2}dS_B= 0.0719497       
\end{align}
The SM value of total branching ratio is $\mathcal{BR}(B\rightarrow K \nu \bar{\nu}) = 4.91688\times10^{-6}$ and Experimental value is $(2.3\pm 0.5)\times10^{-5}$ presented in \cite{Belle-II:2023esi}.

Having derived the analytical expressions for the branching ratios of rare \( B \)-decays and compared them to SM predictions and current experimental measurements, we now proceed to a detailed numerical analysis. In the following chapter, we evaluate these branching ratios using specific values of the NP parameters and compare the results quantitatively with both the SM expectations and experimental data. This includes the use of chi-squared techniques to statistically interpret the impact of NP contributions and to identify the parameter regions most consistent with observations.

\chapter{Numerical Analysis of Branching Ratios in Rare B-Decays}
\label{chap:results}

This chapter presents the numerical evaluation of branching ratios for a set of rare \( B \)-meson decays that serve as sensitive probes of potential contributions from NP, especially in models involving VLQs. The analysis focuses on both inclusive and exclusive flavor-changing neutral current processes, which are forbidden at tree level in the SM and thus particularly susceptible to small deviations induced by NP.

The key decay modes under consideration include:
\begin{itemize}
    \item the purely leptonic decay \( B^0_s \to \mu^+ \mu^- \),
    \item the inclusive semileptonic decay \( B \to X_s \mu^+ \mu^- \),
    \item the inclusive neutrino mode \( B \to X_s \nu \bar{\nu} \), and most importantly,
    \item the exclusive neutrino mode \( B \to K \nu \bar{\nu} \).
\end{itemize}

Among these, the decay \( B \to K \nu \bar{\nu} \) receives particular attention due to its theoretical cleanliness and its enhanced sensitivity to NP contributions. As an exclusive mode with minimal hadronic uncertainties and negligible long-distance effects, it offers a robust testing ground for deviations from SM expectations. Recent measurements by the Belle-II collaboration indicate a branching ratio significantly above the SM prediction, motivating a detailed exploration of this channel within the VLQ framework.

The theoretical groundwork for these decays, including full one-loop amplitude calculations and the effective Hamiltonian formalism, has been laid out in previous chapters. In the present analysis, we incorporate both SM and NP Wilson coefficients into the decay amplitudes, ensuring that interference effects are consistently captured.

To derive meaningful numerical predictions, we first constrain the NP parameter space by introducing the mixing parameters \( r_{sb} \) and \( \theta_{sb} \), and compute the resulting bounds on the flavor-violating coupling \( U_{sb} \). A statistical analysis is performed using a chi-squared (\( \chi^2 \)) contour plots, coded in \texttt{Python}. This method allows for a global fit across a multidimensional parameter space and yields confidence-level contours for the viable NP regions.

Once these constraints are established, we compute the branching ratios for all four decay modes, with particular focus on the exclusive channel \( B \to K \nu \bar{\nu} \). The results are compared with the SM predictions and current experimental measurements of LHCb, Belle II, and CMS. Each comparison is supported with graphical and numerical analysis, highlighting the extent to which VLQ contributions could account for the observed deviations.

This numerical study forms a critical component of the thesis, bridging the gap between theoretical modeling and experimental testing. It provides concrete predictions within the allowed parameter space and illustrates how precision flavor observables can serve as a window into NP scenarios such as vector-like quarks.

 \section{Constraints on  New Physics Parameter $U_{sb}$}
 
 A central feature of the vector-like quark (VLQ) model explored in this thesis is the introduction of new flavor-violating couplings that modify the flavor structure of the Standard Model (SM). In particular, the mixing between the Standard Model down-type quarks and the additional iso-singlet down-type VLQ leads to non-unitarity in the extended quark mixing matrix. This mixing is parametrized by the effective coupling \( U_{sb} \), which plays a crucial role in determining the magnitude of NP contributions to FCNC processes. Its absolute value and complex phase directly affect the Wilson coefficients in the effective Hamiltonian, and consequently the branching ratios of rare $B$ decays. In this work, we constrain \( U_{sb} \) using the well-measured rare decays \( B_s \to \mu^+ \mu^- \) and  $B\rightarrow X_s \mu^+\mu^-$ which are experimentally precise.

The resulting allowed region for \( U_{sb} \), defined by constant contours \( \chi^2 \), is then used in subsequent sections to compute the branching ratios for other rare decays, most notably the exclusive decay \( B \to K \nu \bar{\nu} \). In this way, \( B_s \to \mu^+ \mu^- \) serves as a crucial anchor process for constraining NP in the VLQ scenario and link different observables in a consistent framework.
  To find the constraint on $U_{sb}$, we need the branching ratios of $B_s \rightarrow \mu^+\mu^-$ and $B\rightarrow X_s \mu^+\mu^-$.
  
 \subsection{Branching Ratio of  $B_s\rightarrow \mu^+\mu^-$}
 
 In the SM, decay \( B_s \to \mu^+ \mu^- \) proceeds through electroweak penguin and box diagrams and is highly suppressed. However, in the presence of VLQs, this decay receives an additional tree-level contribution through the FCNC $Z$ exchange induced by \( U_{sb} \). The total amplitude is therefore sensitive to both the magnitude and phase of \( U_{sb} \), allowing experimental measurements to impose stringent bounds on its allowed values.

We perform a numerical analysis by comparing the experimentally measured branching ratio\cite{Santimaria:2018wfw},
\begin{equation}
\mathcal{BR}(B_s \to \mu^+ \mu^-)_\text{exp} = (3.09 \pm 0.46) \times 10^{-9},
\end{equation}
with the theoretical prediction in the presence of VLQs. The modified Wilson coefficients are functions of \( U_{sb} \), and by varying its magnitude \( r_{sb} \) and phase \( \theta_{sb} \), we construct a chi-squared function(\( \chi^2 \)) to quantify the goodness-of-fit across the parameter space.

The branching ratio of the $B_s\rightarrow \mu^+\mu^-$ process in the model with VLQ is given as follows
 \cite{Morozumi:2018cnc},
 \begin{equation}
      \mathcal{BR}(B_s \to \mu^+ \mu^-) 
= \tau_{B_s} \, \frac{G_F^2}{\pi} 
\left( \frac{\alpha}{4\pi } \right)^2 
f_{B_s}^2 \, M_{B_s} \, m_\mu^2 
\sqrt{1 - \frac{4 m_\mu^2}{M_{B_s}^2}}\
\left| V_{tb}^* V_{ts} \right|^2 
\left|\eta_Y C_{10}\right|^2 .
\end{equation}
where $\eta_Y$ is the next-to-leading-order (NLO) QCD correction. The $\tau_{B_s}$ is the lifetime of the $B_s$ meson. These values are shown in Table 2. The Wilson coeﬃcient $C_{10}$ evaluated at the scale $\mu_b$, is then written as
\begin{equation}
    \left| C_{10}(\mu_b) \right|^2= \left|\eta_Y C_{10} \right|^2
\end{equation}
We can write $U_{sb}$ in terms of $r_{sb}$ and $\theta_{sb}$ defined in \cite{Morozumi:2018cnc} as
\begin{equation}
    r_{sb}\equiv\left|\frac{U_{sb}}{\lambda_t}\right|, \quad \theta_{sb}\equiv arg\left[\frac{U_{sb}}{\lambda_t}\right]
\end{equation}
we can write $C_{10}$ as
   \begin{equation}
    \left| C_{10} \right|^2=\left|\frac{Y_0(x_t) }{\sin^2{\theta_W}}\right|^2\left|\Delta(r_{sb},\theta_{sb})  \right|^2
\end{equation}
Hence, where the branching ratio is
 \begin{equation}
      \mathcal{BR}(B_s \rightarrow \mu^+\mu^-)=\frac{G_F^2 \alpha^2 M_{B_s} m_\mu^2}{16 \pi^3}|\lambda_t|^2 f_{B_s}^2 \frac{|\eta_Y Y_0(x_t)|^2}{\sin^2{\theta_W}} \sqrt{1-4(\frac{m_\mu}{M_B})^2} \tau_{B_s} |\Delta (r_{sb},\theta_{sb})|^2
      \end{equation}
 The parameter $\Delta$ with the tree-level VLQ contribution from \cite{Morozumi:2018cnc} is
\begin{equation}
    |\Delta (r_{sb},\theta_{sb})|^2=\left[1-\frac{2\pi\sin^2{\theta_W}}{\alpha_{em}Y_0(x_t)}r_{sb} \cos{\theta_{sb}}+\left(\frac{\pi\sin^2{\theta_W}}{\alpha_{em}Y_0(x_t)}\right)^2 r^2_{sb}\right]
    \label{rsb}
\end{equation}
The parameter $\Delta$ with the non-unitarity+ VLQ (tree+loop level) contribution from
\begin{equation}
\begin{aligned}
|\Delta(r_{sb},\theta_{sb})|^2&=\Bigg[1-\frac{2\sin^2{\theta_W}}{Y_0(x_t)}r_{sb} \cos{\theta_{sb}}\Bigg[\frac{1}{4\sin^2{\theta_W}}+\frac{\pi}{\alpha}+\frac{M^2_D}{M^2_W}\frac{1}{48}(F(x_1)+F(x_2))\\
&+\frac{M^2_D}{M^2_W}\left(\frac{\frac{-1}{2}+\frac{1}{3}\sin^2{\theta_W}}{16\sin^2{\theta_W}}\right)(F'(x_1)+F'(x_2))\Bigg]+\left(\frac{\sin^2{\theta_W}}{Y_0(x_t)}\right)^2 r^2_{sb}\\
    &\times\Bigg[\frac{1}{4\sin^2{\theta_W}}+\frac{\pi}{\alpha}+\frac{M^2_D}{M^2_W}\frac{1}{48}(F(x_1)+F(x_2))\\&+\frac{M^2_D}{M^2_W}\left(\frac{\frac{-1}{2}+\frac{1}{3}\sin^2{\theta_W}}{16\sin^2{\theta_W}}\right)(F'(x_1)+F'(x_2))\Bigg]^2\Bigg]
   \label{rsb}
\end{aligned}
\end{equation}
where,
\begin{equation}
    F(x)=\left(\frac{-2}{\epsilon}-\frac{1}{2}+\frac{x^{2} \log x}{(1-x)^{2}}+\frac{x}{\left(1-x)\right)}\right)
\end{equation}
\begin{equation}
    F^{'}(x)= \left(\frac{1}{\epsilon}+\frac{3}{4}-\frac{x^{2} \log x}{2(1-x)^{2}}-\frac{1}{2(1-x)}\right)
\end{equation}
and $x_1=\frac{M_D^2}{M_H^2}$, and $x_2=\frac{M_D^2}{M_Z^2}$.
 The constraint arising from $\text{Br}(B_s \to X_s \gamma)$ will not be considered, as it is weaker than the one obtained from $\mathcal{BR}(B_s\to \mu^+ \mu^-)$.

\subsection{Branching Ratio of  $B\rightarrow X_s \mu^+\mu^-$}

The branching ratio for the inclusive decay \( B \to X_s \mu^+ \mu^- \) in the presence of vector-like quarks (VLQs) can be expressed as:

\begin{equation}
    \mathcal{BR}(B \to X_s \mu^+ \mu^-) = \frac{\alpha^2\, \mathcal{BR}(B \to X_c e \bar{\nu})}{4\pi^2\, f(\hat{m}_c)\, \kappa(\hat{m}_c)} 
    \cdot \frac{|V_{ts}^* V_{tb}|^2}{|V_{cb}|^2} 
    \int D(z)\, dz,
\end{equation}
where, 
\begin{equation}
  D(z) = (1 - z)^2 (1 + 2z) \left( |C^{\text{tot}}_9|^2 + |C^{\text{tot}}_{10}|^2 \right)
+ 4 \left(1 + \frac{2}{z} \right) |C^{\text{eff}}_7|^2
+ 12\, \text{Re}\left(C^{\text{eff}}_7 {C^{\text{tot}}_9}^* \right)  
\end{equation}

Here $z \equiv \frac{q^2}{m^2_ b} \equiv \frac{(p_{µ^+} + p_{µ^-} )^2}{m^2_b}$ and $\hat{m}_q = \frac{m_q}{m_b}$ for all quarks q. The expressions for the phase-space factor $f(\hat{m}_c)$ and the 1-loop QCD correction factor $\kappa(\hat{m}_c)$ are given in eq.(\ref{fmc}) and eq.(\ref{kmc}).

The integral of  $D(z)$ for the branching ratio of $ B \to X_s \mu^+ \mu^-$in the low-$q^2$($1~\text{GeV}^2 \leq q^2 \leq 6~\text{GeV}^2$) and the high-$q^2$($14.4 ~\text{GeV}^2 \leq  q^2 \leq  m^2_b$ ) regions are given in \cite{Alok:2012xm},
\begin{equation}
  \begin{aligned}
D_{\text{low}} &= \int_{\frac{1}{m_b^2}}^{\frac{6}{m_b^2}} D(z)\, dz 
= \mathcal{BR}(\bar{B} \to X_s \mu^+ \mu^-)_{\text{low}} 
\frac{4\pi^2 f(\hat{m}_c) \kappa(\hat{m}_c)}{\alpha^2 \mathcal{BR}(B \to X_c e \bar{\nu})}
\frac{|V_{cb}|^2}{|V_{ts}^* V_{tb}|^2}\\&
= 5.69947 \pm 1.82522, \\[10pt]
D_{\text{high}} &= \int_{\frac{14.4}{m_b^2}}^{\left(1 - \frac{m_s}{m_b}\right)^2} D(z)\, dz 
= \mathcal{BR}(\bar{B} \to X_s \mu^+ \mu^-)_{\text{high}} 
\frac{4\pi^2 f(\hat{m}_c) \kappa(\hat{m}_c)}{\alpha^2 \mathcal{BR}(B \to X_c e \bar{\nu})}
\frac{|V_{cb}|^2}{|V_{ts}^* V_{tb}|^2}\\&
= 1.56735 \pm 0.635465.
\end{aligned}  
\end{equation}

The experimental result from \cite{Belle:2005fli} is
\begin{align}
    \mathcal{BR}(\bar{B} \to X_s \mu^+ \mu^-)_{\text{low}} =(1.60 \pm 0.50) \times 10^{-6}\\ \quad  \mathcal{BR}(\bar{B} \to X_s \mu^+ \mu^-)_{\text{high}} =(0.44 \pm 0.12) \times 10^{-6}
\end{align}
assuming moderate mixing with SM quarks and dominant decays into third-generation final states. Hence, the New Wilson Coefficient is 
\begin{equation}
    \begin{aligned}
        C_{10}^{'} &=C_{10, SM}-\frac{U^{sb}}{\lambda_t}\left(\frac{1}{4\sin^2{\theta_W}}+\frac{\pi}{\alpha} \right)-\frac{U^{sb}}{\lambda_t}\frac{M_D^2}{M_W^2}\Bigg\{\frac{1}{48} 
        \Bigg(\frac{-2}{\epsilon}-\frac{1}{2}+\frac{x_1^{2} \log x_1}{(1-x_1)^{2}}+\frac{x_{1}}{\left(1-x_{1}\right)})\\
        &+(\frac{-2}{\epsilon}-\frac{1}{2}+\frac{x_2^{2} \log x_2}{(1-x_2)^{2}}+\frac{x_{2}}{\left(1-x_{2}\right)})\Bigg)+\left(\frac{\frac{-1}{2}+\frac{1}{3}\sin^2{\theta_W}}{16\sin^2{\theta_W}}\right)\\&\times\left[\left(\frac{1}{\epsilon}+\frac{3}{4}-\frac{x_1^{2} \log x_1}{2(1-x_1)^{2}}-\frac{1}{2(1-x_1)}\right)+\left(\frac{1}{\epsilon}+\frac{3}{4}-\frac{x_2^{2} \log x_2}{2(1-x_2)^{2}}-\frac{1}{2(1-x_2)}\right)\right]\Bigg\}
        \label{C10'}
    \end{aligned}
\end{equation}
   
For SM we have,  
\begin{equation}
\begin{aligned}
      \mathcal{BR}(B_s \rightarrow \mu^+\mu^-)&=\frac{G_F^2 \alpha^2 M_{B_s} m_\mu^2}{16 \pi^3}|V_{ts}^* V_{tb}|^2 f_{B_s}^2 (C_{10})^2 \sqrt{1-4(\frac{m_\mu}{M_B})^2} \tau_B \\
      &= (3.06\pm 0.14)* 10^{-9}
        \label{Bs}
      \end{aligned}
 \end{equation}
From eq.(\ref{C10'}), we can write
 \begin{equation}
 \begin{aligned}
      (C_{10}^{'})^2&= \frac{\mathcal{BR}(B_s \rightarrow \mu^+\mu^-)}{\frac{G_F^2 \alpha^2 M_{B_s} m_\mu^2}{16 \pi^3}|V_{ts}^* V_{tb}|^2 f_{B_s}^2  \sqrt{1-4(\frac{m_\mu}{M_B})^2} \tau_B}\\
     &= 17.2 \pm 3.5
     \label{c10'}
 \end{aligned}   
 \end{equation}

\begin{figure}[h]
    \centering
    \includegraphics[width=10cm]{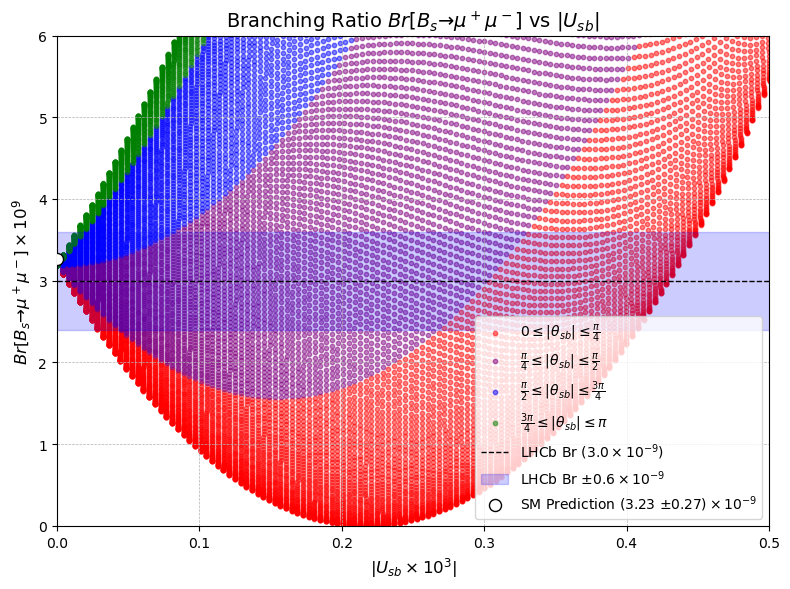}
    \caption{ Branching Ratio of $B_s \rightarrow \mu^+ \mu^-$ with VLQ depending on $U_{sb}$. The different colors represents different ranges for the value of $|\theta|_{sb}$.The experimental allowed values of branching ratios is shown as blue shaded region}
    \label{fig:Branching Ratio and U_sb}
\end{figure}
 Using eq.(\ref{c10'}) and eq.(\ref{Bs}) in order to have constraint on $U_{sb}$, we get
 \begin{equation}
     \begin{aligned}
    U^{sb} = \frac{(C_{10, SM}- C_{10}^{'})\lambda_t}{\frac{\pi}{\alpha}+\frac{m_D^2}{M_W^2}\left(\frac{1}{24}\left[F(x_1)+F(x_2)\right]+\left(\frac{\frac{-1}{2}+\frac{1}{3}\sin^2{\theta_W}}{8\sin^2{\theta_W}}\right)\frac{m_D^2}{M_W^2}\left(F^{'}(x_1)+F^{'}(x_2)\right)\right)}
 \end{aligned}
 \end{equation}

The branching ratio parameter space we used to constrain $U_{sb}$ is given in Fig.(\ref{fig:Branching Ratio and U_sb}), gives $U_{sb}= (4.09 \pm 0.17)\times10^{-4}$ at $\theta_{sb}=0$. As $|U_{sb}|$increases from zero to $(4.09 \pm 0.17)\times10^{-4}$, the branching ratio decreases for $\theta_{sb}=\frac{\pi}{2}$ but increases for $\frac{\pi}{2}<\theta<\pi$. The branching ratio increases regardless of the range of $\theta_{sb}$, as $|U_{sb}|$ becomes larger, since the third term in eq.(\ref{rsb}) is dominant. Hence we find that the stringent constraint on the parameters $r_{sb}$ and $\theta_{sb}$ comes from $B_s \to \mu^{+}\mu^{-}$.

By evaluating the branching ratios across the allowed parameter space, we are able to quantitatively assess how NP contributions modify the Standard Model expectations. The comparison of these NP-modified branching ratios with both SM predictions and experimental measurements allows us to determine the regions of parameter space that provide the best fit to data. This is reflected in updated \( \chi^2 \) values, which serve as a measure of agreement between theory and experiment.

In this way, the chi-squared analysis not only constrains the NP parameter space but also provides a statistically grounded method for identifying which combinations of parameters yield phenomenologically viable predictions. The interplay between constraints and branching ratio predictions is therefore crucial for evaluating the consistency and predictive power of the NP model under investigation.

\section{Numerical values of Branching Ratios}

   In this section, we present the final numerical results for the branching ratios of the rare \( B \)-decays, calculated using the Standard Model contributions supplemented with selected NP parameter values given in Fig.(\ref{fig:Branching Ratio and U_sb}). All computations have been performed using \texttt{Python}, and the resulting branching ratios will later be used for comparison with experimental data and chi-squared analysis.  

\begin{table}[h!]
    \centering
    \resizebox{\textwidth}{!}{%
    \begin{tabular}{|c|c|c|c|}
        \hline
       Parameters& Value  & Parameters & Value \\ \hline
       $\alpha^{-1}_{em}$& $130.3\pm2.3$ &  $M_Z$ & $91.1876\pm0.0021$ GeV \\ \hline
        $\alpha_s(M_Z)$ &  $0.1181 \pm0.0006$ GeV& $M_{B^\pm}$ & $5.27934 \pm 0.00012$ GeV \\ \hline
         $\alpha^{-1}(M_Z)$ & $127.937$ &$m_{b,\overline{\text{MS}}}$ & $4.183\pm 0.007$ GeV \\ \hline
          $\alpha(\mu_t)$ & $0.10805$& $\alpha(\mu_b)$ &$ 0.212$\\ \hline
        $m_t(m_t)$ & $162.5^ {+2.1}_{-1.5}$ GeV & $M_W$ & $80.3779 \pm 0.024$ GeV \\ \hline
        $f_{B_s}$ & $0.225 \pm 0.0015$ & $M_{B^0_s}$ & $5.36689\pm 0.00019$ GeV \\ \hline
        $m_\mu$ & $0.105$ GeV & $M_D$ & 1000 GeV\\ \hline
        $M_W$ & $80.385\pm 0.015$ GeV & $G_F$ & $1.16 \times 10^{-5}\, \text{GeV}^{-2}$ \\ \hline
         $\bar{\eta}$ & $0.87^{+0.12}_{-0.09}$ & $\sin^2{\theta_W}$ & 0.23129 \\ \hline
        $B_s$ & $1.320\pm 0.016$ & $V_{ts}^* V_{tb}$& $0.0403\pm 0.0009$ \\ \hline
        $\tau_{B_s}$&$(1.521\pm 0.005)\,\text{ps}=(2.311\pm 0.008)\times10^{12}\, \text{GeV}^{-1}$& $X(x_t)=(\frac{m_t(m_t)}{M_W})^2$ & $4.09^{+0.11}_{-0.08}$ \\ \hline
      $\eta_B$ & $0.5510\pm 0.0022$ &  $M_K$ &$0.4936$ GeV \\ \hline
        $X_0$ & $1.48^{+0.07}_{-0.05}$&$X^{'}= X_0+\frac{\alpha_s}{4\pi}X_1$ & $1.38^{+0.07}_{-0.05}$\\ \hline
        $\frac{m_c}{m_b}$& $0.29\pm 0.02$& $Br(B\rightarrow X_{s}e\bar{\nu})$ & $(10.61 \pm 0.17)\times 10^{-2}$\\ \hline 
       $V_{ts}^* V_{tb}$& $0.0403\pm 0.0009$ &$\eta_Y$& $1.0113$\\ \hline
        $M_K$ &$0.4936$ GeV& $\tau_{B}$&$2.489\times10^{12}\text{GeV}^{-1}$\\ \hline
    \end{tabular}%
    }
    \caption{Numerical values of input parameters}
    \label{tab:table}
\end{table}

\begin{figure}[h!]
    \centering
    \begin{subfigure}{0.48\textwidth}
        \centering
        \includegraphics[width=\linewidth]{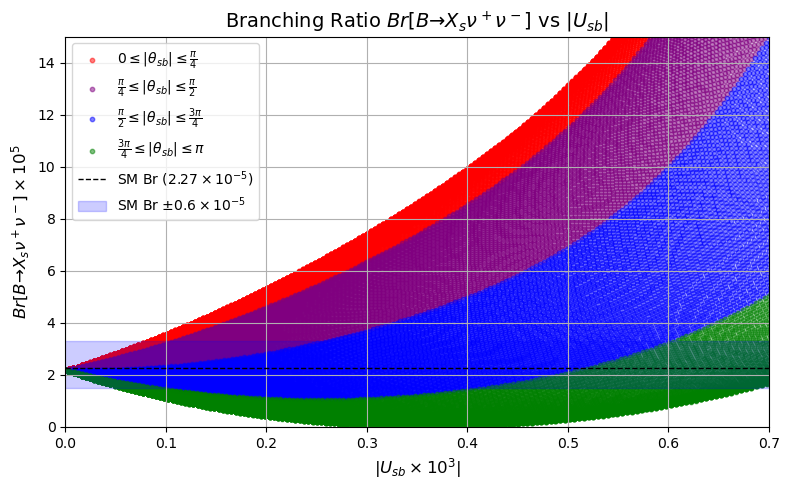}
        \caption{Inclusive decay $B \rightarrow X_s \nu \bar{\nu}$}
        \label{fig:inclusive}
    \end{subfigure}\hfill
    \begin{subfigure}{0.48\textwidth}
        \centering
        \includegraphics[width=\linewidth]{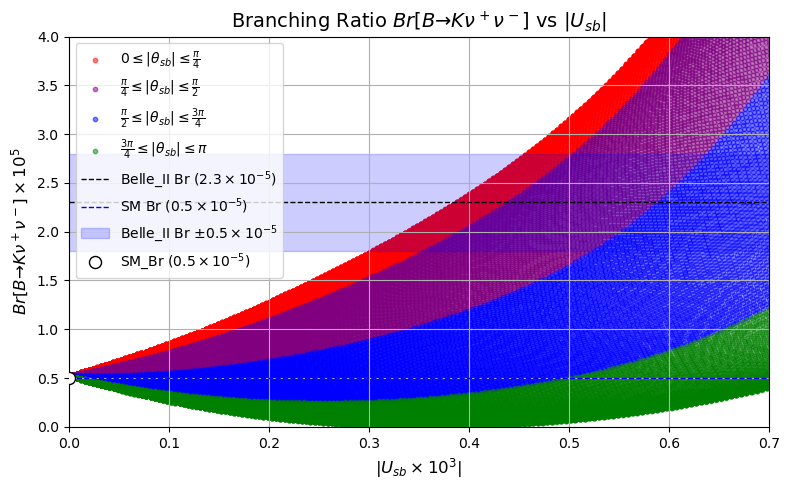}
        \caption{Exclusive decay $B \rightarrow K \nu \bar{\nu}$}
        \label{fig:exclusive}
    \end{subfigure}
    \caption{VLQ predictions for (a) the inclusive decay $B \rightarrow X_s \nu \bar{\nu}$ and (b) the exclusive decay $B \rightarrow K \nu \bar{\nu}$ across different values in the parameter space. The experimental allowed values of branching ratios is shown as blue shaded region}
    \label{fig:BtoVVplots}
\end{figure}

Using $U^{sb}$ at $\theta_{sb}=0$ and input paramters from Table\ref{tab:table}, the branching ratio for inclusive B-decay given in  eq.(\ref{Xvv}) gives 
\begin{equation}
   \mathcal{BR}(B\rightarrow X_s \nu \bar{\nu})_{VLQ}= 10 \times 10^{-5} 
\end{equation}
 see in Fig.(\ref{fig:BtoVVplots}a) while the SM value using the same formula,
 \begin{equation}
     \mathcal{BR}(B\rightarrow X_s \nu \bar{\nu})=2.14\times10^{-5}
 \end{equation}
 whereas, the experimental upper bound is $\mathcal{BR}(B \to X_s \bar{\nu} \nu) < 64 \times 10^{-5}$ at $90\%$ CL \cite{ALEPH:2000vvi}.
 Using eq.(\ref{BRK}), we calculated the branching ratio of $\mathcal{BR}(B\rightarrow K \nu \bar{\nu})$ given in Fig.(\ref{fig:BtoVVplots}b) shows that our calculation overlaps with experimental values, while using $U^{sb}$ and $\theta_{sb}=0$  
 \begin{equation}
     \mathcal{BR}(B\rightarrow K \nu \bar{\nu})_{VLQ}= 2.4\times 10^{-5}
 \end{equation}

The parameter space of $B\rightarrow K \nu \bar{\nu}$ is given Fig.(\ref{fig:BtoVVplots}b), which clearly shows that the addition of Vector-Like quark has expanded the parameter space of the process within the experimental band.

\section{Chi-Squared Analysis and Comparison with the Standard Model}

To quantitatively assess how well the NP contributions align with experimental data, we perform a chi-squared (\( \chi^2 \)) analysis. This statistical method evaluates the goodness of fit between the theoretical predictions computed using the selected NP parameter values and the experimentally measured branching ratios for rare \( B \)-decays.

The \( \chi^2 \) values are calculated for each parameter point by comparing the predicted branching ratios with their corresponding experimental central values and uncertainties. A lower \( \chi^2 \) value indicates better agreement with data. By scanning over a range of NP parameter values, we generate \( \chi^2 \)-contour plots that highlight the regions of parameter space favored by current measurements.

\begin{figure}[h!]
    \centering
    \begin{subfigure}{0.5\textwidth}
        \centering
        \includegraphics[width=\linewidth]{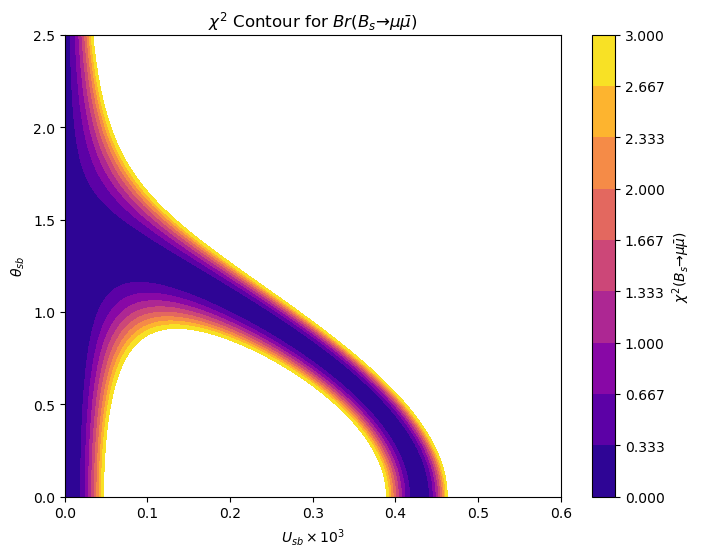}
        \caption{Chi-squared plot of $B_s \rightarrow  \mu^+ \mu^-$}
        \label{fig:contour plot}
    \end{subfigure}
    \hfill
    \begin{subfigure}{0.5\textwidth}
        \centering
        \includegraphics[width=\linewidth]{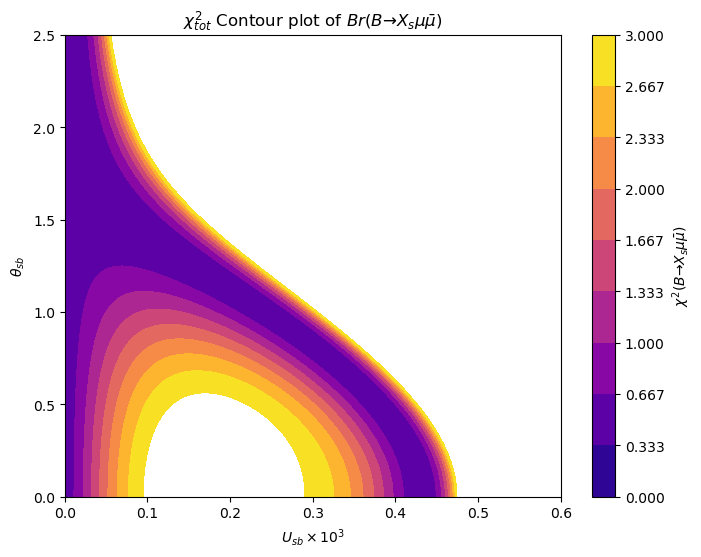}
        \caption{Chi-squared contour plot of $B \rightarrow X_s \mu^+ \mu^-$}
        \label{fig:chi-2}
    \end{subfigure}
    \caption{$\chi^2$ contour plots of (a) $B_s \rightarrow \mu^+ \mu^-$, (b) $B \rightarrow X_s \mu^+ \mu^-$ corresponding to $U_{sb}$ and $\theta_{sb}$. Here the $\chi^2$ varies from ($0-3$), with lower the value $\chi^2$ is dark blue and higher value is yellow in color}
    \label{fig:Bsplots}
\end{figure}
\begin{figure}[h!]
    \centering
    \includegraphics[width=9cm]{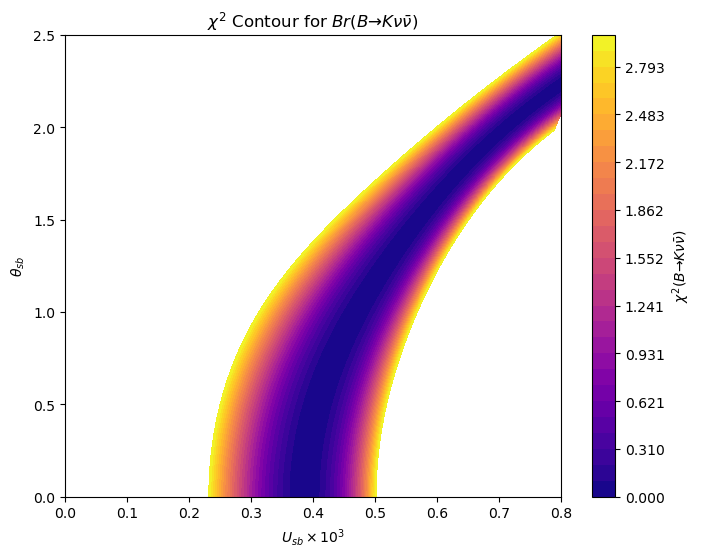}
    \caption{ Chi-squared contour plot of $B \rightarrow K  \nu \bar{\nu} $ }
    \label{fig:chi2BK}
\end{figure}
\begin{figure}[h!]
    \centering
    \includegraphics[width=9cm]{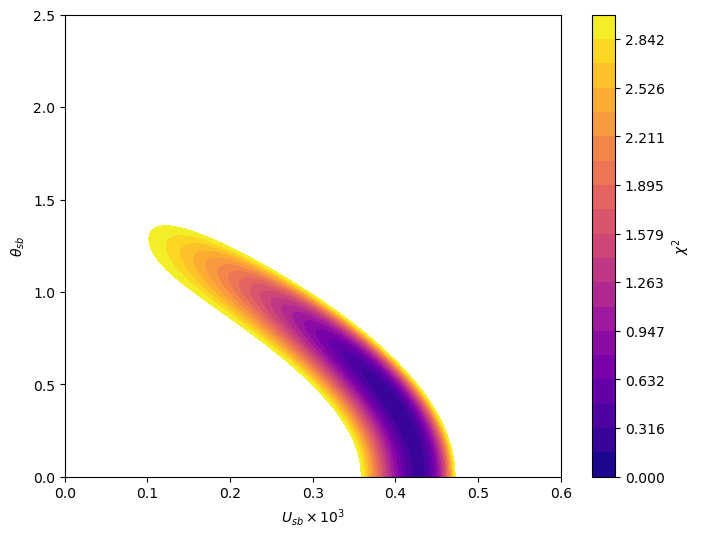}
    \caption{ Total contour plot of B-decays($B_s \rightarrow \mu^+ \mu^-,B \rightarrow X_s \mu^+ \mu^-$ and $B \rightarrow K \bar{\nu} \nu$)}
    \label{fig:chi2}
\end{figure}

Here, the chi-squared analysis has been performed individually for each rare \( B \)-decay channel to evaluate the level of agreement between theoretical predictions including NP contributions and experimental measurements. Each individual \( \chi^2 \) plot highlights the preferred NP parameter regions specific to that decay mode, revealing how sensitive each channel is to variations in the NP parameters. In Fig.(\ref{fig:contour plot}) and Fig.(\ref{fig:chi-2}), we have chi-squared contour plot of $B_s \rightarrow \mu^+ \mu^-$ and $B \rightarrow X_s\mu^+ \mu^-$, that we used to constrain the NP parameter $U_{sb}$. And Fig.(\ref{fig:chi2BK}) is contour plot of $B \rightarrow K \nu\bar{\nu}$.
While these individual plots provide valuable insights into the constraints imposed by each decay separately, the overall viability of the NP model must consider all observables simultaneously. Therefore, we combine the individual contributions into a total chi-squared value by summing over all decay channels. The resulting total \( \chi^2 \) plot in Fig.(\ref{fig:chi2}) presents a global fit, identifying the NP parameter space that best reconciles all measured branching ratios at once.

Interpreting these plots together allows us to determine whether the NP scenario improves the fit relative to the Standard Model (which corresponds to \( \chi^2 \) evaluated at zero NP contributions) and to identify parameter regions favored or excluded by the current data. A significant reduction in total \( \chi^2 \) compared to the SM indicates that the NP model provides a better description of the experimental results. These plots also allow a direct comparison with the Standard Model, which corresponds to the point in parameter space where all NP contributions vanish at $U_{sb}=0$. The relative \( \chi^2 \) values thus indicate whether the inclusion of NP improves or worsens the agreement with experimental observations compared to the SM alone. 
From Fig.(\ref{fig:chi2}) we can see that $\chi^2$ is constraint around the $(4.09 \pm 0.17)\times10^{-4}$ and gives $\chi^2 \sim1$ in combined plot.

 The numerical evaluation of branching ratios combined with the chi-squared analysis provides a comprehensive framework for testing New Physics scenarios against experimental data. The individual and total \( \chi^2 \) results highlight the parameter regions that are most consistent with observations and quantify the extent to which the NP model can improve upon the Standard Model predictions. This analysis forms a crucial step toward identifying viable extensions of the Standard Model and guiding future experimental searches.
\chapter{Conclusion}
\label{chap:conclusion}

The Standard Model shortcoming indicate the presence of NP beyond the SM. This problem can be solved by various new theories, one of them is Vector Quark Model (VQM). Vector-like quarks are heavy and effectively invisible to direct searches at present collider energies, hence the indirect searches such as rare decays become important as VLQs can still influence low energy observables and rare processes.

In this thesis, we have studied rare \( B \)-meson decays through  the framework of Effective Field Theory (EFT), focusing on processes mediated by FCNCs derived by integrating out VLQ fields. The model including down-type SU(2) singlet VLQ in addition to the SM quarks and we assumed the mass of the VLQ is much larger than the electroweak scale. We derived the analytical results of the branching ratios using the Operator Product Expansion (OPE) and the renormalization group evolution of Wilson coefficients. This formalism allowed us to systematically separate short-distance effects from long-distance hadronic physics, ensuring that both SM and NP contributions are parameterized by effective Hamiltonian formalism. We matched the effective field theory with the full theory not only at tree level but also at one-loop level and obtained the effective operators corresponding to the contribution from the diagram including VLQ in the internal line. Our analysis shows that VLQs can play a significant role in addressing the observed tensions in flavor physics, particularly in the context of the so-called \( B \)-anomalies, such as the deviations in the branching ratio of \( B \to K \nu \bar{\nu} \) which is $2.7\sigma$ above Standard Model predictions. These anomalies can be explained through inducing VLQ contributions. This framework thus provides a promising avenue for accommodating NP effects in a way that is consistent with current data while offering testable predictions for future experiments. 

For our numerical analysis we used \texttt{Python}, which scanned over the parameter space of VLQs. We used the process \( B_s \to \mu^+ \mu^- \), \( B \to X_s \mu^+ \mu^- \) to constrain NP parameter $U_{sb}$, since these processes provide the best agreement to SM prediction and the constraint on the model parameters $r_{sb}$ and $\theta_{sb}$ from  $(B_s \to \mu^+ \mu^-)_{VLQ}$ is more stringent than that from \( B \to X_s\gamma \) In particular, we found that for VLQ masses around \( 1.5~\text{TeV} \) and NP parameters $r_{sb},\theta_{sb}$, the fit to the branching ratio of the rare decay \( B \to K \nu \bar{\nu}\) improved compared to the SM prediction. When $U_{sb}$ is of the order of $(4.09 \pm 0.17)\times10^{-4}$ , the  $\mathcal{BR}( B \to K \nu \bar{\nu})_{VLQ}$ approaches the experimental value. For each parameter point, we computed the branching ratios of rare \( B \)-decays and compared them against SM predictions and current experimental constraints from collaborations such as LHCb and Belle II. To quantify the agreement between theory and experiment, we carried out a chi-squared (\( \chi^2 \)) analysis. Individual \( \chi^2 \) evaluations were performed for each decay mode, followed by a fit using the total \( \chi^2 \). The NP parameter at $(4.09 \pm 0.17)\times10^{-4}$, and the resulting plots allowed us to identify favored regions in the NP parameter space and to assess whether the inclusion of NP improves the fit relative to the SM. This plot clearly shows that including VLQ can be a promising NP candidates, which indicates that large enhancement in branching ratio of  \( B \to K \nu \bar{\nu}\) is possible.

This work demonstrated that rare \( B \)-decays offer a powerful tool to probe NP through VLQ scenarios, with sensitivities that are competitive with those from direct collider searches. The interplay of Effective Field Theory (EFT) techniques, numerical computations, and statistical analysis provided a robust framework to identify and quantify the imprints of VLQ-induced New Physics. Future experimental advancements, particularly in the precision measurements of \( B \to K^{(*)} \nu \bar{\nu} \) decays and angular observables, will further sharpen these probes, potentially revealing the first hints of TeV-scale flavor dynamics.

In summary, the combination of analytical methods, numerical tools, and statistical analysis provided a comprehensive understanding of rare \( B \)-decays as precision tests of the Standard Model and sensitive windows to potential New Physics. Future work could extend this framework to include additional observables, such as angular distributions, and explore further constraints from upcoming experimental results.

\appendix

\begin{appendix}

\chapter{Feynman Rules}

The Feynman rules used in this thesis are presented in the 't Hooft-Feynman gauge. This gauge simplifies gauge boson propagators and explicitly includes Goldstone bosons. Couplings include projection operators $P_{L,R} = \frac{1 \mp \gamma_5}{2}$ and mixing matrices $V_{ij}$ (CKM).

\section*{A.1 Propagators}

\subsection*{A.1.1 Gauge Bosons}
\begin{align*}
A_{\mu\nu}(k) &= \frac{-i g_{\mu\nu}}{k^2 + i \epsilon} \quad &\text{(Photon)} \\
W^{\pm}_{\mu\nu}(k) &= \frac{-i g_{\mu\nu}}{k^2 - M_W^2 + i \epsilon} \quad &\text{(Charged W boson)} \\
Z_{\mu\nu}(k) &= \frac{-i g_{\mu\nu}}{k^2 - M_Z^2 + i \epsilon} \quad &\text{(Neutral Z boson)}
\end{align*}

\subsection*{A.1.2 Fermions and Scalars}
\begin{align*}
S_F(p) &= \frac{i (\not{p} + m)}{p^2 - m^2 + i \epsilon} \quad &\text{(Fermion)} \\
H(k) &= \frac{i}{k^2 - M_H^2 + i \epsilon} \quad &\text{(Higgs)} \\
\chi^{\pm}(k) &= \frac{i}{k^2 - M_W^2 + i \epsilon} \quad &\text{(Charged Goldstone)} \\
\chi^{0}(k) &= \frac{i}{k^2 - M_Z^2 + i \epsilon} \quad &\text{(Neutral Goldstone)}
\end{align*}

\section*{A.2 Interaction Vertices}

\subsection*{A.2.1 Electromagnetic Interactions}
\[ \bar{f} A_\mu f = i e Q_f \gamma_\mu \]

\subsection*{A.2.2 Neutral Current (Z) Interactions}
\begin{align*}
\bar{u}_i Z_\mu u_i &= \frac{ig_2}{\cos \theta_W} \gamma_\mu \left[ \left( \frac{1}{2} - \frac{2}{3} \sin^2 \theta_W \right) P_L - \frac{2}{3} \sin^2 \theta_W P_R \right] \\
\bar{d}_i Z_\mu d_i &= \frac{ig_2}{\cos \theta_W} \gamma_\mu \left[ \left( -\frac{1}{2} + \frac{1}{3} \sin^2 \theta_W \right) P_L + \frac{1}{3} \sin^2 \theta_W P_R \right] \\
\bar{\nu}_i Z_\mu \nu_i &= \frac{ig_2}{\cos \theta_W} \gamma_\mu \left[ \frac{1}{2} P_L \right] \\
\bar{\ell}_i Z_\mu \ell_i &= \frac{ig_2}{\cos \theta_W} \gamma_\mu \left[ \left( -\frac{1}{2} + \sin^2 \theta_W \right) P_L + \sin^2 \theta_W P_R \right]
\end{align*}

\subsection*{A.2.3 Charged Current (W) Interactions}
\begin{align*}
\bar{u}_i W^+_\mu d_j &= \frac{ig_2}{\sqrt{2}} V_{ij} \gamma_\mu P_L, &
\bar{d}_j W^-_\mu u_i &= \frac{ig_2}{\sqrt{2}} V^*_{ij} \gamma_\mu P_L \\
\bar{\nu}_i W^+_\mu \ell_j &= \frac{ig_2}{\sqrt{2}} U_{ij} \gamma_\mu P_L, &
\bar{\ell}_j W^-_\mu \nu_i &= \frac{ig_2}{\sqrt{2}} U^*_{ij} \gamma_\mu P_L
\end{align*}

\subsection*{A.2.4 Higgs Interactions}
\[ \bar{f} H f = -i \frac{g_2}{2} \frac{m_f}{M_W} \]

\begin{equation*}
  W^{\pm}_\mu Z_\nu \chi^{\mp} : -ig_2 M_W \frac{\sin^2 \theta_W}{\cos \theta_W} g^{\mu\nu}, \qquad
W^{\pm}_\mu A_\nu \chi^{\mp} : ie M_W g^{\mu\nu},  
\end{equation*}
\begin{equation*}
W^{\pm}_\mu H W^\mp_\nu : ig_2 M_W g_{\mu\nu}, \qquad
Z_\mu H Z_\nu : i \frac{g_2}{\cos \theta_W} M_Z g_{\mu\nu}.
\end{equation*}

\subsection*{A.2.5 Goldstone Boson Interactions}
\begin{align*}
\bar{u}^i \chi^+ d^j &= \frac{ig_2}{\sqrt{2} M_W} \left( m_u^i P_L - m_d^j P_R \right) V_{ij} \\
\bar{d}^j \chi^- u^i &= \frac{ig_2}{\sqrt{2} M_W} \left( m_u^i P_R - m_d^j P_L \right) V^*_{ij} \\
\bar{\nu}^i \chi^+ \ell^j &= -\frac{ig_2 m^j_{\ell}}{\sqrt{2} M_W} P_R U_{ij} \\
\bar{\ell}^j \chi^- \nu^i &= -\frac{ig_2 m^j_{\ell}}{\sqrt{2} M_W} P_L U^*_{ij} \\
\bar{\ell}^i \chi^0 \ell^j &= \frac{g_2 m^i_{\ell}}{2 M_W} \gamma_5 \delta_{ij}
\end{align*}

\chapter{ Loop Integrals}
\section*{B.1 Integrals with Two Propagators}
This list of integrals is sufficient for the calculation of one-loop diagrams with vanishing external momenta.
\begin{align*}
I_1(m, M) &= \int \frac{d^D k}{(2\pi)^D} \frac{m}{[(k + p)^2 - m^2][k^2 - M^2]} = m \frac{i}{16\pi^2} \left[ \frac{2}{\bar{\epsilon}} + \frac{3}{2} + F_1(x) + F_2(x) \right],  \\
I_2(m, M) &= \int \frac{d^D k}{(2\pi)^D} \frac{(p + k)_\mu}{[(k + p)^2 - m^2][k^2 - M^2]} = p_\mu \frac{i}{16\pi^2} \left[ \frac{1}{\bar{\epsilon}} + \frac{3}{4} + F_2(x) \right], 
\end{align*}

where \( p_\mu \) is an external momentum, and we set \( p^2 = 0 \). \( F_{1,2}(x) \) are given by
\begin{align*}
F_1(x) &= -\frac{1}{2(1 - x)^2} \left[ x^2 \log x - 2x \log x - x(1 - x) \right],  \\
F_2(x) &= -\frac{1}{2(1 - x)^2} \left[ x^2 \log x + (1 - x) \right], 
\end{align*}

and \((D = 4 - 2\varepsilon)\)

\begin{align*}
x = \frac{m^2}{M^2}, \quad \frac{1}{\bar{\epsilon}} = \frac{1}{2\varepsilon} + \frac{1}{2} \left[ \log 4\pi - \gamma_E + \log \left( \frac{\mu^2}{M^2} \right) \right]. 
\end{align*}

\section*{B.2 Integrals with Three Propagators}

\begin{align*}
I_3(m_1, m_2, M) &= \int \frac{d^4k}{(2\pi)^4} \frac{1}{[k^2 - m_1^2][k^2 - m_2^2][k^2 - M^2]} \\
&= \frac{i}{16\pi^2} \frac{1}{M^2} \left[ \frac{x_1 \log x_1}{(1 - x_1)(x_1 - x_2)} + \frac{x_2 \log x_2}{(1 - x_2)(x_2 - x_1)} \right], 
\end{align*}
where
\begin{align*}
x_i = \frac{m_i^2}{M^2}, \quad x = \frac{m^2}{M^2}. 
\end{align*}
Special cases:
\begin{align*}
I_3(m, m, M) &= \frac{i}{16\pi^2} \frac{1}{M^2} \left[ \frac{\log x}{(1 - x)^2} + \frac{1}{(1 - x)} \right],  \\
I_3(m, M, M) &= -\frac{i}{16\pi^2} \frac{1}{M^2} \left[ \frac{x \log x}{(1 - x)^2} + \frac{1}{(1 - x)} \right], \label{eq:I3_mMM} \\
I_3(M, M, M) &= -\frac{i}{32\pi^2} \frac{1}{M^2}. 
\end{align*}

\begin{align*}
I_4(m_1, m_2, M) &= \int \frac{d^D k}{(2\pi)^D} \frac{k_{\mu} k_{\nu}}{[k^2 - m_1^2][k^2 - m_2^2][k^2 - M^2]} \\
&= \frac{i g_{\mu \nu}}{32\pi^2} \left[ \frac{1}{\bar{\epsilon}} + \frac{3}{4} + \frac{x_1^2 \log x_1}{2(1 - x_1)(x_1 - x_2)} + \frac{x_2^2 \log x_2}{2(1 - x_2)(x_2 - x_1)} \right].
\end{align*}
 Special cases:
\begin{align*}
I_4(m, m, M) &= \frac{i g_{\mu \nu}}{32\pi^2} \left[ \frac{1}{\bar{\epsilon}} + \frac{3}{4} + F_1(x) \right], \\
I_4(m, M, M) &= \frac{i g_{\mu \nu}}{32\pi^2} \left[ \frac{1}{\bar{\epsilon}} + \frac{3}{4} + F_2(x) \right],  \\
I_4(M, M, M) &= \frac{i g_{\mu \nu}}{32\pi^2} \left[ \frac{1}{\bar{\epsilon}} \right], 
\end{align*}

where \( F_1(x) \) and \( F_2(x) \) are given in (B.1).
\end{appendix}



\addcontentsline{toc}{chapter}{Bibliography}

\begin{thebibliography}{99}

\bibitem{Farnes:2017gbf}
J.~S.~Farnes,
``A unifying theory of dark energy and dark matter: Negative masses and matter creation within a modified $\Lambda$CDM framework,''
Astron. Astrophys. \textbf{620}, A92 (2018)
doi:10.1051/0004-6361/201832898
[arXiv:1712.07962 [physics.gen-ph]].

\bibitem{Peskin:1995ev}
M.~E.~Peskin and D.~V.~Schroeder,
``An Introduction to quantum field theory,''
Addison-Wesley, 1995,
ISBN 978-0-201-50397-5, 978-0-429-50355-9, 978-0-429-49417-8
doi:10.1201/9780429503559

\bibitem{Peskin:1997ez}
M.~E.~Peskin,
``Beyond the standard model,''
[arXiv:hep-ph/9705479 [hep-ph]].
\bibitem{Griffiths:1987tj}
D.~J.~Griffiths,
``Introduction to Elementary Particles ,''
doi:10.1002/9783527618460

\bibitem{Cheng:1984vwu}
T.~P.~[.~0.~1.~0.~Cheng and L.~F.~[.~0.~8.~3.~Li,
``Gauge Theory of Elementary Particle Physics,''
Oxford University Press, 1984,
ISBN 978-0-19-851961-4, 978-0-19-851961-4

\bibitem{cabibbo1963}
 Cabibbo, N. (1963). Unitary symmetry and leptonic decays. \textit{Phys. Rev. Lett.}, 10, 531–533. doi.org/10.1103/Phys Rev Lett.10.531
 
\bibitem{kobayashi1973}
Kobayashi, M., \& Maskawa, T. (1973). CP violation in the renormalizable theory of weak interaction. \textit{Prog. Theor. Phys.}, 49, 652–657. 

\bibitem{Wilson 1969}
 K.G. Wilson, Phys. Rev. 179 (1969) 1499. Wilson, K. G., Zimmermann, W. (1972). Comm. Math. Phys. 24, 87.
 
 \bibitem{zimmermann1971} 
 Zimmermann, W, In Proc. 1970 Brandeis Summer Institute in Theor. Phys, (eds. S. Deser, M. Grisaru and H. Pendleton), MIT Press,1971 p.396. Ann. Phys. 77, 570.
 
 \bibitem{witten1977} 
 Witten, E. (1977). Nucl. Phys. B 120, 189.
 
 \bibitem{Glashow:1970gm}
S.~L.~Glashow, J.~Iliopoulos and L.~Maiani,
``Weak Interactions with Lepton-Hadron Symmetry,''
Phys. Rev. D \textbf{2}, 1285-1292 (1970)
doi:10.1103/PhysRevD.2.1285

\bibitem{Weinberg:1973xwm}
S.~Weinberg,
``New approach to the renormalization group,''
Phys. Rev. D \textbf{8}, 3497-3509 (1973)
doi:10.1103/PhysRevD.8.3497

\bibitem{Weinberg:1967tq}
S.~Weinberg,
``A Model of Leptons,''
Phys. Rev. Lett. \textbf{19}, 1264-1266 (1967)
doi:10.1103/PhysRevLett.19.1264

\bibitem{Salam:1968rm}
A.~Salam,
``Weak and Electromagnetic Interactions,''
Conf. Proc. C \textbf{680519}, 367-377 (1968)
doi:10.1142/9789812795915\_0034
\bibitem{Weinberg:1973ew}
S.~Weinberg,
``General Theory of Broken Local Symmetries,''
Phys. Rev. D \textbf{7}, 1068-1082 (1973)
doi:10.1103/PhysRevD.7.1068

\bibitem{Higgs:1964pj}
P.~W.~Higgs,
``Broken Symmetries and the Masses of Gauge Bosons,''
Phys. Rev. Lett. \textbf{13}, 508-509 (1964)
doi:10.1103/PhysRevLett.13.508

\bibitem{Goldstone:1962es}
J.~Goldstone, A.~Salam and S.~Weinberg,
``Broken Symmetries,''
Phys. Rev. \textbf{127}, 965-970 (1962)
doi:10.1103/PhysRev.127.965

\bibitem{tHooft:1972tcz}
G.~'t Hooft and M.~J.~G.~Veltman,
``Regularization and Renormalization of Gauge Fields,''
Nucl. Phys. B \textbf{44}, 189-213 (1972)
doi:10.1016/0550-3213(72)90279-9

\bibitem{Feynman:1950ir}
R.~P.~Feynman,
``Mathematical formulation of the quantum theory of electromagnetic interaction,''
Phys. Rev. \textbf{80}, 440-457 (1950)
doi:10.1103/PhysRev.80.440

\bibitem{T2K:2023smv}
K.~Abe \textit{et al.} [T2K],
``Measurements of neutrino oscillation parameters from the T2K experiment using $3.6\times 10^{21}$ protons on target,''
Eur. Phys. J. C \textbf{83}, no.9, 782 (2023)
doi:10.1140/epjc/s10052-023-11819-x
[arXiv:2303.03222 [hep-ex]].


\bibitem{Wolfenstein:1983yz}
L.~Wolfenstein,
``Parametrization of the Kobayashi-Maskawa Matrix,''
Phys. Rev. Lett. \textbf{51}, 1945 (1983)
doi:10.1103/PhysRevLett.51.1945

 \bibitem{inami1981} 
 Inami, T., Lim, C.S. (1981). Progr. Theor. Phys. 65, 297.
 
\bibitem{Bardeen:1978yd}
W.~A.~Bardeen, A.~J.~Buras, D.~W.~Duke and T.~Muta,
``Deep Inelastic Scattering Beyond the Leading Order in Asymptotically Free Gauge Theories,''
Phys. Rev. D \textbf{18}, 3998 (1978)
doi:10.1103/PhysRevD.18.3998

\bibitem{Buras:2020xsm}
A.~Buras, Gauge Theory of Weak Decays,
Cambridge University Press, 2020, ISBN 978-1-139-52410-0, 978-1-107-03403-7
doi:10.1017/9781139524100

\bibitem{ParticleDataGroup:2024cfk}
S.~Navas \textit{et al.} [Particle Data Group],
``Review of particle physics,''
Phys. Rev. D \textbf{110}, no.3, 030001 (2024)
doi:10.1103/PhysRevD.110.030001

 \bibitem{Buras:1998raa}
A.~J.~Buras, Weak Hamiltonian, CP violation and rare decays,
[arXiv:hep-ph/9806471 [hep-ph]].

\bibitem{Langacker:2017uah}
P.~Langacker,
`The Standard Model and Beyond,'
Taylor \& Francis, 2017,
ISBN 978-1-4987-6322-6, 978-1-4987-6321-9, 978-0-367-57344-7, 978-1-315-17062-6
doi:10.1201/b22175

\bibitem{Descotes-Genon:2013vna}
S.~Descotes-Genon, T.~Hurth, J.~Matias and J.~Virto,
``Optimizing the basis of $B\to K^*ll$ observables in the full kinematic range,''
JHEP \textbf{05}, 137 (2013)
doi:10.1007/JHEP05(2013)137
[arXiv:1303.5794 [hep-ph]].

\bibitem{Alok:2012xm}
A.~K.~Alok and S.~Gangal,
$b\to s$Decays in a model with Z-mediated flavor changing neutral current,''
Phys. Rev. D \textbf{86}, 114009 (2012)
doi:10.1103/PhysRevD.86.114009
[arXiv:1209.1987 [hep-ph]]

\bibitem{ALEPH:2000vvi}
R.~Barate \textit{et al.} [ALEPH],
``Measurements of BR ($b \to \tau \bar{\nu_{\tau}}$ X) and BR ($b \to \tau \bar{\nu_{\tau}}$D*+- X) and upper limits on BR ($b \to \tau \bar{\nu_{\tau}}$) and BR ($b \to s \nu \bar{\nu}$),''
Eur. Phys. J. C \textbf{19}, 213-227 (2001)
doi:10.1007/s100520100612
[arXiv:hep-ex/0010022 [hep-ex]].

\bibitem{LHCb:2014cxe}
R.~Aaij \textit{et al.} [LHCb],
``Differential branching fractions and isospin asymmetries of $B \to K^{(*)} \mu^+ \mu^-$ decays,''
JHEP \textbf{06}, 133 (2014)
doi:10.1007/JHEP06(2014)133
[arXiv:1403.8044 [hep-ex]].

\bibitem{LHCb:2022qnv}
R.~Aaij \textit{et al.} [LHCb],
``Test of lepton universality in $b \rightarrow s \ell^+ \ell^-$ decays,''
Phys. Rev. Lett. \textbf{131}, no.5, 051803 (2023)
doi:10.1103/PhysRevLett.131.051803
[arXiv:2212.09152 [hep-ex]].

\bibitem{LHCb:2022vje}
R.~Aaij \textit{et al.} [LHCb],
``Measurement of lepton universality parameters in $B^+\to K^+\ell^+\ell^-$ and $B^0\to K^{*0}\ell^+\ell^-$ decays,''
Phys. Rev. D \textbf{108}, no.3, 032002 (2023)
doi:10.1103/PhysRevD.108.032002
[arXiv:2212.09153 [hep-ex]].

\bibitem{Gubernari:2022hxn}
N.~Gubernari, M.~Reboud, D.~van Dyk and J.~Virto,
``Improved theory predictions and global analysis of exclusive $b \to s\mu^+\mu^-$ processes,''
JHEP \textbf{09}, 133 (2022)
doi:10.1007/JHEP09(2022)133
[arXiv:2206.03797 [hep-ph]].

\bibitem{Czaja:2024the}
M.~Czaja and M.~Misiak,
``Current Status of the Standard Model Prediction for the B$_{s}$ {\textrightarrow} {\ensuremath{\mu}}$^{+}${\ensuremath{\mu}}$^{-}$ Branching Ratio,''
Symmetry \textbf{16}, no.7, 917 (2024)
doi:10.3390/sym16070917
[arXiv:2407.03810 [hep-ph]].

\bibitem{Handoko:1994xw}
L.~T.~Handoko and T.~Morozumi,
`b ---\ensuremath{>} s (d) gamma with a vector-like quark as the fourth generation,''
Mod. Phys. Lett. A \textbf{10}, 309-322 (1995)
[erratum: Mod. Phys. Lett. A \textbf{10}, 1733 (1995)]
doi:10.1142/S021773239500034X
[arXiv:hep-ph/9409240 [hep-ph]].

\bibitem{Buras:2014fpa}
A.~J.~Buras, J.~Girrbach-Noe, C.~Niehoff and D.~M.~Straub,
``$ B\to {K}^{\left(\ast \right)}\nu \overline{\nu} $ decays in the Standard Model and beyond,''
JHEP \textbf{02}, 184 (2015)
doi:10.1007/JHEP02(2015)184
[arXiv:1409.4557 [hep-ph]]

\bibitem{Bobeth2017}
C.~Bobeth, A.~J.~Buras, A.~Celis and M.~Jung,
``Patterns of Flavour Violation in Models with Vector-Like Quarks,''
JHEP \textbf{04}, 079 (2017)
[arXiv:1609.04783 [hep-ph]].

 \bibitem{Ahmady:2001qh}
M.~R.~Ahmady, M.~Nagashima and A.~Sugamoto,
`Inclusive dileptonic rare B decays with an extra generation of vector-like quarks,
Phys. Rev. D \textbf{64}, 054011 (2001)
doi:10.1103/PhysRevD.64.054011
[arXiv:hep-ph/0105049 [hep-ph]].

 \bibitem{Bardeen:1978yd}
W.~A.~Bardeen, A.~J.~Buras, D.~W.~Duke and T.~Muta,
``Deep Inelastic Scattering Beyond the Leading Order in Asymptotically Free Gauge Theories,''
Phys. Rev. D \textbf{18}, 3998 (1978)
doi:10.1103/PhysRevD.18.3998

\bibitem{Vatsyayan:2020jan}
D.~Vatsyayan and A.~Kundu,
`Constraints on the quark mixing matrix with vector-like quarks,''
Nucl. Phys. B \textbf{960}, 115208 (2020)
doi:10.1016/j.nuclphysb.2020.115208
[arXiv:2007.02327 [hep-ph]].

\bibitem{Parrott:2022zte}
W.~G.~Parrott \textit{et al.} [HPQCD],
`Standard Model predictions for B\textrightarrow{}K\ensuremath{\ell}+\ensuremath{\ell}-, B\textrightarrow{}K\ensuremath{\ell}1-\ensuremath{\ell}2+ and B\textrightarrow{}K\ensuremath{\nu}\ensuremath{\nu}\textasciimacron{} using form factors from Nf=2+1+1 lattice QCD,''
Phys. Rev. D \textbf{107}, no.1, 014511 (2023)
[erratum: Phys. Rev. D \textbf{107}, no.11, 119903 (2023)]
doi:10.1103/PhysRevD.107.014511
[arXiv:2207.13371 [hep-ph]].

\bibitem{Belle-II:2023esi}
I.~Adachi \textit{et al.} [Belle-II],
`Evidence for $B^{+}\to K^{+}\nu\bar{\nu}$ Decays,''
[arXiv:2311.14647 [hep-ex]].

\bibitem{Branco:2022gja}
G.~C.~Branco and M.~N.~Rebelo,
``Vector-like Quarks,''
PoS \textbf{DISCRETE2020-2021}, 004 (2022)
doi:10.22323/1.405.0004
[arXiv:2208.07235 [hep-ph]].

\bibitem{Altmannshofer:2017wqy}
W.~Altmannshofer, C.~Niehoff and D.~M.~Straub,
`$B_s\to\mu^+\mu^-$ as current and future probe of new physics,'
JHEP \textbf{05}, 076 (2017)
doi:10.1007/JHEP05(2017)076
[arXiv:1702.05498 [hep-ph]].

\bibitem{Alok:2014yua}
A.~K.~Alok, S.~Banerjee, D.~Kumar and S.~Uma Sankar,
`Flavor signatures of isosinglet vector-like down quark model,''
Nucl. Phys. B \textbf{906}, 321-341 (2016)
doi:10.1016/j.nuclphysb.2016.03.012
[arXiv:1402.1023 [hep-ph]].

\bibitem{Alok:2022pjb}
A.~K.~Alok, N.~R.~Singh Chundawat, S.~Gangal and D.~Kumar,
``A global analysis of $b \rightarrow s \ell \ell $ data in heavy and light $Z'$ models,''
Eur. Phys. J. C \textbf{82}, no.10, 967 (2022)
doi:10.1140/epjc/s10052-022-10816-w
[arXiv:2203.13217 [hep-ph]].


\bibitem{Morozumi:2018cnc}
T.~Morozumi, Y.~Shimizu, S.~Takahashi and H.~Umeeda,
``Effective theory analysis for vector-like quark model,''

PTEP \textbf{2018}, no.4, 043B10 (2018)
doi:10.1093/ptep/pty042
[arXiv:1801.05268 [hep-ph]].

\bibitem{Becirevic:2023aov}
D.~Be\v{c}irevi\'c, G.~Piazza and O.~Sumensari,
`Revisiting $B\rightarrow K^{(*)} \nu {\bar{\nu }}$ decays in the Standard Model and beyond,''
Eur. Phys. J. C \textbf{83}, no.3, 252 (2023)
doi:10.1140/epjc/s10052-023-11388-z
[arXiv:2301.06990 [hep-ph]].

\bibitem{Ishiwata:2015cga}
K.~Ishiwata, Z.~Ligeti and M.~B.~Wise,
``New Vector-Like Fermions and Flavor Physics,''
JHEP \textbf{10}, 027 (2015)
doi:10.1007/JHEP10(2015)027
[arXiv:1506.03484 [hep-ph]].

\bibitem{GlashowIliopoulosMaiani1970}
S.~L.~Glashow, J.~Iliopoulos and L.~Maiani,
``Weak Interactions with Lepton-Hadron Symmetry,''
Phys.\ Rev.\ D \textbf{2}, 1285-1292 (1970).

\bibitem{Botella:2016ibj}
F.~J.~Botella, G.~C.~Branco, M.~Nebot, M.~N.~Rebelo and J.~I.~Silva-Marcos,
``Vector-like Quarks at the Origin of Light Quark Masses and Mixing,''
Eur. Phys. J. C \textbf{77}, no.6, 408 (2017)
doi:10.1140/epjc/s10052-017-4933-3
[arXiv:1610.03018 [hep-ph]].

\bibitem{Aoki:1982ed}
K.~I.~Aoki, Z.~Hioki, M.~Konuma, R.~Kawabe and T.~Muta,
``Electroweak Theory. Framework of On-Shell Renormalization and Study of Higher Order Effects,''
Prog. Theor. Phys. Suppl. \textbf{73}, 1-225 (1982)
doi:10.1143/PTPS.73.1

\bibitem{Capdevila:2023yhq}
B.~Capdevila, A.~Crivellin and J.~Matias,
``Review of Semileptonic $B$ Anomalies,''
Eur. Phys. J. ST \textbf{1}, 20 (2023)
doi:10.1140/epjs/s11734-023-01012-2
[arXiv:2309.01311 [hep-ph]].

\bibitem{Mahmoudi:2023upg}
F.~Mahmoudi, T.~Hurth, D.~Mart{\'\i}nez Santos and S.~Neshatpour,
``Model independent analysis for B anomalies,''
EPJ Web Conf. \textbf{289}, 01002 (2023)
doi:10.1051/epjconf/202328901002

\bibitem{Mahmoudi:2024zna}
F.~Mahmoudi and Y.~Monceaux,
``Overview of B{\textrightarrow}K$^{(*)}${\ensuremath{\ell}}{\ensuremath{\ell}} Theoretical Calculations and Uncertainties,''
Symmetry \textbf{16}, no.8, 1006 (2024)
doi:10.3390/sym16081006
[arXiv:2408.03235 [hep-ph]].


\bibitem{Buchalla:1990qz}
G.~Buchalla, A.~J.~Buras and M.~K.~Harlander,
``Penguin box expansion: Flavor changing neutral current processes and a heavy top quark,''
Nucl. Phys. B \textbf{349}, 1-47 (1991)
doi:10.1016/0550-3213(91)90186-2

\bibitem{Buchalla:1998ba}
G.~Buchalla and A.~J.~Buras,
``The rare decays $K\to \pi \nu\bar\nu$, $B \to X \nu\bar\nu$ and $B \to l^+ l^-$: An Update,''
Nucl. Phys. B \textbf{548}, 309-327 (1999)
doi:10.1016/S0550-3213(99)00149-2
[arXiv:hep-ph/9901288 [hep-ph]].

\bibitem{Banerjee:2024zvg}
A.~Banerjee, E.~Bergeaas Kuutmann, V.~Ellajosyula, R.~Enberg, G.~Ferretti and L.~Panizzi,
``Vector-like quarks: Status and new directions at the LHC,''
SciPost Phys. Core \textbf{7}, 079 (2024)
doi:10.21468/SciPostPhysCore.7.4.079
[arXiv:2406.09193 [hep-ph]].


\bibitem{Ali:1999mm}
A.~Ali, P.~Ball, L.~T.~Handoko and G.~Hiller,
``A Comparative study of the decays $B \to$ ($K$, $K^{*)} \ell^+ \ell^-$ in standard model and supersymmetric theories,''
Phys. Rev. D \textbf{61}, 074024 (2000)
doi:10.1103/PhysRevD.61.074024
[arXiv:hep-ph/9910221 [hep-ph]].

\bibitem{Kim:1989ac}
C.~S.~Kim and A.~D.~Martin,
``On the Determination of $V (u b$) and $V (c b$) From Semileptonic $B$ Decays,''
Phys. Lett. B \textbf{225}, 186-190 (1989)
doi:10.1016/0370-2693(89)91033-2

\bibitem{Bailey:2015dka}
J.~A.~Bailey, A.~Bazavov, C.~Bernard, C.~M.~Bouchard, C.~DeTar, D.~Du, A.~X.~El-Khadra, J.~Foley, E.~D.~Freeland and E.~G{\'a}miz, \textit{et al.}
``$B\to Kl^+l^-$ Decay Form Factors from Three-Flavor Lattice QCD,''
Phys. Rev. D \textbf{93}, no.2, 025026 (2016)
doi:10.1103/PhysRevD.93.025026
[arXiv:1509.06235 [hep-lat]].

\bibitem{Ball:2004ye}
P.~Ball and R.~Zwicky,
``New results on $B \to \pi, K, \eta$ decay formfactors from light-cone sum rules,''
Phys. Rev. D \textbf{71}, 014015 (2005)
doi:10.1103/PhysRevD.71.014015
[arXiv:hep-ph/0406232 [hep-ph]].

\bibitem{Handoko:1994xw}
  L.~T.~Handoko and T.~Morozumi,
  ``b ---> s (d) gamma with a vector - like quark as fourth generation,''
  Mod.\ Phys.\ Lett.\ A {\bf 10} (1995) 309,~
   Erratum: [Mod.\ Phys.\ Lett.\ A {\bf 10} (1995) 1733]
  doi:10.1142/S021773239500034X
  [hep-ph/9409240].
  
 \bibitem{Bednyakov:2015penguin}
A.~V.~Bednyakov and Ş.~H.~Tanyıldızı,
``A Mathematica package for calculation of one-loop penguins in FCNC processes,''
Int.\ J.\ Mod.\ Phys.\ C {\bf 26} (2015) no.04, 1550042,
doi:10.1142/S0129183115500424

\bibitem{Belle:2005fli}
M.~Iwasaki \textit{et al.} [Belle],
``Improved measurement of the electroweak penguin process $B \to X_s l^+ l^-$,''
Phys. Rev. D \textbf{72}, 092005 (2005)
doi:10.1103/PhysRevD.72.092005
[arXiv:hep-ex/0503044 [hep-ex]].

\bibitem{Hou:2024vyw}
B.~F.~Hou, X.~Q.~Li, M.~Shen, Y.~D.~Yang and X.~B.~Yuan,
``Deciphering the Belle II data on $ B\to K\nu \overline{\nu} $ decay in the (dark) SMEFT with minimal flavour violation,''
JHEP \textbf{06}, 172 (2024)
doi:10.1007/JHEP06(2024)172
[arXiv:2402.19208 [hep-ph]].

\bibitem{Descotes-Genon:2012isb}
S.~Descotes-Genon, J.~Matias, M.~Ramon and J.~Virto,
``Implications from clean observables for the binned analysis of $B -> K*\mu^+\mu^-$ at large recoil,''
JHEP \textbf{01}, 048 (2013)
doi:10.1007/JHEP01(2013)048
[arXiv:1207.2753 [hep-ph]].

\bibitem{CMS:2014xfa}
V.~Khachatryan \textit{et al.} [CMS and LHCb],
``Observation of the rare $B^0_s\to \mu^+\mu^-$ decay from the combined analysis of CMS and LHCb data,''
Nature \textbf{522}, 68-72 (2015)
doi:10.1038/nature14474
[arXiv:1411.4413 [hep-ex]].

\bibitem{Santimaria:2018wfw}
M.~Santimaria,
``Search for the $B^0_d \to \mu^+ \mu^-$ decay and measurement of the $B^0_s \to \mu^+ \mu^-$ branching fraction and effective lifetime,''
[arXiv:1703.05747 [hep-ex]]

\bibitem{Altmannshofer:2023hkn}
W.~Altmannshofer, A.~Crivellin, H.~Haigh, G.~Inguglia and J.~Martin Camalich,
``Light new physics in B\textrightarrow{}K(*)\ensuremath{\nu}\ensuremath{\nu}\textasciimacron{}?,''
Phys. Rev. D \textbf{109}, no.7, 075008 (2024)
doi:10.1103/PhysRevD.109.075008
[arXiv:2311.14629 [hep-ph]].

\bibitem{Altmannshofer:2009ma}
W.~Altmannshofer, A.~J.~Buras, D.~M.~Straub and M.~Wick,
"New strategies for New Physics search in $B \to K^{*} \nu \bar{\nu}$, $B \to K \nu \bar{\nu}$ and $B \to X_{s} \nu \bar{\nu}$ decays,''
JHEP \textbf{04}, 022 (2009)
doi:10.1088/1126-6708/2009/04/022
[arXiv:0902.0160 [hep-ph]].

\bibitem{Chang:1998sk}
C.~H.~V.~Chang, D.~Chang and W.~Y.~Keung,
``Vector quark model and B meson radiative decay,''
[arXiv:hep-ph/9811354 [hep-ph]]


\end{thebibliography}
\end{document}